%% file: paper.tex
\documentstyle[mnextra]{mn}
\input{epsf}
\input{defs.tex}

\nobrackets    % no brackets in cite macros
\seceqnum      % equations numbered by section
\begin{document}

\title[Galaxy Formation]{Hierarchical Galaxy Formation}
\author[S. Cole et al.]{Shaun Cole$^{1,4}$, Cedric G. Lacey$^{1,2,3,5}$, 
Carlton M. Baugh$^{1,6}$ and Carlos S. Frenk$^{1,7}$ \\
$^1$Department of Physics, University of Durham, 
Science Laboratories, South Rd, Durham DH1 3LE. \\
$^2$Theoretical Astrophysics Center, Juliane Maries Vej 30, DK-2100 Copenhagen, Denmark. \\
$^3$SISSA, via Beirut 2-4, 34014 Trieste, Italy\\
$^4$Shaun.Cole@durham.ac.uk\\
$^5$lacey@sissa.it \\
$^6$C.M.Baugh@durham.ac.uk\\
$^7$C.S.Frenk@durham.ac.uk\\
}

\maketitle

%%%%%%%%%%%%%%%%%%%%%%%%%%%%%%%%%%%%%%%%%%%%%%%%%%%%%%%%%%%%%%%%%%%%%%%%%%
\begin{abstract}
We describe the GALFORM semi-analytic model for calculating the formation and
evolution of galaxies in hierarchical clustering cosmologies. It
improves upon, and extends, the earlier scheme developed by Cole \etal
(\shortcite{cafnz}). 
The model employs a new Monte-Carlo algorithm
to follow the merging evolution of dark matter halos with
arbitrary mass resolution. It incorporates realistic descriptions of
the density profiles of dark matter halos and the gas they contain; it
follows the chemical evolution of gas and stars, and the associated
production of dust; and it includes a detailed calculation of the
sizes of disks and spheroids. Wherever possible, our prescriptions for
modelling individual physical processes are based on results of
numerical simulations. They require a number of adjustable parameters
which we fix by reference to a small subset of local galaxy data. This
results in a fully specified model of galaxy formation which can be
tested against other data. We apply our methods to the $\Lambda$CDM
cosmology ($\Omega_0=0.3$, $\Lambda_0=0.7$), and find good agreement
with a wide range of properties of the local galaxy population: the
B-band and K-band luminosity functions, the distribution of colours
for the population as a whole, the ratio of ellipticals to spirals,
the distribution of disk sizes, and the current cold gas content of disks.
%and the observed metallicity-luminosity relation for star-forming gas
%in disk-dominated galaxies, and for stars in bulge-dominated galaxies. 
In spite of the overall success of the model, some interesting
discrepancies remain: the colour-magnitude relation for ellipticals in
clusters is significantly flatter than observed at bright magnitudes
(although the scatter is about right), and the model predicts galaxy
circular velocities, at a given luminosity, that are about 30\% larger
than is observed.  It is unclear whether these discrepancies represent
fundamental shortcomings of the model or whether they result from the
various approximations and uncertainties inherent in the
technique. Our more detailed methods do not change our earlier
conclusion that just over half the stars in the universe are expected to
have formed since $z\lsim 1.5$.
\end{abstract}


\input{intro}

\input{model}

\input{method}

\input{lcdm}

\input{results}

\input{disc}

%%%%%%%%%%%%%%%%%%%%%%%%%%%%%%%%%%%%%%%%%%%%%%%%%%%%%%%%%%%%%%%%%%%%%%%%%%%%%%
\section*{Acknowledgements}

SMC acknowledges the support of a PPARC Advanced Fellowship and CSF a
PPARC Senior Fellowship and a Leverhulme Research Fellowship. CGL
acknowledges the support of the Danish National Research Foundation
through its establishment of the Theoretical Astrophysics Center, and
a PPARC Visiting Fellowship.  This work was partially supported by the
PPARC rolling grant for extragalactic astronomy and cosmology at
Durham, and by the EC TMR Network on ``Galaxy Formation and
Evolution''. We thank Stephane Charlot for providing us with his
stellar population synthesis models, and Andrea Ferrara and Simone
Bianchi for providing us with their dust models in advance of
publication. We thank Jon Gardner for providing us with his redshift
survey data, and Bepi Tormen for providing us with his data on the
orbits of satellite halos in simulations.  Finally, we thank Simon
White for many useful discussions, and he, Eric Bell, Fabio Governato
and David Weinberg for their detailed comments on an earlier draft of
this paper.

\appendix
\input{app_halo}

\input{app_sf}
\input{app_con}

\end{document}

%% file: defs.tex
\def\gsim{\ga}
\def\lsim{\la}
\def\logh{\log h}
\def\lambdah {\lambda_{\rm H}}
\def\BJ {${\rm b_{\rm J}}$}
\def\Jh {J_{\rm H}}
\def\Msol{M_\odot}
\def\Mh {M_{\rm H}}
\def\Eh {E_{\rm H}}
\def\Th {T_{\rm H}}
\def\Wh {W_{\rm H}}
\def\Vh { V_{\rm H}}
\def\Vhot { V_{\rm hot}}
\def\Vhalo { V_{\rm halo}}
\def\Vd { V_{\rm disk}}
\def\Vdisk { V_{\rm disk}}
\def\taud { \tau_{\rm disk}}

\def\G {{\rm G}}
\def\vrot { V_{\rm rot}}
\def\rvir { {r_{\rm vir}}}
\def\Tvir { {T_{\rm vir}}}
\def\Deltavir { {\Delta_{\rm vir}}}

\def\cm{{\rm cm}}
\def\pc{{\rm pc}}
\def\kms{{\rm km\,s^{-1}}}
\def\mag{{\rm mag\,}}
\def\eg{{\rm e.g.\ }}
\def\ie{{\rm i.e.\ }}

\def\etal{{\rm et al.\ }}
\def\Omegab{\Omega_{\rm b}}

\def\rcool{r_{\rm cool}}
\def\crate{\dot M_{\rm cool}}
\def\macc{M_{\rm acc}}
\def\tauf{\tau_{\rm eff}}
\def\zhalo{z_{\rm halo}}
\def\nbody{N-body\ }
\def\anfw{a_{\rm NFW}}
\def\rnfw{r_{\rm NFW}}
\def\rhocrit{\rho_{\rm crit}}
\def\grad{\nabla}
\def\tdyn{\tau_{\rm dyn}}
\def\taumerge{\tau_{\rm mrg}}
\def\fdf{f_{\rm df}}
\def\vhot{V_{\rm hot}}
\def\alphahot{\alpha_{\rm hot}}
\def\alphastar{\alpha_{\star}}
\def\taudf{\Theta_{\rm orbit}}

\def\fellip{f_{\rm ellip}}
\def\CAFNZ{Cole \etal} 
\def\hD{h_{\rm D}}
\def\MD{M_{\rm D}}
\def\MB{M_{\rm B}}
\def\rD{r_{\rm D}}
\def\rDz{r_{\rm D0}}
\def\rBz{r_{\rm B0}}
\def\rB{r_{\rm B}}
\def\jD{j_{\rm D}}
\def\jB{j_{\rm B}}
\def\kD{k_{\rm D}} 
 
\def\kh{k_{\rm h}} 
\def\MHz {M_{\rm H0}}
\def\MH {M_{\rm H}}
\def\VcHz {V_{\rm c0}}
\def\VcH {V_{\rm c}}
\def\VcD{V_{\rm cD}}
\def\VcB{V_{\rm c}}
\def\fH{f_{\rm H}}

\def\Mdisk{M_{\rm disk}}
\def\Mbulge{M_{\rm bulge}}
\def\rdisk{r_{\rm disk}}
\def\rbulge{r_{\rm bulge}}
\def\simprop{ \lower .75ex \hbox{$\sim$} \llap{\raise .27ex \hbox{$\propto$}} }

%% file: intro.tex
\section{Introduction}

The past few years have been a remarkably rich period in observational
studies of galaxy formation. Major advances have resulted from
observations at many wavelengths, from the far ultraviolet to the
sub-millimeter. Breakthroughs include the discovery and measurement of the
clustering of ``Lyman-break'' galaxies, a population of luminous,
star-forming galaxies at redshifts $z\sim 3-4$ (Steidel \etal 
\shortcite{steidel86}; Adelberger \etal \shortcite{adelberger96});
estimates of the history of star formation and the attendant
production of metals, from $z \sim 5$ to the present (Madau \etal
\shortcite{madau96},\shortcite{madau98}); 
measurements of the galaxy luminosity function
at $z \sim 0.5-1$ (\cite{lilly96,ellis96}) and $z\sim 3-4$
(\cite{steidel99}); the discovery of a population of bright
sub-millimeter sources, some of which, at least, appear to be dusty,
star-forming galaxies at $z \gsim 2$ (\cite{ivison98}).  All of these
and many other observations are beginning to sketch out an empirical
picture of galaxy evolution.

On their own, the data provide only a partial description of specific
stages of galaxy evolution. To develop a physical understanding of the
processes at work, and to relate observations to cosmological theory,
requires detailed modelling that exploits our current understanding of
astrophysical processes in their cosmological context. The theoretical
infrastructure required for this programme has been in place for over
a decade (\eg \cite{bfpr84,defw85}). 
In its standard form, it assumes that galaxies grew out of
primordial Gaussian density fluctuations generated during inflation
and amplified by gravitational instability acting on cold dark matter,
the dominant mass component of the Universe. Gas is initially mixed in
with the dark matter, and when dark matter halos collapse, the visible
component of galaxies accumulates as stars condense out of gas that
has cooled onto a disk.

To construct a theory of galaxy formation that can be tested against
observations requires combining the theory of the evolution of
cosmological density perturbations with a description of various
astrophysical processes such as the cooling of gas in halos, the
formation of stars, the feedback effects on interstellar gas of energy
released by young stars, the production of heavy elements, 
the evolution of stellar populations, the effects
of dust, and the merging of galaxies. The most appropriate methodology is to carry out {\it ab
initio} calculations that follow directly the development of
primordial density fluctuations into luminous galaxies. Within the
standard cosmological model, the initial conditions are very well
defined. They are specified by the power spectrum of primordial density
perturbations, whose shape is fixed by the cosmological parameters:
the mean mass density, $\Omega_0$, the mean baryon density, $\Omegab$,
the cosmological constant, $\Lambda_0$, and the Hubble constant, 
$H_0$ (which, throughout this paper, we express as $H_0=100 
h {\rm\ km s}^{-1} {\rm Mpc}^{-1}$.)

The subsequent evolution of the dark matter and baryons is best
calculated by Monte Carlo simulation. Two different approaches have
been developed for this purpose.  In the first, direct simulations,
the gravitational and hydrodynamical equations in the expanding
universe are solved explicitly, using one or more of a variety of
numerical techniques that have been specifically developed for this
purpose over the past twenty years
(e.g. \cite{katz92,evrard94,frenk96},
\shortcite{frenk99}; \cite{katz96,ns97,pearce00,thacker99,blanton99}).
In the second approach, now commonly known as ``semi-analytic
modelling of galaxy formation'' (\cite{wr78,wf91,kwg93,cafnz}), the
evolution of the baryonic component is calculated using simple
analytic models, while the evolution of the dark matter is calculated
either directly, using N-body methods, or using a Monte-Carlo
technique that follows the formation of dark matter halos by
hierarchical merging. It is this second approach that we discuss in
this paper.

The two modelling techniques have complementary strengths. The major
advantage of direct simulations is that the dynamics of the cooling
gas are calculated in full generality, without the need for
simplifying assumptions. The main disadvantage is that even with the
best codes and fastest computers available today, the attainable
resolution is still some orders of magnitude below that required to
resolve the formation and internal structure of individual galaxies in
cosmological volumes. In addition, a phenomenological model, similar
to that employed in semi-analytic modelling, is required to include
star formation and feedback processes in the simulation. These processes
are, in fact, much more difficult to treat and much more uncertain
than the dynamics of the diffuse gas.

Semi-analytic modelling does not suffer from resolution limitations,
particularly when Monte-Carlo methods are used to generate the halo merger
histories. In this case, the resolution can be made arbitrarily high
at a relatively small computational cost. The major disadvantage is the
need for simplifying assumptions in the calculation of gas properties,
such as spherical symmetry or a particular flow structure.  It is
encouraging that detailed comparisons between direct and semi-analytic
simulations show good agreement (Pearce
\etal \shortcite{pearce00}, Benson \etal \shortcite{benson00}). 
An important advantage of semi-analytic
modelling is its flexibility. This allows the effects of varying
assumptions or parameter choices to be readily investigated and makes it
possible to calculate a wide range of observable galaxy properties, such as
luminosities in any waveband, sizes, mass-to-light ratios,
bulge-to-disk ratios, circular velocities, etc.

Semi-analytic modelling has its roots in the work of White \& Rees
(\shortcite{wr78}), 
Cole (\shortcite{cole91}), Lacey \& Silk ({\shortcite{ls91}), 
and White \& Frenk (\shortcite{wf91}) who
laid out the overall philosophy and basic methodology of this
approach. Throughout most of the 1990s, this technique was developed and
promoted primarily by two collaborations, one currently based at Munich
(e.g. Kauffmann \etal \shortcite{kwg93},\shortcite{kauffmann94}
Kauffmann \shortcite{k95a},b,
Kauffmann, Nusser \& Steinmetz \shortcite{kns97},
Mo, Mao \& White \shortcite{mmw98},b,\shortcite{mmw99},
Kauffmann \etal \shortcite{kcdw98} 
%,Diaferio \etal \shortcite{dkcw99}
), and
the other at Durham (e.g. Cole \etal \shortcite{cafnz}, Heyl \etal 
\shortcite{hcfn}, 
Baugh \etal \shortcite{bcf96a},b, \shortcite{bcfl98}, Benson \etal
\shortcite{bcfbl99a}; see also Lacey \etal \shortcite{lgrs93}).  In
the past two years, several other groups have begun to apply this
technique to study various aspects of galaxy formation (e.g. Avila-Reese \&
Firmani \shortcite{a-rf99}, Guiderdoni \etal \shortcite{guiderdoni99},
Wu, Fabian \& Nulsen \shortcite{wfn98}, van Kampen, Jimenez \& Peacock 
\shortcite{kjp99}, Somerville \& Primack \shortcite{sp98}).  
This body of work has demonstrated the
usefulness of semi-analytic modelling as a means for fleshing out the
observable consequences of current cosmological theories and for the
interpretation of observational data, particularly at high redshift.

A growing body of galaxy properties has been analysed using
semi-analytic methods.  Examples of noteworthy successes include the
ability to reproduce the local field galaxy luminosity function, the
slope and scatter of the Tully-Fisher relation for spiral galaxies,
and the counts and redshift distributions of faint galaxies (e.g. see
\cite{kwg93,cafnz,kauffmann94}).  Nevertheless, some important
properties have remained obstinately difficult to reproduce, most
notably the colour-magnitude relation for cluster ellipticals
(but see Kauffmann \& Charlot \shortcite{kc98}) and a simultaneous fit to
the local luminosity function and the zero-point of the Tully-Fisher
relation (e.g. Heyl \etal \shortcite{hcfn}).

\begin{figure*}
\centering
\centerline{\epsfxsize=15 truecm \epsfbox[10 20 600 760]{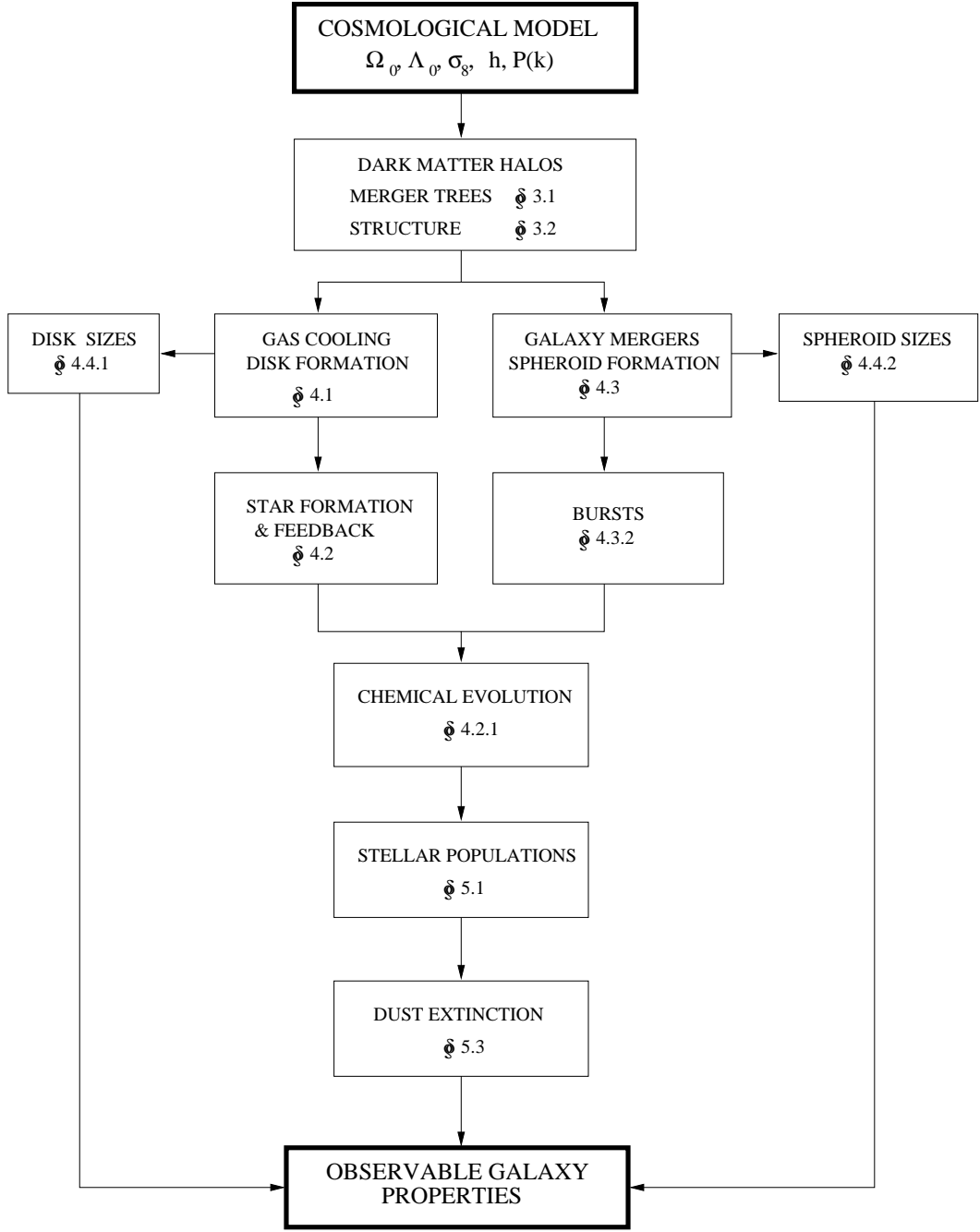}}
\caption{A schematic showing how different physical processes are
combined to make predictions for the observable properties of
galaxies, starting from initial conditions specified by the cosmology. The
numbers in each box indicate the subsection of the paper in which our
method for modelling that process is described.}
\label{fig:model}
\end{figure*}

\begin{figure*}
\centering
\centerline{\epsfxsize=17 truecm  \epsfbox[42 220 560 576]{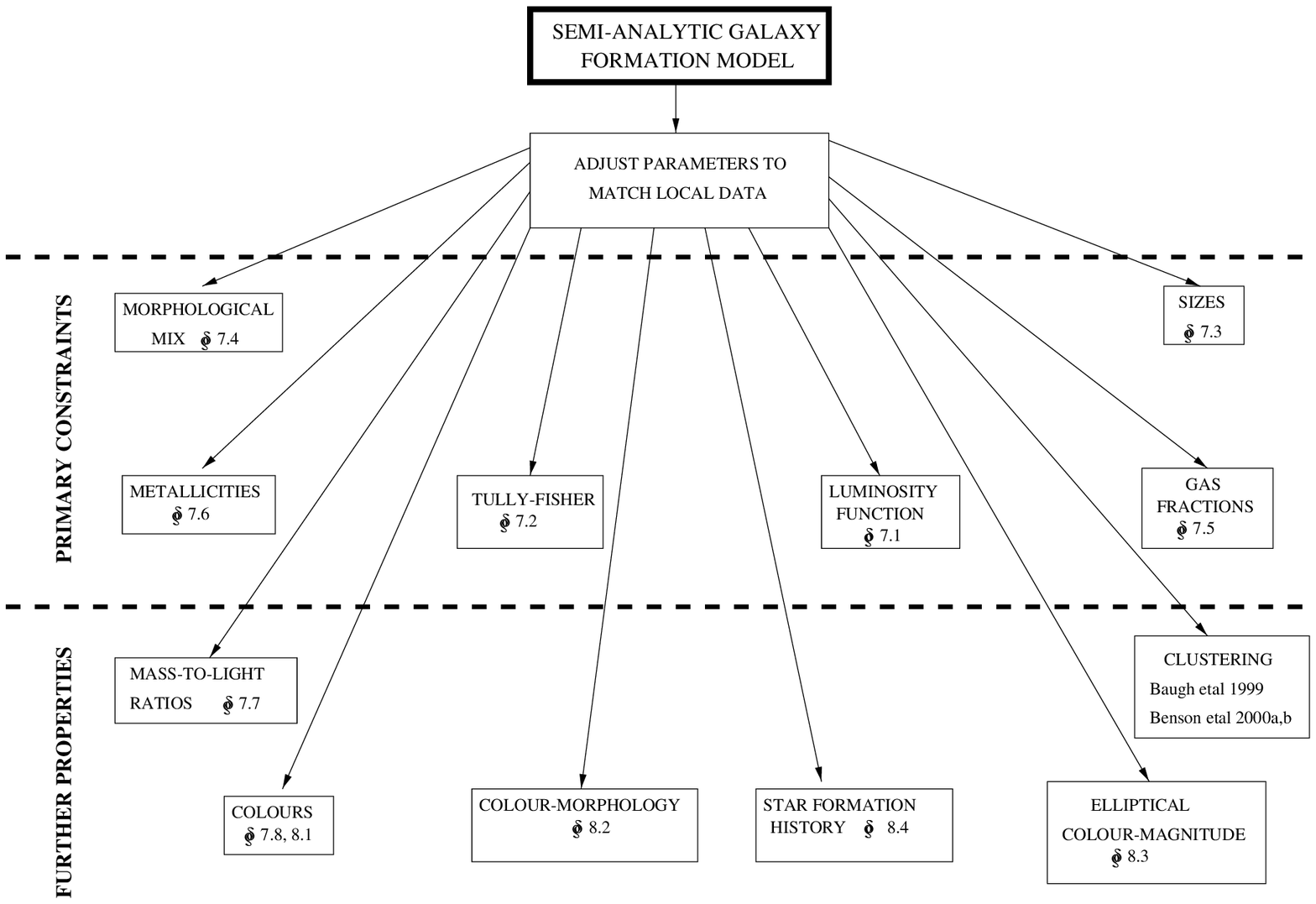}}
\caption{A schematic showing the observable galaxy properties predicted by
the model. The numbers in each box indicate the subsection of the
paper in which the comparison of model predictions with observations
of that property are described. The predictions for galaxy clustering
are described in separate papers, as indicated in the box.}
\label{fig:outputs}
\end{figure*}

A wide variety of physical processes are involved in the formation of
galaxies. Some of them, like star formation, are very poorly understood.
Modelling galaxy formation therefore inevitably requires making
approximations and adopting simplified descriptions of some of these
processes. Most often, an incomplete understanding of a physical ingredient is
subsumed within a simple scaling law that contains free parameters. A
remarkable facet of modern semi-analytic modelling is that a realistic
picture of galaxy evolution can be formulated with a relatively small
number of such parameters, typically four or five in the simplest
versions. A strategy that has proved useful is to fix the values of these
parameters by trying to match a subset of local data (for example, the
luminosity functions in two passbands or the Tully-Fisher relation). This
leads to a completely specified model that has predictive power and may be
used to calculate theoretical expectations for other local properties or
for properties at high redshift. This approach has met with considerable
success. For example, Cole \etal (\shortcite{cafnz}) predicted that
most of the stars in the universe formed at relatively low redshift
($z\lsim 1$ for an $\Omega_0=1$ standard CDM cosmology);
Kauffmann (\shortcite{k96}) predicted a sharply declining
number of bright elliptical galaxies at high redshift; and 
Baugh \etal (\shortcite{bcfl98})  and Governato \etal (\shortcite{gov98})
predicted strong clustering for Lyman-break galaxies at
redshift $z\simeq 3$.

In this paper we present a new semi-analytic model which builds upon the
scheme described by Cole \etal (\shortcite{cafnz}) which we used for a
number of applications (Heyl \etal \shortcite{hcfn}, Baugh \etal
\shortcite{bcf96a}b).  Our new model differs from the earlier one
primarily in its greater scope and richness, but also in the manner in
which certain key physical properties are calculated. These improvements
are called for both by recent theoretical developments and, most
importantly, by the increase in the quantity and quality of observational
data.  The main additions to our new code are the inclusion of chemical
enrichment and dust processes, prescriptions for calculating the sizes of
disks and spheroids, the use of more realistic density profiles for
dark matter halos and gas, and the ability to follow the mergers of halos
with fine mass and time resolution. It should be noted that despite all 
these improvements, when one adopts the same cosmological parameters 
and also the same galaxy formation parameters (\eg for stellar feedback),
then the main  predictions of the model, including the galaxy luminosity
function, Tully-Fisher relation and overall star formation history,
are practically identical to those in Cole \etal (\shortcite{cafnz}).
The one exception is the inclusion of dust extinction which makes the galaxy
colours somewhat redder than in the Cole \etal (\shortcite{cafnz}) models.
Thus, the changes to the model may be viewed as refinements that allow
more properties of the galaxy population, such as galaxy sizes and
metallicities, to be calculated.

The main aim of this paper is to lay out the methods that we use in our new
semi-analytic model and to compare results with a restricted set
of observational data. This is a long paper containing a mixture of
technical descriptions and results of more general interest. In the
following, brief section, we present an overview, together with schematics
illustrating how different parts of the model fit together.  Non-specialist
readers may wish to skip the more detailed passages of the paper on a first
reading and we recommend how this might done in Section~\ref{sec:overview}.

%%%%%%%%%%%%%%%%%%%%%%%%%%%%%%%%%%%%%%%%%%%%%%%%%%%%%%%%%%%%%%%%%%%%%%%%%%%
%%%%%%%%%%%%%%%%%%%%%%%%%%%%%%%%%%%%%%%%%%%%%%%%%%%%%%%%%%%%%%%%%%%%%%%%%%%

\section{Overview}
\label{sec:overview}

Our galaxy formation model is a synthesis of many techniques, 
each of which has been developed to treat particular aspects of
the complex process of galaxy formation. Its backbone 
is a Monte Carlo method for generating ``merger trees'' to
describe the hierarchical growth of dark matter halos. 
The full range of properties and processes that we model within 
this framework are:

\begin{enumerate}
\item The gravitationally-driven formation and merging of dark matter halos.

\item The density and angular momentum
profiles of dark matter and shock heated
gas within dense non-linear halos.

\item The radiative cooling of gas and its collapse to form centrifugally
supported disks.

\item The scalelengths of disks based on angular
momentum conservation and including the effect of the adiabatic 
contraction of the surrounding halo during the formation of the disk.

\item Star formation in disks. 

\item Feedback, \ie the regulation of the star formation rate
resulting from the injection of
supernova (SN) energy into the interstellar medium (ISM).

\item Chemical enrichment of the ISM and hot halo gas
and its influence on both the gas cooling rates and the
properties of the stellar populations that are formed.

\item The frequency of galaxy mergers resulting from 
dynamical friction operating
on galaxies as they orbit within common dark matter halos.

\item The formation of galactic spheroids, accompanied
by bursts of star formation, during violent galaxy-galaxy 
mergers, and estimates of their effective radii.

\item Spectrophotometric evolution of the stellar populations.

\item The effect of dust extinction on galaxy luminosities
and colours, and its dependence on galaxy inclination.

\item The generation of emission lines from interstellar gas ionized
by young stars.

\end{enumerate}
Our treatment of each of these processes is described in the
following sections. The model is summarized schematically in
Fig.~\ref{fig:model}, which also shows in which subsection of the
paper each of the processes is described.

The scheme we present is largely modular, and within each module one
has the choice of selecting various options as well as certain
parameter values. The options (for example, including or ignoring the
baryonic mass of the galaxy when computing its rotation curve) are not
degrees of freedom within the model. Instead they allow us to vary the
complexity of the description in order to gain physical
understanding. By switching processes on and off and changing certain
assumptions we are able to determine which physical processes are
directly responsible for a particular galaxy property.

The observable galaxy properties predicted by the model are shown
schematically in Fig.~\ref{fig:outputs}. This figure separates
the predicted quantities into two categories. Some aspects of
the observational quantities in the first category
are used as primary constraints on the model parameters. 
The boxes in the figure also indicate in which 
subsection of the paper the observational comparison for that 
quantity is presented, or in the case of galaxy clustering, the
related papers in which they are presented. Predictions for the
redshift evolution of galaxy properties will be presented in future papers.

The layout of the paper is as follows: Section~\ref{sec:halos}
presents techniques for generating merger trees to describe the
gravitational growth of dark matter halos and models for their 
internal structure.
Section~\ref{sec:dsksphform} describes how we calculate disk
and spheroid formation, including star formation, feedback and
chemical evolution, and the scalelengths of galactic disks and
bulges. Section~\ref{sec:lum} presents our methods for calculating the
luminosities of stellar populations, and the effects of dust
extinction.  These three sections may be skipped on a first reading,
simply noting the definition of the parameters that describe our star
formation and feedback model, equations~(\ref{eq:sfr}), (\ref{eq:betadef}),
(\ref{eqn:sflaw}), and~(\ref{eqn:feedback}). An overview of the model
and the strategy we adopt to constrain parameters and to obtain a
well-specified, predictive model are presented in
Section~\ref{sec:method}, which should be of general interest.  This
procedure is implemented in Section~\ref{sec:cons}, where we
illustrate the effects of varying each model parameter. The general
reader may simply wish to study the figures in this section and read
the summary in subsection~\ref{sec:sumcons}.  In
Section~\ref{sec:results} we test our fully specified model against
further properties of the observed galaxy population. Again, the
general reader may wish to study only the figures in this
section. Finally, in Section~\ref{sec:disc} we restate the general
philosophy of our approach and discuss the strengths and weaknesses of
the specific model that we have presented. This, and the final section
presenting a summary of our results, are both of general interest.

%% file: model.tex
\section{Formation of Dark Matter Halos}
\label{sec:halos}

Galaxies are assumed to form inside dark matter halos, and their
subsequent evolution is controlled by the merging histories of the
halos containing them. It is therefore essential to have an accurate
description of how dark halos form and evolve through hierarchical
merging, and of the internal structure of these halos. These are both
described in this section.

\subsection{Dark Matter Halo Merger Trees}

We use a new Monte-Carlo algorithm to generate merger
trees that describe the formation paths of randomly selected dark
matter halos. It is an improvement over the ``block model'' that we
used previously (\cite{cafnz,hcfn,bcf96a},b).  The new algorithm is
directly based on the analytic expression for halo merger rates
derived by Lacey \& Cole (\shortcite{lc93}).  At each branch in the
tree, a halo splits into two progenitors, but unlike in the ``block
model'', the mass ratio of the progenitors can take any value. Below,
we briefly describe this new algorithm, and the way in which a
population of merger trees is set up to provide a framework
for modelling the processes of galaxy formation.

It is possible to generate merger trees directly by following the evolution 
of dark matter halos in collisionless cosmological N-body simulations (\eg 
Roukema \shortcite{roukema97}; 
Kauffmann \etal \shortcite{kcdw98}; 
van Kampen \etal \shortcite{kjp99}).
Combining semi-analytic modelling and N-body simulations certainly
provides a very powerful technique to investigate small scale galaxy
clustering (\eg Kauffmann, Nusser \& Steinmetz \shortcite{kns97},
Governato \etal \shortcite{gov98}, 
Kauffmann \etal \shortcite{kcdw98},b,
Benson \etal \shortcite{bcfbl99a},b). However, 
extracting merger trees directlty from  N-body simulations carries the
high price of a limited dynamic range in mass and much greater
computational complexity.  Also, for many applications this appears 
to be unnecessary, as 
the properties of halo
merger trees are uncorrelated with environment (Lemson \& Kauffmann
\shortcite{lk99}), and the Monte Carlo merger trees agree well,
statistically,  with
those extracted from N-body simulations
(\cite{kw93,lc94,slkd99}; Lacey \& Cole in preparation).

\subsubsection{A New Algorithm} 

Our starting point for generating a merger tree to describe the 
history of mergers experienced by an individual dark matter halo is 
equation (2.15) of Lacey \& Cole (\shortcite{lc93}):
\begin{eqnarray}
f_{12}(M_1,M_2) dM_1 \negthinspace &=& \negthinspace \negthinspace 
\frac{1}{\sqrt{2 \pi}} \ 
\frac{(\delta_{\rm c 1}-\delta_{\rm c 2})^{{\hphantom{3/2}}}} 
{(\sigma_1^2 -\sigma_2^2)^{3/2}} \nonumber \\ 
&\times& \negthinspace \negthinspace 
\exp\left( - \frac{(\delta_{\rm c 1}-\delta_{\rm c 2})^2}
{2(\sigma_1^2 -\sigma_2^2)\hphantom{2}} \right) 
\frac{d\sigma_1^{2}}{dM_1} dM_1 .
\label{eqn:eps}
\end{eqnarray}
This equation, derived from the extension of the Press \& Schechter
(\shortcite{ps74}) theory proposed by Bond \etal (\shortcite{bond91})
and Bower (\shortcite{bower91}), gives the fraction of mass,
$f_{12}(M_1,M_2)dM_1$, in halos of mass $M_2$, at time $t_2$, which at
an earlier time, $t_1$, was in halos of mass in the range $M_1$ to
$M_1+dM_1$.  Here, the quantities $\sigma_1$ and $\sigma_2$ are the
linear theory rms density fluctuations in spheres of mass $M_1$ and
$M_2$. The $\delta_{\rm c 1}$ and $\delta_{\rm c 2}$ are the critical
thresholds on the linear overdensity for collapse at times $t_1$ and
$t_2$ respectively. (Specifically, $\sigma(M)$ and $\delta_{\rm c
}(t)$ are the values extrapolated to $z=0$ according to linear
theory.) For a critical density ($\Omega=1$) universe, we adopt
$\delta_{\rm c }= 1.686(1+z)$, while for low $\Omega_0$, open and
flat, universes we adopt the appropriate expressions in the appendices
of Lacey \& Cole (\shortcite{lc93}) and Eke, Cole \& Frenk
(\shortcite{ecf96}) respectively.

Equation (\ref{eqn:eps}) can be used to estimate the recent merger
histories of a set of halos which at time $t_2$ have mass
$M_2$. Taking the limit of equation (\ref{eqn:eps}) as $t_1
\rightarrow t_2$, we obtain an expression for the average mass fraction of a
halo of mass $M_2$ which was in halos of mass $M_1$ at the slightly
earlier time $t_1$,
\begin{equation}
\frac{ d f_{12}}{d t_1} \Big|_{t_1=t_2} \negthinspace
\negthinspace \negthinspace \negthinspace
\negthinspace \negthinspace \negthinspace
dM_1 dt_1 =  \frac{1}{\sqrt{2 \pi}} 
\frac{1}{(\sigma_1^2 -\sigma_2^2)^{3/2}} 
\frac{d \delta_{\rm c 1}}{dt_1}   
\frac{d \sigma_1^2}{dM_1} dM_1  dt_1.
\end{equation}
Thus, the mean number of objects of mass $M_1$ that a halo of mass
$M_2$ ``fragments into'' when one takes a step $dt_1$ back in time is
given by
\begin{equation}
\frac{dN}{dM_1} = \frac{ d f_{12}}{d t_1} \frac{M_2}{M_1}  dt_1 
\qquad\qquad\qquad (M_1<M_2).
\label{eqn:split}
\end{equation}
This expression gives the average number of progenitors as a function
of the fragment mass, $M_1$.  It is this simple expression that our
algorithm uses to build a binary merger tree.

The quantities that must be specified in order to define the merger tree
are the density fluctuation power spectrum, which gives the function
$\sigma(M)$, and the cosmological parameters, $\Omega_0$ and $\Lambda_0$,
which enter through the dependence of $\delta_c(t)$ on the cosmological
model.  There is also one numerical parameter involved, the mass
resolution, $M_{\rm res}$. Having specified these
parameters one can compute the two quantities,
\begin{equation}
P = \int_{M_{\rm res}}^{M_2/2} \frac{dN}{dM_1} dM_1,
\label{eqn:merge}
\end{equation}
which is the mean number of fragments
with masses, $M_1$, in the range $M_{\rm res}<M_1<M_2/2$, and
\begin{equation}
F = \int_0^{M_{\rm res}} \frac{dN}{dM_1} \frac{M_1}{M_2} dM_1,
\label{eqn:accr}
\end{equation}
which is the fraction of mass in fragments with mass below the
resolution limit.  Note that both these quantities are proportional to
the timestep $dt_1$, through the dependence in equation
(\ref{eqn:split}).

Once these quantities have been defined, the algorithm to generate the
merger trees is simple.  First, choose a timestep, $dt_1$, such that
$P\ll 1$, to ensure that multiple fragmentation is unlikely. Next,
generate a random number, $R$, drawn uniformly from the interval 0 to
1. If $R>P$, then the main halo does not fragment at this
timestep. However, the original mass is reduced to account for mass
accreted in the form of halos with masses below the resolution limit,
to produce a new halo of mass $M_2(1-F)$.  If, however, $R<P$, then a
random value of $M_1$ in the range $M_{\rm res}<M_1<M_2/2$ is
generated, consistent with the distribution given by equation
(\ref{eqn:split}), to produce two new halos with masses $M_1$ and
$M_2(1-F)-M_1$. The same operation is repeated on each fragment at
successive timesteps going back in time, and thus a merger tree is built up.

The main advantage of this new algorithm over the ``block model'' that
we used previously (\cite{cafnz,hcfn,bcf96a}b), and which has been used
recently by Wu \etal (\shortcite{wfn98},\shortcite{wfn99}), is that
there is no quantization of the progenitor halo masses.  The algorithm
also enables the merger process to be followed with high time
resolution, as timesteps are not imposed on the tree but rather are
controlled directly by the frequency of mergers.  It is similar in
spirit to the method used by Kauffmann \etal (\shortcite{kwg93}), but
has several advantages, including smaller timesteps and not having to
store large tables of progenitor distributions.  Somerville \& Kolatt
(\shortcite{sk99}) investigated a similar algorithm also based on
equation (\ref{eqn:eps}), which they referred to as binary mergers
without accretion. They rejected that algorithm as it over-predicted
the number of massive halos at high redshift, and instead opted for a
more elaborate algorithm which they compared with N-body simulations
in Somerville \etal (\shortcite{slkd99}). Our algorithm, which we first used
in Baugh \etal (\shortcite{bcfl98}), differs from the one they
rejected in two important respects. First, we explicitly account for
accretion of objects below the mass resolution
using the expression (\ref{eqn:accr}). Second, we make the
rather subtle choice of selecting the first progenitor mass, $M_1$,
only in the range $M_{\rm res}<M_1<M/2$ of the distribution defined by
(\ref{eqn:split}).  We have found, that together, these choices produce an
algorithm that successfully produces distributions of progenitors
which, on average, agree quite accurately with the analytic expressions
given by the extended Press-Schechter theory. Moreover, statistics that
are not predicted by the extended Press-Schechter theory, such as the
frequency distribution of progenitors of a given mass (the extended
Press-Schechter theory only predicts the mean of the distribution), we
find to be in excellent agreement with the same statistics extracted
from N-body simulations.  This detailed investigation of the behaviour
of the algorithm will be presented in a future paper (Lacey \& Cole in
preparation).

\subsubsection{Utilizing the Merger Trees} 

Although the merger trees described above have very high time
resolution, the nature of the galaxy formation rules that we implement
below require placing the merger tree onto a predefined grid of
timesteps.  The original binary merger tree is used to find which
halos exist at each timestep and to identify which of them merge
together between timesteps. As a consequence, mergers that are
actually rapid, consecutive binary mergers in the original tree will
appear as simultaneous, multiple mergers in the discretized tree.  The
loss of information involved is not significant since, in reality,
mergers are not instantaneous events and our discrete timesteps are
typically much smaller than the dynamical timescales of the merging
halos. Each merger tree thus starts from a single halo of a specified
mass $M$ at $z=\zhalo$, where $\zhalo$ is the redshift for which we
want to calculate the galaxy properties. It extends up to some earlier
redshift $z_{\rm start} > \zhalo$, where the tree has split into many
branches. The generation of the halo merger tree proceeds backwards in
time, starting from the trunk at $z=\zhalo$, but the calculation of
galaxy formation and evolution through successive halo mergers
proceeds forwards in time, moving down from the top of the tree. The
appropriate grid of $N_{\rm steps}$ timesteps, the starting redshift
$z_{\rm start}$ and also the mass resolution, $M_{\rm res}$,
depend on the problem of interest. The timesteps 
can, for example, be chosen to be uniform either in time or in the 
logarithm of the cosmological expansion factor. For the models
presented here, we typically set $M_{\rm res}=5\times 10^9 h^{-1} \Msol$
and use $100$~timesteps  logarithmically spaced in expansion factor
between $z=0$ and~$7$. In our models, stellar feedback 
prevents significant amounts of star formation occurring in halos
of very low circular velocity and so provided $M_{\rm res}$ is sufficiently
low, the model results are not affected by its value.

The second way in which we manipulate the merger trees before applying our
galaxy formation rules is by chopping each tree into branches that define
the formation time and lifetime of each halo. So far we have done this
using a simple algorithm.  We start, at the first timestep, at the top of
the tree (corresponding to the lowest mass halos in the merging hierarchy),
and define each halo present as a new halo that formed at that timestep.
We then follow each of these halos through their subsequent mergers until
they have become part of a halo with mass greater than $f_{\rm form}$ times
the original mass. We normally set $f_{\rm form}=2$. This point defines the
end of the original halo's lifetime.  Consistent with this definition, 
the point at which a new halo life begins is defined by the point when 
mergers produce a halo whose mass exceeds $f_{\rm form}$ times the formation 
mass of its largest progenitor. When applying the galaxy formation
rules detailed in the following sections, we always treat the halos as if
they retained, throughout their lifetime, the mass and other properties,
(mean density, angular momentum, etc.) with which they formed.  The mass
accreted prior to the final merger, which can, in extreme cases, be as large
as $f_{\rm form}-1$ times the original halo mass, is effectively treated as
if it were all accreted at the end of the halo's lifetime.
We have chosen $f_{\rm form}=2$ for consistency with our earlier work 
(Cole \etal \shortcite{cafnz}) in which a factor of two was built into
the ``block model'' that we used to generate merger trees. With this
choice the two models produce near identical results when the content
and parameters of the galaxy formation model are the same. In Table~3 of
Section~\ref{sec:lumfn}, we include one variant model with $f_{\rm form}=1.5$
which demonstrates that the model is not very sensitive to the choice
of this parameter. This is natural as most halos end their lives when 
they are accreted onto much more massive halos and thus their lifetimes are
robust to the choice of $f_{\rm form}$.

In order to investigate the statistical properties of galaxy
populations, we generate a set of merger trees starting from a grid of
parent halo masses specified at some redshift, $\zhalo$. For each
given halo mass, we generate many realizations of the merger tree. The
resulting model galaxies can then be sampled, taking account of the
abundance of the parent halos
at $z=\zhalo$, to construct galaxy catalogues with any
desired selection criteria such as an absolute or apparent magnitude
limit.  Alternatively, properties such as the galaxy luminosity
function or number counts can be estimated directly by a weighted sum
over the model galaxies. We have used the
Press-Schechter mass function to estimate the halo abundance, but it 
is well known that this formula overestimates the abundance of
$M_\star$ objects somewhat (\eg Efstathiou \etal \shortcite{efwd88}; Lacey \&
Cole \shortcite{lc94}). Recently,
Sheth, Mo \& Tormen  (\shortcite{smt99}) and Jenkins \etal 
(\shortcite{jfwccey00}) have presented fitting formulae that match the 
results of N-body simulations to high accuracy. In future it will be
preferable to use these formulae, but here we simply note that adopting
the Jenkins \etal (\shortcite{jfwccey00}) mass function would make little
difference to our model predictions.

\subsection{Halo Properties}

In order to calculate the properties of the galaxies that form within
the dark matter halos produced by the merger tree, we need a model for
the internal structure of the halos. This must specify the halo
rotation velocity required to calculate the angular momentum of the gas that
cools to form disks, and the halo density profile required 
to calculate the sizes and rotation speeds of the galaxies.

The properties of dark matter halos formed in cosmological, collisionless,
\nbody simulations have been extensively studied (\eg Frenk \etal
\shortcite{frenk85},
\shortcite{frenk88}; 
Barnes \& Efstathiou \shortcite{be87}; Warren \etal
\shortcite{wqsz92}; Cole \& Lacey
\shortcite{cl96}; Navarro, Frenk \& White \shortcite{nfw95}, 
\shortcite{nfw96}, \shortcite{nfw97}; Moore \etal
\shortcite{moore}, Jing \shortcite{jing99}). 
The models detailed below are designed to be
consistent with the results from these simulations.

\subsubsection{Spin Distribution}

Dark matter halos gain angular momentum from tidal torques
operating during their formation. The magnitude of this angular
momentum is conventionally quantified by the dimensionless
spin parameter
\begin{equation}
\lambdah = \frac{ \Jh \vert \Eh \vert^{1/2}} {\G \Mh^{5/2}},
\label{eqn:lambda}
\end{equation}
where $\Mh$, $\Jh$ and $\Eh$ are the total mass, angular momentum and
energy of the halo. The distributions of $\lambdah$ found in various
\nbody studies (\cite{be87,efwd88,wqsz92,cl96,lk99}) agree very well with one
another. They depend only very weakly on halo mass and on the form of the
initial spectrum of density fluctuations.

A good fit to the results of Cole \& Lacey (\shortcite{cl96}) 
is provided by the log-normal distribution, 
\begin{equation}
P(\lambdah) d\lambdah =  \frac{1}{\sqrt{2\pi}\sigma_\lambda} \exp \left( 
- \frac{( \ln \lambda
-\ln \lambda_{\rm med})^2}{2\sigma_\lambda^2}
\right) \frac{d\lambdah}{\lambdah},
\label{eqn:spin_prob}
\end{equation}
with $\lambda_{\rm med}=0.039$ and $\sigma_\lambda=0.53$.  This fit was
obtained specifically for halos with $M_\star<M_{\rm H}< 2M_\star $ in the case
of an $n=-2$ power spectrum, which is the most relevant for CDM models
on galaxy scales, but we stress that the fit parameters depend only
very weakly on mass and on the slope of the power spectrum. For example,
this fit also reproduces quite accurately the distribution
plotted in Fig.4 of Lemson \& Kauffmann (\shortcite{lk99}),
which is for galactic mass halos in a $\tau$CDM simulation.
We use this distribution to assign, at random, a value of $\lambdah$ 
to each newly formed halo. Note that we do not take account of a 
possible correlation between the angular momenta of merging halos.
It would be necessary to do this if one wanted to follow the angular
momenta of galaxy merger products, but we currently do not attempt this.

\subsubsection{Halo density Profile}
\label{sec:halo_strc}

Our standard choice is to model the dark matter density profile using
the NFW model (\cite{nfw95}):
\begin{equation}
\rho(r) = \frac{\Deltavir \rhocrit }{f(\anfw)} \
\frac{1}{r/\rvir \; (r/\rvir+\anfw)^2}  \quad (r\le\rvir),
\label{eqn:nfw}
\end{equation}
with $f(\anfw)= \ln(1+1/\anfw) -1/(1+\anfw) $, truncated at the virial
radius, $\rvir$.  We define the virial radius as the radius at which
the mean interior density equals $\Deltavir$ times the critical
density, $\rhocrit = 3 H^2/(8\pi G)$.  Here the virial overdensity,
$\Delta_{\rm vir}$, is defined by the spherical collapse model which
yields $\Delta_{\rm vir}=178$ for $\Omega_0=1$. Expressions for
$\Delta_{\rm vir}$ in low $\Omega_0$, open and flat, universes can be
found in the appendices of Lacey \& Cole (\shortcite{lc93}) and Eke
\etal (\shortcite{ecf96}) respectively. Confirmation that this
definition of the virial radius is physically sensible is provided by Fig.13 of
Cole \& Lacey (\shortcite{cl96}) and Fig.10 of Eke, Navarro \&
Frenk (\shortcite{enf98}).  These show that on average the transition between
dynamical equilibrium and the surrounding infall occurs close to this
radius.  
The NFW profile has one free parameter, $\anfw$, which is
a scalelength measured in units of the virial radius.  Allowing this
one parameter (equivalent to the inverse of the concentration in the
terminology of Navarro \etal ) to vary, the density profile accurately
fits the profiles of isolated halos grown in cosmological N-body
simulations for a wide range of masses and initial conditions (Navarro
\etal \shortcite{nfw96},\shortcite{nfw97}),
including simulations that contain adiabatic gas as well as collisionless dark
matter (\cite{enf98,frenk99}).  Furthermore, there is a 
correlation between the best fit value of $\anfw$ and halo mass. This
can be understood in terms of how the typical formation time of a halo
depends on mass (\cite{cl96}).  
A simple analytic model for this
relation has been presented in the appendix of Navarro \etal
(\shortcite{nfw97}), and it is this that we use to set the values
of $\anfw$ for our halos.
Jing (\shortcite{jing99}) and
Bullock \etal (\shortcite{bul99}) have found that there is considerable 
scatter about the mean correlation, which is presumably
related to the differing dynamical states and formation histories
of the halos. We do not take this into account,
but we note that simply including this scatter
by randomly perturbing the $\anfw$ values
has little effect of the resulting distributions of galaxy properties.
For instance, the distribution of galaxy sizes is already broad as
a result of its dependence on the very broad distribution of halo
angular momenta.

Subsequently to the Navarro \etal (\shortcite{nfw97}) paper, there
has been some debate as to the accuracy with which the NFW profile
fits the very central regions of dark matter halos simulated at
very high resolution (Moore \etal \shortcite{moore98}; Kravtsov \etal
\shortcite{kravtsov98}). When a concensus is reached it may be
possible to improve this aspect of our modelling by adopting 
a slightly modified density profile. We note that the
most recent simulations by Moore \etal (\shortcite{moore99})  
yield density profiles which are slightly more centrally concentrated
than the Navarro \etal (\shortcite{nfw97}) result. To an accuracy of
20\% they can be fitted by NFW profiles, but with the scalelengths,
$\anfw$, reduced by a factor of 2/3. Such a change 
has only a relatively small effect on the galaxy properties that we examine 
below. The largest changes are to the disk scalelengths, which decrease
by 10\%, and to the disk circular velocities, which increase by 7.5\%.

\subsubsection{Halo  Rotation Velocity}
\label{sec:halo_rot}

To compute the angular momentum of that fraction of the halo gas that cools 
and is involved in forming a galaxy, we need a model of the 
rotational structure of the halo. We
assume that the mean rotational velocity, $V_{\rm rot}$, of concentric
shells of material is constant with radius and always aligned in the
same direction. This simple description is broadly consistent with the
behaviour seen in the simulations of Warren \etal (\shortcite{wqsz92})
and Cole \& Lacey (\shortcite{cl96}).  The appropriate value of
$V_{\rm rot}$ can be related to the halo spin parameter, $\lambdah$,
by evaluating, for the adopted halo model, the quantities defining
$\Jh$ and $\Eh$ in equation (\ref{eqn:lambda}). This calculation is
described in Appendix~\ref{app:halo}. We obtain
\begin{equation}
	\vrot =   A(\anfw) \lambdah \Vh ,
\label{eqn:vrot}
\end{equation}
where $\Vh \equiv (G M /\rvir)^{1/2}$ is the circular velocity of the
halo at the virial radius.  The dimensionless coefficient $A(\anfw)$
is a weak function of $\anfw$, varying from $A \approx 3.9$ for
$\anfw=0.01$ to $A \approx 4.5$ for $\anfw=0.3$. 

Our code allows us to explore the effects of using alternative dark
matter density profiles. In particular, we have included the case of a
singular isothermal density profile, $\rho(r) \propto r^{-2}$, and a
non-singular isothermal density profile, $\rho(r) \propto
1/((r/\rvir)^2+a^2)$. We find $A=8\sqrt{2}/\pi \approx 3.6$ for the
singular isothermal sphere (see Appendix~\ref{app:halo}). The value of
$A$ decreases very slowly as a core radius is introduced, falling to
$A \approx 3.4$ for $a=0.3$.

Our model of the distribution of hot gas in the halo is described in
Section~\ref{sec:hotgas}. As the hot gas is less centrally
concentrated than the dark matter, if we were to assume they had identical 
rotation velocities this would result in the gas having a
slightly higher mean specific angular momentum than the dark matter. 
We therefore take the rotation velocity of the gas to also be constant
with radius, but with a value $V^{\rm gas}_{\rm rot}$ such the gas and
dark matter have the same mean specific angular momentum within the
virial radius.
%We
%therefore take the rotation velocity of the gas to be slightly less
%than that of the halo, and set their ratio $V^{\rm gas}_{\rm
%rot}/V_{\rm rot}$ such that mean specific angular momentum of the gas
%equals that of the dark matter. 
This simple model seems to be in
reasonable accord with the properties of clusters in the
high-resolution, gas-dynamic simulations of Eke \etal
(\shortcite{enf98}, Eke private communication).

%%%%%%%%%%%%%%%%%%%%%%%%%%%%%%%%%%%%%%%%%%%%%%%%%%%%%%%%%%%%%%%%%%%%%%%%%%%
%%%%%%%%%%%%%%%%%%%%%%%%%%%%%%%%%%%%%%%%%%%%%%%%%%%%%%%%%%%%%%%%%%%%%%%%%%%

\section{Formation of Disks and Spheroids}
\label{sec:dsksphform}

In this section we describe how disks and spheroids form, how we model
star formation, feedback and chemical evolution, and how we calculate
galaxy sizes.

\subsection{Disk Formation}
\label{sec:diskform}

We assume that disks form by cooling of gas initially in the halo.
Tidal torques impart angular momentum to all material in the halo,
including the gas, so that gas which has cooled and lost its pressure
support will naturally settle into a disk.  Below, we detail how we
compute the mass of the forming disk based on the radiative cooling
rate of the halo gas, and how we compute its angular momentum.

\subsubsection{Hot Gas Distribution}
\label{sec:hotgas}

Diffuse gas which is not part of galaxies is assumed to be
shock-heated during halo collapse and merging events. We will refer to
this halo gas as ``hot'', to distinguish it from the gas in galaxies,
which we call ``cold''. To calculate how much of this hot gas cools to
form a disk, we need to know its initial temperature and density
profile. In contrast to most previous work, we will not assume that
the hot gas has the same density profile as the dark matter.

High-resolution hydrodynamical simulations of the formation of galaxy
clusters (Navarro \etal \shortcite{nfw95}; Eke, Navarro \& Frenk
\shortcite{enf98}, Frenk \etal \shortcite{frenk99}) show that, in the
absence of radiative cooling, the resulting dark matter distribution
is well-modelled by an NFW profile, but that the shock-heated gas is
less centrally concentrated. The gas distribution is well fit by the
$\beta$-model (\cite{cf76}), $ \rho_{\rm gas}(r) \propto (r^2+r_{\rm
core}^2)^{-3 \beta_{\rm fit}/2} $, traditionally used to model the
hot X-ray emitting gas in galaxy clusters.  The simulations of Eke
\etal (\shortcite{enf98}), which span a narrow range of halo mass in an
$\Omega_0=0.3$, $\Lambda_0=0.7$ cosmology, indicate that the typical
cluster gas profile is accurately described by a $\beta$-model with
$\beta_{\rm fit}
\approx 2/3$ and $r_{\rm core}/r_{\rm NFW} \approx 1/3$.
Here, $r_{\rm NFW}$  is the NFW scalelength, equal to $\anfw \rvir$,
and so for these clusters $r_{\rm core}/\rvir \approx 1/20$. 
A similar result was found for
clusters in an $\Omega_0=1$ cosmology by Navarro \etal
(\shortcite{nfw95}).  In both cases, the simulations produce cluster
gas temperature profiles that vary slowly with radius, consistent with
hydrostatic equilibrium. The mean temperature of the gas is close to
the virial temperature, defined by
\begin{equation}
\Tvir = \frac{1}{2} \, {\mu m_{\rm H} \over k} \Vh^2,
\label{eqn:Tvir}
\end{equation}
where $m_{\rm H}$ is the mass of the hydrogen atom and $\mu$ the mean
molecular mass.

Motivated by these simulation results, we assume that any diffuse gas
present in the progenitors of a forming halo is shock-heated during
the halo formation process and then settles into a spherical
distribution with density profile, 
\begin{equation}
\rho_{\rm gas}(r) \propto 1/(r^2+r_{\rm core}^2). 
\label{eqn:rhogas}
\end{equation} 
For simplicity, we assume the gas temperature to be constant and equal
to the virial temperature, $\Tvir$.  The effect this assumption has on the
cooling radii and masses, computed below, is generally very small, as
the cooling time of the gas depends more strongly on density than
temperature and the density gradient is typically much larger
than the temperature gradient.  Guided also by the numerical
simulations, we assume that, for the first generation of halos, $r_{\rm
core}= \rnfw/3$.  However, this result is for simulations which do not
include radiative cooling, and we expect this relationship to be
modified for halos formed from progenitors in which gas has already
been removed by cooling.  The gas that is able to cool most
efficiently in any halo is the densest gas with the lowest entropy.
Thus, the remaining gas involved in the formation of a new halo will
have a higher minimum entropy than if cooling had not occured.  The
analytic work of Evrard \& Henry (\shortcite{evrard91}), Kay \& Bower
(\shortcite{kb99}) and Wu \etal (\shortcite{wfn99}) suggests that
increasing the minimum entropy of the halo gas has the effect of
increasing its core radius.  Further out, where cooling has had
little effect, the gas properties will be less affected and, in
particular, the pressure at the virial radius, which is ultimately
maintained by shocks from infalling material, will remain unchanged.

As a qualitative description of the behaviour described above we have
constructed the following simple model.  When a new halo is formed in a
merger, if the hot gas fraction in the halo is less than the global value
of $\Omegab/\Omega_0$ (indicating that some gas has already cooled), we
increase the gas core radius, $r_{\rm core}$, until we recover the same
density at the virial radius that we would have obtained had no gas
cooled. In principle, this ceases to be possible once the gas fraction is
so low that even if it were placed in the halo at constant density, this
density would be below the target value. To deal with this contingency, we
set an upper limit of $r_{\rm core}= 10 \rvir$, but in practice this
extreme is rarely reached.  The result of this procedure, when applied to
the models discussed in Section~\ref{sec:cons}, is that at high redshift
the core radii start with values close to $r_{\rm core}= \rnfw/3 $, which
for isolated bright galaxies, groups and clusters are approximately
$20$,$30$,$50 h^{-1}$ kpc respectively.  As gas cools and galaxy formation
proceeds, the core radii grow until, at the present day, the corresponding
median core radii for newly formed halos are $85$, $125$ and $175 h^{-1}$
kpc. The distributions of core radii typically span a factor of two in
scale.

As alternatives to this standard description, our code also allows us
to keep the core radius fixed, either as a fixed fraction of the
virial radius or of the NFW scalelength, or even simply to assume that
the gas traces the dark matter density profile. These options allow us
to gauge directly the effects of our model assumptions.

\subsubsection{Cooling}
\label{sec:cooling}

Assuming that the shock-heated halo gas is in collisional ionization
equilibrium, the cooling time, defined as the ratio of the thermal
energy density to the cooling rate per unit volume, $\rho_{\rm gas}^2
\Lambda(T_{\rm gas},Z_{\rm gas})$, is
\begin{equation}
\tau_{\rm cool}(r) = \frac{3}{2} \frac{1}{\mu m_{\rm H}}
\frac{k T_{\rm gas}}{\rho_{\rm gas}(r) \Lambda(T_{\rm gas},Z_{\rm gas})} .
\label{eqn:tcool}
\end{equation}
Here $\rho_{\rm gas}(r)$ is the density of the gas at radius $r$,
$T_{\rm gas}$ is the temperature and $Z_{\rm gas}$ the metallicity. We
use the cooling function $\Lambda(T_{\rm gas},Z_{\rm gas})$ tabulated
by Sutherland \& Dopita (\shortcite{sd93}). We estimate the amount of
gas that has cooled by time $t$ after the halo has formed by defining
a cooling radius, $r_{\rm cool}(t)$, at which $\tau_{\rm cool} =t$.
Note that for the purpose of computing this cooling radius, the 
gas density profile is kept fixed throughout the halo lifetime.

The gas that cools is assumed to be accreted onto a disk at the centre
of the halo. We estimate the time taken for this material to be
accreted onto the disk as the free-fall time in the halo with the
assumed density profile. Conversely, we can define a free-fall radius
$r_{\rm ff}(t)$ beyond which, at time $t$, material has not yet had
sufficient time to fall into the central disk.  Thus, to compute the mass 
that cools and is added to the disk in one timestep, $\Delta t$, we 
compute $r_{\rm min}(t) = \min[r_{\rm cool},r_{\rm ff}]$ 
at the begining and the end of the timestep,  and set $\crate \Delta
t$ equal to 
the mass of hot gas originally in the spherical shell
defined by the two values of $r_{\rm min}$. For one timestep, this
defines the cooling rate, $\crate$, that enters into the
differential equations (\ref{eqn:sff}) to (\ref{eqn:sfl}) of
Section~\ref{sec:sf_disk}
describing the star formation, chemical enrichment and feedback.

\subsubsection{Angular momentum}
\label{sec:ang_mom}

We assume that when the halo gas cools and collapses down to a disk,
it conserves its angular momentum. Thus, the specific angular momentum
of the material added to the disk by cooling since the formation of
the halo is equal to that of the
gas originally within $r_{\rm min} = \min[r_{\rm cool},r_{\rm ff}]$ . 
As described in 
Section~\ref{sec:halo_rot} we take the rotation velocity of the
hot halo gas, $V^{\rm gas}_{\rm rot}$,  to be 
constant with radius, which implies that the specific angular momentum 
increases linearly with radius in the halo.

The assumption of angular momentum conservation during the collapse is
not a trivial one. In fact, numerical hydrodynamical simulations of
galaxy formation including radiative cooling have, up to now, found
that the cold gas loses most of its angular momentum (e.g. White 
\& Navarro \shortcite{wn93}, Navarro,
Frenk \& White \shortcite{nfw95b}, Navarro \& Steinmetz
\shortcite{ns99}). However, these simulations have either not
included star formation and feedback, or only included it in a very
simple way which may not be accurate. In the absence of stellar feedback
the gas distribution in a forming galactic halo is very clumpy. These
clumps are efficient at losing angular momentum  to the dark matter halo 
via dynamical friction. It is precisely this process that we model,
in Section~\ref{sec:dynfric}, to follow the merging of galaxies.
However, if feedback keeps the gas that is not in galaxies diffuse,
then the loss of angular momentum will be much reduced. 
This has been investigated by Weil, Eke, \&  Efstathiou 
(\shortcite{wee98}), Sommer-Larsen, Gelato \& Vedel
(\shortcite{sgv99}) 
and Eke, Efstathiou \& Wright (\shortcite{eew99})
who found that delaying the cooling of the gas considerably reduces the
loss of angular momentum.
We also note that strong angular momentum loss results in galaxy
disk sizes much smaller than observed. In contrast, as we show later,
our assumption of angular momentum conservation leads to disk sizes
very similar to observed values.

In the following section we will introduce a model
of stellar feedback whereby gas can be ejected from the disk. When
this occurs, we assume that the specific angular momentum of the
remaining material is unaffected.  In Section \ref{sec:sizes} we give
details of how we relate the size of the disk to its mass and angular
momentum.

\begin{figure*}
\centering
\centerline{\epsfxsize=14 truecm \epsfbox[100 470 474 720]{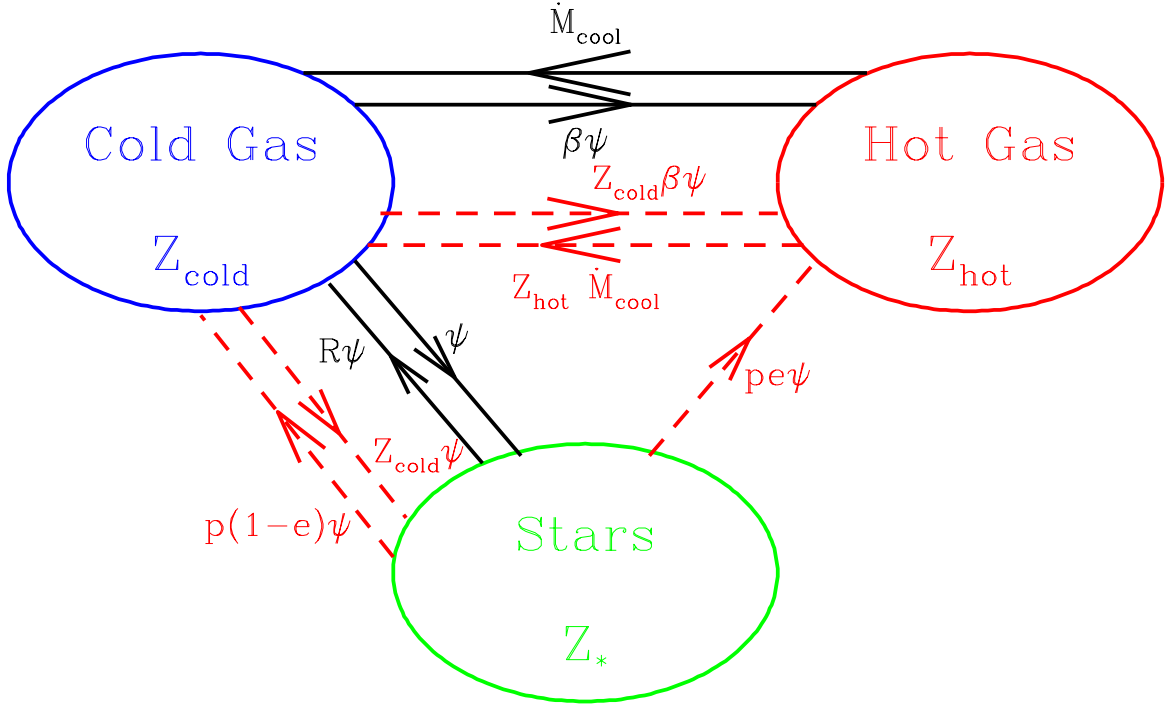}}
\caption{A schematic diagram showing the transfer of mass
and metals between stars and the hot and cold gas phases 
during a single timestep. The solid lines indicate
the routes and rates by which mass is transferred between the
three reservoirs, while the dashed lines refer only to the 
exchange of metals. The instantaneous rate of star formation is $\psi$
and the cooling rate is $\crate$. The metallicities of the cold gas,
stars and hot halo gas are $Z_{\rm cold}$, $Z_*$ and $Z_{\rm hot}$ 
respectively. The yield of the assumed IMF is $p$ and the parameters
$\beta$ and $e$ describe the effect of SN feedback and the direct ejection
of SN metals into the hot halo gas.
}
\label{fig:sf_schematic}
\end{figure*}

\subsection{Star Formation in Disks}
\label{sec:sf_disk}

We now turn to the important process of star formation within
disks. Star formation not only converts cold gas into luminous stars,
but it also affects the physical state of the surrounding gas, as SNe
and young stars inject energy and metals back into the ISM. The energy
that is released can be sufficient to drive gas and metals out of the
galactic disk in the form of a hot wind. The removal of material from
the disk acts as a feedback process which regulates the star formation
rate. Also, the injected metals enrich both the cold star-forming gas
and the surrounding diffuse hot halo gas.  Enrichment of the halo gas
decreases the cooling time defined in (\ref{eqn:tcool}), allowing more
gas to cool at late times, while stellar enrichment affects the colour
and luminosity of the stellar populations. Early semi-analytic models
were unable to include these effects accurately, as stellar population 
synthesis models
with a wide range of metallicities were not available. Now that such
models are widely available, simple chemical enrichment models have been
included in several semi-analytic models, \eg Kauffmann (\shortcite{k96}),
Kauffmann \& Charlot(\shortcite{kc98}), 
Guiderdoni \etal (\shortcite{guiderdoni99}),
and Somerville \& Primack (\shortcite{sp98}). 
%and van Kampen \etal (\shortcite{kjp99}).

\subsubsection{Chemical Enrichment and Feedback}

Our basic model of star formation assumes that stars are formed in the
disk at a rate directly proportional to the mass of cold gas.  Thus,
the instantaneous star formation rate, $\psi$, is given by
\begin{equation}
\psi=M_{\rm cold}/\tau_\star ,
\label{eq:sfr}
\end{equation}
where the star formation timescale is $\tau_\star$.  To model the
feedback effects of energy input from young stars and SNe into the
gas, we assume that cold gas is reheated and ejected from the
disk at a rate 
\begin{equation}
\dot  M_{\rm eject} = \beta \psi.
\label{eq:betadef}
\end{equation}
In general, both $\tau_\star$ and $\beta$ are functions of the
properties of the surrounding galaxy and halo.  We will return later
(Section~\ref{sec:sflfb}) to the way in which we model these
dependencies.

The processes of gas cooling from the reservoir of hot halo gas and
accreting onto the disk, star formation from the cold gas, and the
reheating and ejection of gas all occur simultaneously. For each halo,
we estimate the rate at which gas cools and is accreted by the central
galaxy by computing the cooling radius, as described in
Section~\ref{sec:cooling}, at each discrete timestep at which the halo
merger tree is stored.  Within one of these discrete steps we
approximate the cooling rate as a constant, $\crate$, and use a simple
instantaneous recycling approximation to model star formation,
feedback and chemical enrichment (\cite{tinsley80}). Note that for
satellite galaxies $\crate=0$, as their hot gas is assumed to be 
stripped. Fig.~\ref{fig:sf_schematic} depicts the various
channels by which mass and metals are transferred between the
three phases.  Note that we always compute $\crate$ from the initial
density profile of the hot gas and so we are implicitly assuming 
that gas reheated by SNe plays no role until it is incorporated into
a new halo as a result of a merger.
Under the instantaneous recycling approximation, the
rate of flow down each channel is simply proportional to the instantaneous
star formation rate, $\psi$, or the cooling rate, $\crate$. The
labels in Fig.~\ref{fig:sf_schematic} give the rates in terms of these
quantities.  The solid lines refer to total rates and the dashed lines
to the metal component. Note that we have allowed for the possibility
that some fraction of the metals produced by stars may be directly
transferred to the hot halo gas, but we have neglected the
corresponding transfer of mass.  This is a good approximation, since
the directly ejected material would be very metal rich and the mass
transferred by this route will always be small compared to that
transferred by reheating of the cold gas by SN feedback.

In Fig.~\ref{fig:sf_schematic} and below, $p$ denotes the yield
(the fraction of mass converted into stars that is returned to the
ISM in the form of metals), $R$
the fraction of mass recycled by stars (winds and SNe), $e$ the fraction of
newly produced metals ejected directly from the stellar disk to the
hot gas phase, $Z_{\rm cold}$ the metallicity of the cold gas, and
$\beta$ the efficiency of stellar feedback.  Each of the arrows in
Fig.~\ref{fig:sf_schematic} gives rise to a term in the following
differential equations that describe the evolution of the mass and
metal content of the three reservoirs:
\begin{eqnarray}
\dot  M_{\star{\hphantom{col}}}     &=&  (1-R) \psi  \label{eqn:sff} \\
\dot  M_{\rm hot{\hphantom{l}}}   &=& - \crate + \beta \psi \\
\dot  M_{\rm cold}^{\hphantom{Z}}  &=& \crate - (1-R+\beta) \psi \\
\dot  M_{\star{\hphantom{olj}}}^Z   &=& (1-R)  Z_{\rm cold} \psi \\
\dot  M_{\rm hot{\hphantom{l}}}^Z &=& -\crate Z_{\rm hot} + 
                       (p e + \beta Z_{\rm cold}) \psi \\
\dot M_{\rm cold}^Z &=& \crate Z_{\rm hot} \nonumber \\
 &+& (p(1-e) - (1+\beta-R)Z_{\rm cold}) \psi, 
\label{eqn:sfl}
\end{eqnarray}
where $Z_{\rm cold}=M_{\rm cold}^Z/M_{\rm cold}$ and $Z_{\rm
hot}=M_{\rm hot}^Z/M_{\rm hot}$. The values of $R$ and $p$ in these
equations are related to the IMF, as discussed in
Section~\ref{sec:yield}.

We assume that over one timestep the cooling rate, $\crate$, and the 
metallicity of the hot gas, $Z_{\rm hot}$, can be taken to be constant. 
This set of first-order, coupled differential equations can be 
straightforwardly solved to give the change in mass and metal content of
cold gas, hot gas and stars since the start of the timestep (Appendix
\ref{app:sf}). The model is quite flexible: its behaviour is
determined by specifying how the functions $\tau_\star$, $\beta$ and
$e$ depend on the properties of the galaxy and its surrounding halo.
We note that compared to the simple, ``closed-box'' chemical
enrichment model, the yield is modified by the metal ejection and
feedback to produce an effective yield $p_{\rm eff} = (1-e)
p/(1-R+\beta) $ (equation \ref{eqn:peff}), 
which is therefore a function of the potential-well
depth of the galaxy. The evolution of the stellar metallicity differs
from the closed-box model because it is affected by both the ejection
of reheated gas and the  accretion of cold gas and associated
metals.

\subsubsection{Star Formation Law and Feedback Parameterization}
\label{sec:sflfb}

In our previous work (\eg \cite{cafnz}), we specified the star
formation timescale and feedback efficiency in terms of the circular
velocity of the halo in which each galaxy formed, $\Vh$. T
he relations we adopted were
\begin{equation}
  \tau_\star = \tau_\star^{0\prime} \, (\Vh/ 300 \, \kms)^{\alpha_\star^\prime}
\label{eqn:sflawold}
\end{equation}
and
\begin{equation}
	\beta = (\Vh/V_{\rm hot}^\prime)^{-\alpha_{\rm hot}^\prime} .
\end{equation}
The parameter $\tau_\star^{0\prime}$, we treated as a free parameter,
while the other three parameters, $\alpha_\star^\prime$, $V_{\rm
hot}^\prime$ and~$\alpha_{\rm hot}^\prime$, we constrained by
comparing our models to the numerical simulations of galaxy formation
of Navarro \& White (\shortcite{nw93}). These simulations had only one
free parameter, the fraction of SN energy injected as kinetic energy
into the interstellar medium.  In order to suppress the formation of
low luminosity galaxies, and thus produce a galaxy luminosity function
with a reasonably shallow faint end slope, as observed, we adopted a
fiducial model with very strong feedback for low circular velocity
halos, which we obtained by setting
the parameter values $\alpha_\star^\prime=-1.5$, $V_{\rm
hot}^\prime=140 \, \kms$ and~$\alpha_{\rm hot}^\prime=5.5$.

The more detailed modelling that we now perform of the structure of
our model galaxies allows us to specify the star formation timescale
and feedback efficiency more naturally in terms of the properties of
the galaxy disk, namely its circular velocity, $\Vd$, and dynamical time
$\taud \equiv r_{\rm disk}/ \Vd$. $\Vd$ and $\rdisk$ are both taken at
the disk half-mass radius. The relations that we adopt are
\begin{equation}
\tau_\star = \epsilon_\star^{-1} \,\taud \, (\Vd/200 \, \kms)^{\alpha_\star} 
\label{eqn:sflaw}
\end{equation}
and
\begin{equation}
	\beta = (\Vd/V_{\rm hot})^{-\alpha_{\rm hot}} ,
\label{eqn:feedback}
\end{equation}
where $\epsilon_\star$, $\alpha_\star$ and $\alpha_{\rm hot}$ are
dimensionless parameters, and the parameter, $V_{\rm hot}$, has the
dimensions of velocity.  If $\alphastar=0$, then our star formation
law, (\ref{eqn:sflaw}), simply gives a star formation timescale
proportional to the galaxy dynamical time, broadly consistent with the
observational data compiled by Kennicutt (\shortcite{kennicutt98}).  The
inclusion of the velocity dependent term allows us to explore models
that have a similar dependence on velocity as our previous, quite
successful, model which had $\alpha_\star^\prime=-1.5$ in
(\ref{eqn:sflawold}). It should be noted that because the cold gas
reservoir is depleted both by the formation of stars and by reheating
due to SN feedback, the timescale on which the reservoir is depleted 
(in the absence of any further gas cooling) is shorter than $\tau_\star$.  In
Appendix~\ref{app:sf}, where the analytic solutions of (\ref{eqn:sff}) to
(\ref{eqn:sfl}) are discussed, it is shown that this timescale, which in turn
determines the effective star formation timescale, is given by
$\tauf=\tau_\star/(1-R+\beta)$.  The feedback equation is the same as
we used previously, but now expressed in terms of the galaxy circular
velocity rather than the halo circular velocity. This is physically
more realistic as it is the depth of the potential at the point where
the stars are forming which is most relevant.  To constrain these four
parameters we now prefer to take a more empirical approach and use a
wider range of observational data, rather than to fix the parameters to
emulate one particular set of numerical simulations of galaxy
formation, as we did before.

\subsection{Spheroid Formation}

In our model, the primary route by which bright elliptical galaxies and
the bulge components of spiral galaxies form is through galaxy
mergers. When dark matter halos merge, we assume that the most massive
galaxy automatically becomes the central galaxy in the new halo, while
all the other galaxies become satellite galaxies orbiting within the
dark matter halo. The orbits of these satellite galaxies will
gradually decay as energy and angular momentum are lost via dynamical
friction to the halo material. Thus, eventually, the satellite
galaxies spiral in and merge with the central galaxy. We now describe
how we estimate the times at which such galaxy-galaxy mergers occur
and what they produce.

\subsubsection{Dynamical Friction}
\label{sec:dynfric}

When a new halo forms, we assume that each satellite galaxy enters the
halo on a random orbit.  The most massive pre-existing galaxy on the
other hand is assumed to become the central galaxy in the new halo,
where it will act as the focus for any gas that may cool within the new
halo. The time for a satellite's orbit to decay due to the effects of
dynamical friction depends on the initial energy and angular momentum
of the orbit.  Lacey \& Cole (\shortcite{lc93}) estimated the time for
an orbit to decay in an isothermal halo, based on the standard
Chandrasekhar formula for the dynamical friction. Their formula (B4) can be
written in the form,
\begin{equation}
\taumerge =   \fdf \,  \taudf \,  \tdyn \,
\frac{0.3722}{\ln(\Lambda_{\rm Coulomb})} \frac{\Mh}{M_{\rm sat}}.
\label{eqn:taumerge}
\end{equation}
Here, $\Mh$ is the mass of the halo in which the satellite orbits, and we
take $M_{\rm sat}$ to be the mass of the satellite galaxy {\it including\ }
the mass of the dark matter halo in which it formed (\cite{nfw95}).  Note
that we deliberately count the mass of the satellite's halo in the
definition of both $M_{\rm sat}$ and $\Mh$.  The Coulomb logarithm, we take
to be $\ln(\Lambda_{\rm Coulomb})=\ln(\Mh/M_{\rm sat})$. The dynamical time
of the new halo is $\tdyn \equiv \pi \rvir/ \Vh$, defined equivalently as
either the half period of a circular orbit at the virial radius, or as $(G
\rho_{\rm vir})^{-1/2}$, where $\rho_{\rm vir}$ is the mean density within
the virial radius, or, for an isothermal sphere, as the full orbital period
of a circular orbit at the half-mass radius.  

The dependence of this merger timescale, $\taumerge$, on the orbital
parameters is contained in the factor $\taudf$, defined as,
\begin{equation}
\taudf = [J/J_{\rm c}(E)]^{0.78} [r_{\rm c}(E) /\rvir]^2, 
\label{eqn:df_circ}
\end{equation}
where $E$ and $J$ are the initial energy and angular momentum of the
satellite's orbit, and $r_{\rm c}(E)$ and $J_{\rm c}(E)$ are the
radius and angular momentum of a circular orbit with the same energy
as that of the satellite. The power-law dependence on the circularity,
$J/J_{\rm c}(E)$, is an accurate fit to the result of numerical
integration of the orbit-averaged equations describing the effect of
dynamical fiction in the range $0.01<J/J_{\rm c}(E)<1$ (\cite{lc93}).
The distribution of initial orbital parameters of infalling satellites
in cosmological \nbody simulations has been studied by Tormen
(\shortcite{tormen97}).  We find from his results that the
distribution of $\taudf$ is well modelled by a log normal with
$\langle \log_{10} \taudf \rangle=-0.14$ and dispersion $\langle (
\log_{10} \taudf - \langle \log_{10} \taudf
\rangle)^2 \rangle^{1/2}= 0.26$. 

The merger timescale computed in this manner is based on several
approximations, \eg treating the satellite as a point mass with mass
equal to the sum of the galaxy mass plus that of its original dark
matter halo.  We therefore allow ourselves some freedom by inserting
the dimensionless parameter $\fdf$, which is greater than unity if the
infalling satellite's halo is efficiently stripped off early on. We
note that recent analytical and numerical investigations by van den
Bosch \etal (\shortcite{vdb99}) and Colpi, Mayer \& Governato
(\shortcite{cmg99}) suggest a weaker dependence of the merger
timescale on the orbital circularity, with the exponent 0.78 in
equation (\ref{eqn:df_circ}) being replaced by a value of 0.4 or 0.5,
but these results were also derived using a somewhat different halo
density profile from the singular isothermal sphere assumed by Lacey
\& Cole (\shortcite{lc93}).
%We
%note that recent sophisticated analytic and numerical investigations
%of this dynamical process by Colpi, Mayer \& Governato
%(\shortcite{cmg99}) suggest a minor modification to the above
%formula. They find that the dependence of the merger timescale on
%angular momentum is weaker and fitted with an exponent of 0.4 rather
%than 0.78 in equation (\ref{eqn:df_circ}). Also, for a specific model
%of the density profile of the infalling satellite, they find that the
%effect of tidal stripping produces $f_{\rm df} \approx 2.7$. 
In this
work we have retained the model defined by equations
(\ref{eqn:taumerge}) and (\ref{eqn:df_circ}), but we note that it 
may soon be possible to have a fully specified and calibrated model for
dynamical friction-driven mergers.

 The procedure for determining the fate of satellite galaxies within dark
matter halos is straightforward.  When a new halo forms, each of the
satellite galaxies that it contains is assigned a random value of $\taudf$
according to the log-normal distribution described above. Then, for each
satellite, we compute $\taumerge$ from equation~(\ref{eqn:taumerge}). 
The satellite is assumed to merge with the central galaxy after this time
interval has elapsed, provided this occurs during the lifetime of the halo,
\ie before the halo has merged to become part of a much larger system.
Satellites that do not merge are assigned a new random value of $\taudf$
when the halo in which they reside is incorporated into a new, more massive 
halo.

\subsubsection{Galaxy Mergers and Bursts}
\label{sec:galmergers}

Our method for modelling galaxy mergers produces, at each timestep, a
list of satellite galaxies which merge with the central galaxy in each
halo. If our grid of timesteps were sufficiently fine, then these
lists would always contain just one or zero satellite galaxies, but in
practice there is often one large satellite and several smaller
satellites merging with the central galaxy at a single timestep.  We
deal with this by ranking the merging satellites by mass and 
then, starting with the most massive one, merge them sequentially with 
the central galaxy.

The outcome of each merger depends on the ratio of the mass of the
merging satellite, $M_{\rm sat}$, to that of the central galaxy,
 $M_{\rm cen}$, and has been studied recently by Walker, Mihos \& Hernquist
(\shortcite{walker96}) and Barnes (\shortcite{barnes98}), using
numerical simulations. As a
simplified description of the outcome of these mergers, we adopt the
prescription used in Kauffmann
\etal (\shortcite{kwg93}) and Baugh \etal (\shortcite{bcf96a}):
\begin{itemize}
\item[a)] If the mass ratio of merging galaxies, defined
in terms of stars and cold gas only, is $M_{\rm sat}/M_{\rm cen}
\geq\fellip$, then the merger is said to be ``violent'' or ``major'',
and a single bulge or elliptical galaxy is produced.  Any gas present in the
disks of the merging galaxies is converted into stars in a burst. We
use the standard star formation and feedback rules, but now based on the
circular velocity and dynamical time of the spheroid that is formed
rather than the disk, and with a very much shorter timescale, similar
to the dynamical timescale of the spheroid. 

\item[b)] Alternatively, if $M_{\rm
sat}/M_{\rm cen} < \fellip$, then the merger is classed as ``minor'',
and, unless explicity stated otherwise, 
the stars of the accreted satellite are added to the bulge of the
central galaxy, while any accreted gas is added to the main gas
disk without changing the disk's specific angular momentum. 
\end{itemize}
The merger simulations mentioned above have not been run for a wide
enough range of initial conditions to determine $\fellip$ exactly, but
suggest a value in the range $0.3 \lsim \fellip \lsim 1$.
The way in which we calculate the size of the spheroid which
forms from a merger is described in Section~\ref{sec:merge}. In
the case of minor mergers, we also have the option of adding the
accreted stars to the disk of the central galaxy. 
If we do this, we assume that the specific angular momentum of 
the disk is unchanged by the accretion.

\subsubsection{Disk Instabilities}
\label{sec:stability}

An issue we have not yet addressed is whether the disks in our model
galaxies are dynamically stable. In particular, strongly
self-gravitating disks are likely to be unstable to the formation of a bar
(\eg Binney \& Tremaine \shortcite{BT87}, \S6; Efstathiou, Lake \&
Negroponte \shortcite{eln82}; Christodoulou, Shlosman \& Tohline
\shortcite{cst95}; Sellwood \shortcite{sellwood99}; Syer, Mao \& Mo
\shortcite{sm99}). Recently, the incidence of unstable disks has been
considered in the context of the hierarchical formation of galaxies by
Mo, Mao \& White (\shortcite{mmw98}).  Our disk model is similar to
theirs, except that we explicitly follow the formation and structure
of a bulge component and, more
importantly, we follow the complete merging history of both the
bulge and the disk.
The stability criterion considered by Mo, Mao \& White
(\shortcite{mmw98}) is based on the quantity:
\begin{equation}
\epsilon_{\rm m} \equiv \frac{V_{\rm max} }{ (G M_{\rm disk} 
/ r_{\rm disk})^{1/2}}.
\label{eq:diskinstab}
\end{equation}
According to Efstathiou, Lake \& Negroponte (\shortcite{eln82}), for
disks to be stable requires $\epsilon_{\rm m} \gsim 1.1$.  In the
original formulation, $V_{\rm max}$ was the rotation velocity at the
maximum of the rotation curve, but in our models we use instead the
circular velocity at the disk half-mass radius. 

We have an option in our code to include the effect of such disk
instabilities on galaxy evolution. In that case, we check the
criterion (\ref{eq:diskinstab}) for each galaxy disk at each timestep.
If at any point a disk is unstable according to this condition, we
assume that the instability results in the stellar disk evolving into
a bar and then into a spheroid (\cite{combes90,combes99}).
We also assume that bar instability
causes any gas present in the disk to undergo a burst of star
formation subject to our standard feedback prescription.  

We do not include the effects of disk instabilities in our reference
model. We briefly present the effect it has on the distribution of
disk scalelengths and the morphological mix of galaxies in 
Sections~\ref{sec:dsize} and~\ref{sec:morph_mix}}, but we postpone to a future
paper a more detailed exploration of their consequences.

\subsection{Galaxy Sizes}
\label{sec:sizes}

The two basic principles upon which we base our estimates of galaxy sizes
are: 
\begin{itemize}
\item[i)] the size of a disk is determined by centrifugal equilibrium
and conservation of angular momentum 
\end{itemize}
 and
\begin{itemize} 
\item[ii)]
the size of a stellar spheroidal 
remnant produced by mergers or disk instability is
determined by virial equilibrium and energy conservation. 
\end{itemize}
The application of these simple principles is complicated by the
gravitational interaction of the galaxy disk, spheroid and surrounding
dark matter halo. Because of this, to determine either the disk or
bulge radius, we must solve for the simultaneous dynamical equilibrium
of all three components. We use the following approach: 

\begin{itemize}
\item[a)] The disk
is assumed to have an exponential surface density profile, with
half-mass radius $\rdisk$. 

\item[b)] The spheroid is assumed to follow an
$r^{1/4}$ law in projection, with half-mass radius (in 3D)
$\rbulge$. 

\item[c)] The dark halo has a specified initial density profile
(NFW in the standard case), but this is spherically deformed in
response to the gravity of the disk and spheroid. 

\item[d)] The mass distribution in the halo and the lengthscales
of the disk and bulge are
assumed to adjust adiabatically in response to each other: for the
disk, we assume that the total angular momentum is conserved; for the
halo, we assume that $r V_{\rm c}(r)$ is conserved for each spherical shell;
for the spheroid, we assume that $r V_{\rm c}(r)$ is conserved at
$\rbulge$. 
\end{itemize}
The task is then to solve for $\rdisk$, $\rbulge$ and
the deformed halo profile in dynamical equilibrium, subject to these
constraints. The method is described in detail in
Appendix~\ref{app:contract}. This adiabatic invariance method for
calculating the response of a halo or spheroid to the disk was
originally developed and applied by Barnes \& White (\shortcite{barnes84}),
Blumenthal \etal (\shortcite{blumenthal86}) and
Ryden \& Gunn (\shortcite{rg87}).

\subsubsection{Disk Sizes}

As already stated, the size of a disk is basically determined by the
angular momentum of the halo gas which cools to form it. Many previous
papers have used a version of the following argument: if the dark
halo and the gas it contains are modelled as a singular isothermal
sphere ($\rho\propto r^{-2}$), then, from the results of
Section~\ref{sec:halo_rot} and Appendix~\ref{app:halo}, the mean
specific angular momentum of the gas which cools is $j_{\rm cool} =
\frac{\pi}{8} \, \rcool V^{\rm gas}_{\rm rot} 
= \sqrt{2} \lambdah \rcool \Vh$. On the other hand, if the self-gravity
of the disk is also neglected, it rotates at constant circular
velocity $\Vh$, and so has mean specific angular momentum $j_{\rm disk}= 2
h_D \Vh$, for an exponential disk with scalelength $h_D$. Equating
$j_{\rm cool}$ and $j_{\rm disk}$ gives $\rdisk = 1.68 h_D = 1.19 \lambdah
\rcool$. This simple relation was originally derived by Fall
(\shortcite{fall83}). It was used to calculate disk sizes in galaxy
formation models (with a fixed $\lambdah$) by Lacey \etal
(\shortcite{lgrs93}), Kauffmann \& Charlot (\shortcite{kc94}),
Kauffmann (\shortcite{k96}) and Somerville \& Primack (\shortcite{sp98}). 
In this paper, we improve on this simple
calculation by including (a) non-isothermal halo profiles for the dark
matter and gas; (b) an initial distribution of $\lambdah$; (c) disk
self-gravity and (d) gravity of the halo and spheroid, and their
contraction in response to the disk. Most of these improvements were
also included in the work on disk sizes by Mo, Mao \& White
(\shortcite{mmw98}), using similar techniques to those used
here. However, their work did not include a physical model for galaxy
formation, so that they were forced to treat the disk-to-halo mass
ratio, the disk-to-halo angular momentum ratio, and the
disk $M/L$ ratio as free parameters. If for a given halo we adopt
the same disk angular momentum and mass, then our model produces
disk scalesizes that agree very accurately with the Mo, Mao \& White
model.

\subsubsection{Sizes of Spheroids Formed by Mergers}
\label{sec:merge}

Spheroids can form either in major mergers (when any pre-existing
disks are destroyed) or in minor mergers (when the disk of the larger
galaxy survives).  To estimate the size of the spheroid formed, we
assume that the merging components spiral together under the action of
dynamical friction until their separation equals the sum of their
half-mass radii. At this point, we assume that the systems merge
together, and we use energy conservation and the virial theorem to
compute the size of the remnant. These considerations lead to:
\begin{equation}
\frac{(M_1+M_2)^2}{ r_{\rm new}} =
\frac{M_1^2}{r_1} + \frac{M_2^2}{r_2} + \frac{ f_{\rm orbit}}{c} 
\frac{M_1 M_2}{r_1+r_2} ,
\label{eq:rm}               
\end{equation}
which relates the half-mass radius of the remnant, $r_{\rm new}$, to
the masses, $M_1$ and $M_2$, and half-mass radii, $r_1$ and $r_2$, of
the merging components. Defining $M_1 \geq M_2$, $M_1$ is the total
galaxy mass for a major merger and the bulge mass for a minor
merger, while $M_2$ is the total galaxy mass for a major merger and
the total stellar mass of galaxy 2 for a minor merger. The masses
$M_1$ and $M_2$ include contributions from the respective dark matter
halos, which are taken to be twice the halo mass within the half-mass
radii $r_1$ or $r_2$. 

The form factor, $c$, and the constant, $f_{\rm
orbit}$, are related to the gravitational self-binding energy of each
galaxy,
\begin{equation}
E_{\rm bind}=  -c  \, \frac{{\rm G}M^2}{r} 
\label{bind}
\end{equation}
and their mutual orbital energy,
\begin{equation}
E_{\rm orbit}= - \frac{f_{\rm orbit}}{2} \, \frac{{\rm G}M_1M_2}{r_1+r_2} 
\end{equation}
at the point at which the merger occurs.  The value of $c$ depends
weakly on the density profile of the galaxy; $c=0.49$ for an
exponential disk and $c=0.45$ for an $r^{1/4}$-law spheroid.  For
simplicity, we adopt $c=0.5$.  For the orbital energy, we adopt $f_{\rm
orbit}=1.0$, which corresponds to the orbital energy of two point
masses in a circular orbit with separation $r_1+r_2$. These
assumptions lead to the result that, for a merger of two identical,
equal mass galaxies, the half-mass radius of the remnant increases by
a factor $r_{\rm new}/r_1=4/3$, which agrees reasonably well with the
factor of $1.42$ found in the simulated galaxy mergers of Barnes
(\shortcite{barnes92}). 

Having solved equation~(\ref{eq:rm}) for
$\rbulge=r_{\rm new}$, we then adiabatically adjust the spheroid,
disk (if any) and halo to find the new dynamical equilibrium, as
described in Appendix~\ref{app:contract}. Typically, this leads to
little change in $\rbulge$, showing that our treatment of the dark
matter during the merger is approximately self-consistent.

%Our modelling of the physics that determines the scalelengths of galactic
%disks and bulges is a significant advance over previous treatments. 
%Some previous semianalytic models have made crude estimates of disk
%sizes based on a simple collapse factor motivated
%by the typical spin parameter of dark matter halos and the
%conservation of angular momentum ( \eg Kauffmann \etal \shortcite{kwg93}, 
%Lacey \etal \shortcite{lgrs93}), but neglecting
%the effect of any central spheroid and of the contraction of
%the dark matter halo induced by the condensation of a galaxy at its centre.
%The work of Mo, Mao \& White (\shortcite{mmw98}) improved over this
%by using a distribution of halo spin parameters and modelling the change
%of the halo density profile that occurs as the result
%of the formation of the disk, but they were forced
%to make crude assumptions about the ratio of the masses of the disk
%and halo and the mass of any central spheroid. 
%Our model combines the techniques of 
%Mo, Mao \& White (\shortcite{mmw98}) with a semianalytic model of
%galaxy formation which predicts the masses and bulge-to-disk ratio
%of each galaxy. We also self-consistently compute the scalelength
%of the central bulge or spheroid formed in a merger. 

\subsubsection{Sizes of Spheroids formed by Disk Instabilities}
\label{sec:rinstab}

As mentioned in the previous section, our code has an option to form
spheroids through bar instabilities in disks. In this case, we compute
the size of the resulting spheroid using virial equilibrium and energy
conservation in much the same way as for the spheroids produced by
mergers. If the mass of the unstable disk is $\Mdisk$, the mass of any
pre-existing central stellar bulge is $\Mbulge$, and their respective
half-mass radii are $\rdisk$ and $\rbulge$, then we calculate the
final bulge half-mass radius, $r_{\rm new}$, from the relation
\begin{eqnarray}
\frac{c_{\rm B} (\Mdisk+\Mbulge)^2}{ r_{\rm new}} & = &
\frac{c_{\rm B} \Mbulge^2}{\rbulge}  +  \frac{c_{\rm D} 
\Mdisk^2}{\rdisk} \nonumber \\
 & + & f_{\rm int} \frac{\Mbulge \Mdisk}{\rbulge+\rdisk}.
\label{eq:rinstab}
\end{eqnarray}
Here we adopt $c_{\rm D}=0.49$ and $c_{\rm B}=0.45$, the form factors
appropriate for an exponential disk and $r^{1/4}$-law spheroid respectively
(see (\ref{bind})). The last term represents the gravitational interaction
energy of the disk and bulge which is reasonably well approximated for a
range of $\rbulge/\rdisk$ by this form with $f_{\rm int}=2.0$. After
we have calculated the new spheroid radius $r_{\rm new}$ from
equation~(\ref{eq:rinstab}), we adiabatically adjust the spheroid and
halo to a new dynamical equilibrium, as for the case of a spheroid
formed by a merger.

%%%%%%%%%%%%%%%%%%%%%%%%%%%%%%%%%%%%%%%%%%%%%%%%%%%%%%%%%%%%%%%%%%%%%%%%%%%
%%%%%%%%%%%%%%%%%%%%%%%%%%%%%%%%%%%%%%%%%%%%%%%%%%%%%%%%%%%%%%%%%%%%%%%%%%%

\begin{figure}
\centering
\centerline{\epsfxsize=8.5 truecm \epsfbox{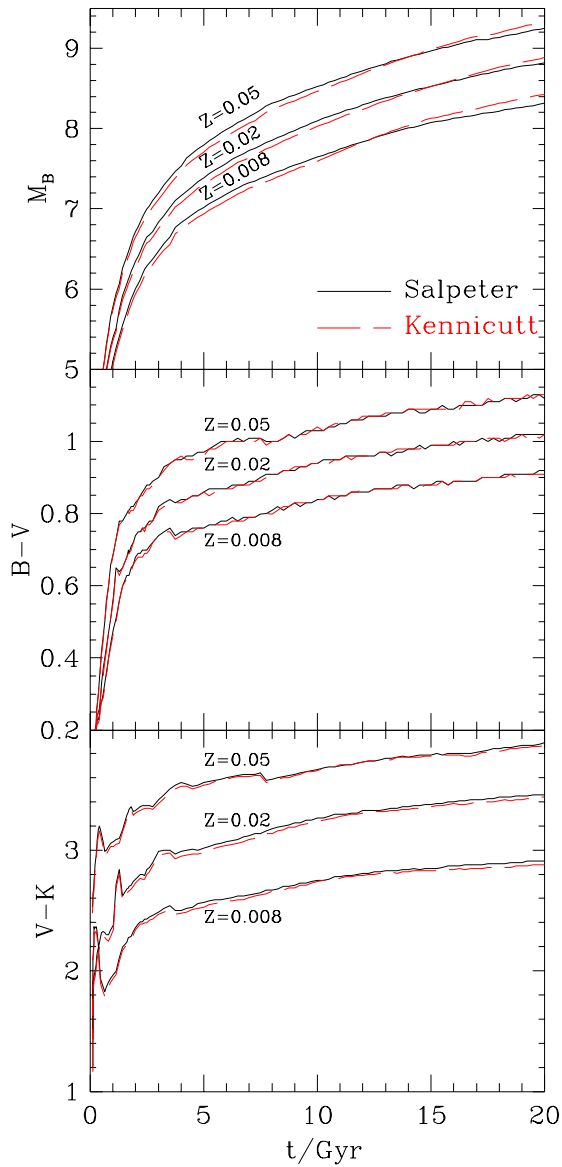}}
\caption{The evolution of the B-band luminosity and the
$B$--$V$  and $V$--$K$ colours for a single age stellar population. 
The solid lines show  results for a stellar population with a Salpeter IMF
for three different metallicities. The middle curves are for
solar metallicity, $Z=0.02$, and the lower and upper curves for
$Z=0.008$ and $0.05$ respectively. 
The absolute magnitudes are normalized to $1 \Msol$ of stars.
The corresponding dashed curves show results assuming
the Kennicutt IMF. In this case, the luminosities in each band have
been reduced by a factor of $1.69$ to make the solar metallicity curves 
for the two IMFs cross at an age of $t= 15$Gyr.
}
\label{fig:imf}
\end{figure}

\section{Galaxy Luminosities and Spectra}
\label{sec:lum}

%\subsection{Galaxy Luminosities, Colours and Spectra}

The aspects of the model described so far enable us to follow the star
formation history, chemical enrichment and size evolution of each
galaxy. In order to convert this information into observable
properties, we must model the spectrophotometric properties of the
stars that are formed, and the effects of dust and ionized gas within
each galaxy on the emerging integrated galaxy spectrum. The models we
adopt for each of these processes are outlined below.

\subsection{Stellar Population Synthesis}
\label{sec:imf}

The technique of stellar population synthesis, pioneered by Tinsley
(\shortcite{tinsley72,tinsley80}) and developed by Guiderdoni \&
Rocca-Volmerange (\shortcite{grv87}), Bruzual \& Charlot
(\shortcite{bc93}), Bressan, Chiosi \& Fagotto (\shortcite{bressan94})
and others, enables the observable properties of a stellar population
to be computed given an assumption about the stellar Initial Mass
Function (IMF) and the star formation history. The latest models of
Bruzual \& Charlot (\shortcite{bc98}) provide the spectral energy
distribution (SED), $l_\lambda(t,Z)$, of a single population of stars
formed at the same time with the same metallicity, as a function of
both age, $t$, and metallicity, $Z$. These can be convolved with the
star formation history of a galaxy to yield its SED:
\begin{equation}
L_\lambda(t) = \int_0^t l_\lambda(t-t^\prime,Z(t^\prime)) 
\  \psi(t^\prime) 
\ dt^\prime,
\end{equation}
where $Z(t^\prime)$ is the metallicity of the stars forming at time
$t^\prime$, and $\psi(t^\prime)$ is the star formation rate at that
time. In the case of a galaxy which formed by merging, we also sum the
contributions to $L_\lambda$ from the different progenitor galaxies,
each with their own star formation and chemical enrichment history.
In performing the convolution integral, we interpolate the
grid of SEDs, $l_\lambda(t,Z)$, provided by Bruzual \& Charlot, to
intermediate ages and metallicities using linear interpolation in $t$
and $\log Z$.  Broad-band colours can then be extracted by integrating
over these spectra weighted by the appropriate filter response
function.

In our models, we always assume that the IMF is universal in time and
space. Observationally, the IMF is best constrained in the solar
neighbourhood.  However, even here there is significant uncertainty
arising mainly from ambiguity in the past star formation
history. Because of this, we consider two possible choices of IMF, the
form proposed by Salpeter (\shortcite{Salpeter}) and the form proposed
by Kennicutt (\shortcite{kenn83}), both of which produce reasonable
agreement with the solar neighbourhood data.  The Salpeter IMF has
$dN/d \ln m \propto m^{-x}$ with $x=1.35$, while the Kennicutt IMF has
$x=0.4$ for $m< \Msol$ and $x=1.5$ for $m>\Msol$. In both cases,
visible stars have $0.1 \Msol< m < 125 \Msol$. The Salpeter IMF has
been widely used in modelling galaxy evolution because of its
simplicity and the fact that it fits the observational data on high
mass stars fairly well. However, there is now considerable
observational evidence, as reviewed by Scalo (\shortcite{Scalo86},
\shortcite{Scalo98}), that the IMF slope at low masses is flatter than
the Salpeter form.  The ``best'' IMF proposed by Scalo
(\shortcite{Scalo98}), which supersedes that of Scalo
(\shortcite{Scalo86}), is actually quite close to that of Kennicutt
(\shortcite{kenn83}). We therefore adopt the Kennicutt IMF as our
standard choice.

We also include in our assumed IMF brown dwarfs ($m<0.1\Msol$), which
contribute mass but no light to the stellar population. The fraction of
brown dwarfs is specified by the parameter $\Upsilon$, defined as 
\begin{equation}
\Upsilon = \frac{({\rm mass\, in\, visible\, stars + brown\,
dwarfs})} {({\rm mass\, in\, visible\, stars})} 
\label{eqn:upsilon}
\end{equation}
at the time a stellar population forms, i.e. before taking account of
the fraction $R$ of the mass that is returned to the ISM by
recycling. Thus, by definition, $\Upsilon\geq 1$. The effect of
including brown dwarfs is to reduce all stellar population
luminosities by a factor $1/\Upsilon$. We will see
in Section~\ref{sec:m_l} that observational estimates of
the mass-to-light ratios of stellar populations constrain viable models
to have modest values of $\Upsilon$ in the range
$1<\Upsilon \lsim 2$.

The way in which the predicted luminosity and colour of a stellar
population depend on age, metallicity and choice of IMF is illustrated
in Fig.~\ref{fig:imf}. A number of properties which affect the
behaviour of our galaxy formation models are worth noting.  The
overall stellar mass-to-light ratio depends significantly on the
choice of IMF. This dependence has been explicitly scaled out of the
curves shown in Fig.~\ref{fig:imf} by reducing all the luminosities in
the Kennicutt IMF case by a factor of $\Upsilon=1.69$, so as to force
the solar metallicity curves for the two IMFs to agree at $t= 15$Gyr.
The slope of the absolute magnitude versus time curve has some
dependence on the choice of IMF. For example, as the age is reduced
from $15$Gyr to around $3$Gyr, the stellar population with the
Kennicutt IMF brightens more rapidly than that with the Salpeter
IMF. The difference is even larger for an IMF such as the Miller-Scalo
IMF (\cite{miller-scalo}) which contains a greater fraction of stars
of a few solar masses. In spite of the dependence of luminosity on the
IMF, the $B$--$V$ and $V$--$K$ colours, both as a function of age and
metallicity, are quite insensitive to the choice of IMF. The colours
do depend strongly on metallicity, with increasing amounts of metals
producing redder stellar populations. A young stellar population is
very blue but rapidly reddens during its first $5$Gyr. At later times
the dependence of colour on age is much weaker.

There are many approximations and assumptions involved in constructing
stellar population synthesis models such as those of Bruzual \&
Charlot. Because of this, the accuracy of the model predictions is
difficult to quantify.  This issue has been addressed by Charlot,
Worthey \& Bressan (\shortcite{cwb96}) by comparing model predictions
from different codes and for varying sets of assumptions. Their
results indicate that for the same choices of IMF and star formation
history, the resulting broad-band colours can differ by a few tenths
of a magnitude, and this could give rise to 20-30\% uncertainties in
either the inferred age or metallicity. While efforts have been made,
and continue to be made, to improve these models, it should be noted
that the uncertainties in the population synthesis model are
sufficiently small that, for our purposes, the dominant source of
uncertainty in modelling galaxy formation is, instead, the choice of
IMF and its associated yield.

\subsection{Yield and Recycled Fraction}
\label{sec:yield}

There are two further quantities related to the IMF that significantly
affect galaxy formation.  These are the recycled fraction, $R$, and
the yield, $p$. They appeared in equations~(\ref{eqn:sff}) 
to~(\ref{eqn:sfl}) for the
evolution of gas and star masses and metallicities. The material which
goes to form massive stars is mostly released back into the ISM via
stellar winds and SN explosions. The returned gas is an important
source of fuel for forming further generations of stars. SN explosions
also enrich the ISM with metals giving rise to subsequent generations
of redder, more metal rich stars.  The recycled fraction and the yield
are defined so that for each mass, $\Delta M$, formed in new stars
(including brown dwarfs), a mass $R\Delta M$ is returned to the ISM,
and a mass $p \Delta M$ of newly synthesized metals is released. These
quantities are given respectively by integrating the total ejected mass and the
ejected mass in newly synthesized metals over the IMF. We
recall that in a closed-box model of chemical evolution, the mean
metallicity of the stars asymptotes to a value of $p/(1-R)$ as the gas
is exhausted (\eg Tinsley \shortcite{tinsley80}).

The values of $R$ and $p$ for any specific IMF can be estimated from
stellar evolution theory and models of supernova explosions. We have
used two different compilations of stellar evolution calculations to
set these parameters: (i) Renzini \& Voli's (\shortcite{rv81}) 
for intermediate mass stars ($1<m\lsim 8\Msol$), and Woosley \&
Weaver's (\shortcite{ww95}) for massive stars ($m\gsim 8\Msol$) which
produce Type~II supernovae; and (ii) results from Marigo \etal
(\shortcite{mbc96}) for intermediate mass stars and from Portinari
\etal (\shortcite{pcb98}) for massive stars. The more recent
calculations in (ii) include the effects of convective overshooting
and quiescent mass loss. However, they rely on the supernova
calculations of Woosley \& Weaver. The contribution of SNII to the
yield is sensitive to the assumed explosion energy (Woosley \&
Weaver's cases A,B,C); we give below the corresponding range in $p$ for case
(i), but Portinari \etal only calculated results for Woosley
\& Weaver's case A (case C would give larger yields). Type~I
supernovae make only a small contribution to the net production of
heavy elements, and are not included here. The results for solar
metallicity are as follows: for the Kennicutt IMF, case (i) gives
$R_1=0.42$, $p_1=0.013-0.023$, and case (ii) gives $R_1=0.44$,
$p_1=0.022$; for the Salpeter IMF, case (i) gives $R_1=0.28$,
$p_1=0.010-0.020$, and case (ii) gives $R_1=0.30$, $p_1=0.018$. These
values assume that $\Upsilon=1$. If $\Upsilon>1$, the appropriate
values become $p=p_1/\Upsilon$ and $R=R_1/\Upsilon$.  As may be seen
from these values, for a given IMF, the recycled fraction, $R$, is
fairly accurately known, but the theoretically predicted yield, $p$,
is uncertain by at least a factor of $2$. In our modelling, we have
chosen to set $R$ according to the above estimates. However, as
the yield is more uncertain, we use these estimates only as a guide
for what is reasonable and instead rely on observed galaxy
metallicities to constrain the value of $p$.

\subsection{Extinction by Dust}

Absorption of starlight by dust has a significant effect on the
optical luminosities and colours of galaxies, and a large effect on
the far-UV luminosities which are used as the main tracer of star
formation rates at high redshift. We model the effects of dust in a
physically self-consistent way, using the models of Ferrara \etal
(\shortcite{fbcg}). Ferrara \etal have calculated radiative transfer
of starlight through dust, including both absorption and scattering by
dust grains, for a realistic 3D distribution of stars and dust, giving
the net attenuation of the galaxy luminosity as a function of
wavelength and inclination angle. In their model, stars are
distributed in both a bulge and a disk, and dust is distributed
smoothly in a disk. The bulge follows a Jaffe (\shortcite{jaffe83})
distribution (which is very similar to an $r^{1/4}$-law) with
projected half-light radius, $r_{\rm e}$. The stars and dust in the
disk both have radially and vertically exponential distributions. The
dust is assumed to have the same radial scalelength, $h_R$, as the
stars, but its scaleheight, $h_z$, is in general different. The total
dust content is parameterized by the central $V$-band optical depth,
$\tau_{V0}$, defined as the extinction optical depth looking
vertically through the whole disk at $r=0$. The dust properties are
chosen to match observations of the extinction law and albedo of dust
in either the Milky Way (MW) or Small Magellanic Cloud (SMC).

Ferrara \etal tabulate separately the attenuations of disk and bulge
light, as functions of wavelength, $\lambda$, inclination, $i$,
central optical depth, $\tau_{V0}$, ratio of bulge-to-disk
scalelengths, $r_{\rm e}/h_R$, and ratio of dust-to-stellar vertical
scaleheights, $h_{z,{\rm dust}}/h_{z,{\rm stars}}$.  We choose a fixed
value for $h_{z,{\rm dust}}/h_{z,{\rm stars}}$, and calculate
$\tau_{V0}$ and $r_{\rm e}/h_R$ for each galaxy directly from the
output of our model. We assign our galaxies random inclination angles,
and then calculate the attenuation factors for the disk and bulge
luminosities at the wavelengths of each of the filters (\eg $B$, $K$)
we are using by interpolating in the tables.

We calculate $\tau_{V0}$ for our model galaxies by assuming that it 
scales as the dust mass per unit area which, in turn, is assumed to
scale with the total mass of metals per unit area in the cold gas:
\begin{equation}
\tau_{V0} \propto {M_{\rm dust} \over  r_{\rm disk}^2} 
\propto {M_{\rm cold} Z_{\rm cold} \over r_{\rm disk}^2}.
\label{eqn:dust}
\end{equation}
The metallicity $Z_{\rm cold}$ is obtained from our chemical evolution
calculation. We normalize equation~(\ref{eqn:dust}) by assuming that gas
with solar metallicity, $Z=0.02$, has the local ISM dust-to-gas ratio. Savage
\& Mathis (\shortcite{sm79}) 
find $A_V/N{\rm _H} = 3.3\times 10^{-22} \mag \cm^{2}$ for the local
ratio of $V$-band extinction, $A_V$, to hydrogen column density
$N_{\rm H}$. This then implies
\begin{equation}
\tau_{V0} = 0.043 \left( {M_{\rm cold}/(2\pi h_R^2) \over \Msol\pc^{-2}}
\right)
\left( {Z_{\rm cold} \over 0.02} \right).
\label{eqn:dustnorm}
\end{equation}

Our standard choice is to use an MW extinction curve and to assume
$h_{z,{\rm dust}}/h_{z,{\rm stars}}=1$. We have investigated
variations in $h_{z,{\rm dust}}/h_{z,{\rm stars}}$ over the range 0.4
to 2.5, and find that most results are very insensitive to this. Most
results also do not change significantly if an SMC rather than an MW
extinction curve is used. The SMC and MW extinction curves differ
significantly only in the far-UV, but even here, the effects on our
results are fairly small, as the net attenuation of galaxy light
calculated using these radiative transfer models has a much weaker
(``greyer'') dependence on wavelength than in a simple foreground
screen model. Our model for dust absorption thus has essentially no
significant free parameters. Results are sensitive mostly to the value
of $\tau_{V0}$, which is calculated directly from our other model
quantities.

Our modelling of dust extinction is a major improvement over what has
been done previously in semi-analytic models, both in terms of
including a realistic 3D distribution for the stars and dust, and in
terms of calculating the dust optical depth in a physically
self-consistent way. The first semi-analytic models to include dust
were those of Lacey \etal (\shortcite{lgrs93}), using the dust and
stellar population model of Guiderdoni \& Rocca-Volmerange
(\shortcite{grv87}), and Guiderdoni \etal (\shortcite{guiderdoni99}).
They modelled the star and dust distributions as
a uniform 1D slab, but calculated the dust content self-consistently
from a closed-box chemical evolution model.  Kauffmann \etal
(\shortcite{kcdw98}) and Somerville \& Primack (\shortcite{sp98}) also
use the 1D slab model, but instead of predicting the slab optical depth,
they use a power-law relation between dust optical depth and galaxy
luminosity that is estimated from observations of $z=0$ galaxies.

The main deficiencies of our current dust model are that it does not
allow for clumping of the dust and stars or deal well with bursts, and
that it only calculates absorption by dust, but not the spectrum of
dust emission. However, Silva \etal (\shortcite{silva98}) have
developed a more sophisticated dust model which includes both clumped
and smooth components of the dust, deals accurately with bursts and is
able to predict not only the extinction of starlight, but also the
spectrum of the energy re-radiated by the dust in the far infra-red
and sub-mm. Granato \etal (\shortcite{granato99}) combine this dust
model with our galaxy formation model to predict galaxy luminosity
functions in the far infra-red and sub-mm and Lacey \etal
(\shortcite{lacey99}) investigate the high redshift behaviour and
predict number counts and integrated radiation backgrounds.

\subsection{Emission line Modelling}

We model the emission lines from photo-ionized gas in our galaxies by
calculating the luminosity in Lyman continuum photons from the stellar
population using the Bruzual \& Charlot models, and combining this
with HII region models to calculate line luminosities and equivalent
widths. We have calculated results for important lines used as star
formation indicators, such as H$\alpha$ and OII. This is described
in detail in a separate paper (\cite{empaper}).

%% file: method.tex
\section{Methodolgy}
\label{sec:method}

\subsection{Model Parameters}

The complete hierarchical model of galaxy formation described in the
previous section contains a significant number of parameters.
However, relatively few should be considered as {\it free} parameters.
These fall into three distinct
categories: numerical parameters, parameters of the cosmological model
and, finally, parameters related directly to our modelling of the
physics of galaxy formation.

The parameters that fall in the first set include the mass resolution,
$M_{\rm res}$, the number of timesteps in the merger tree, $N_{\rm
steps}$ and the starting redshift, $z_{\rm start}$.  These do not
represent freedoms of the model, and we must simply choose values such
that the quantities of interest have converged and are insensitive to
further improvements. Also, there are options such as adopting singular
isothermal spheres to describe the dark matter and gas density
profiles, or varying the distribution of halo spin parameters, which
are not to be viewed as viable alternatives. Instead, we have included
them simply in order to be able to vary our assumptions so as to gain
insight into why the model behaves in a particular way. In these
examples, any viable model should employ the options that are
consistent with results of the high-resolution simulations that
we are trying to emulate.

The second set are parameters that specify the background cosmological
model. These include the density parameter, $\Omega_0$; the
cosmological constant, $\Lambda_0$; the Hubble constant, $h$; the
baryon density, $\Omegab$; and the shape and amplitude of the linear
theory mass
power spectrum, $P(k)$.  In principle, each of these can be determined
from observations that do not depend on galaxy properties. For
example, most of these cosmological parameters are likely to be
determined accurately from microwave background anisotropy
measurements to be carried out by the MAP and Planck satellite
missions (\eg \cite{bet97}).  Alternatively, the power spectrum
amplitude, $\sigma_8$, may be fixed by reference to the abundance of
galaxy clusters, while the baryon density may be constrained by models
of primordial nucleosynthesis and the observed abundance of the 
light elements, like deuterium, at low and  high redshift.  
Our general approach is to set these parameters
according to such external constraints.  However, some properties of
the galaxy formation models are particularly sensitive to the baryon
density, $\Omegab$, and to the normalization, $\sigma_8$. For this
reason, we sometimes allow some variation of these parameters around
the values otherwise indicated by the external observational
constraints.

The final set of parameters are those with which we directly
characterize our physical model of galaxy formation. Firstly, there is
the IMF and its associated yield of metals, $p$, and the fraction,
$R$, of stellar mass that is liberated in stellar winds and SNe. In
principle, $p$ and $R$ are fixed by the choice of IMF, but, in
practice, although $R$ is quite well constrained, $p$ is quite
uncertain.  Secondly, we have the parameters, $\epsilon_\star$,
$\alpha_\star$, $\vhot$ and~$\alphahot$ in the star formation and
feedback laws.  Then, there is the parameter, $e$, the fraction of the
metals produced in SNe which escape directly to the hot diffuse halo
gas.  Also, we require the vertical scale-height of dust relative to
that of the disk stars (although our results are very insensitive to
this parameter). Finally, there are the parameters $\fdf$
and~$\fellip$, which modulate the frequency of galaxy-galaxy mergers
and determine when a merger results in the formation of a spheroid.
Although the number of model parameters is not small, we shall see
that the resulting freedoms of the model are still quite limited and
that only a small subset of observed properties of low redshift
galaxies are needed to constrain a model fully.

\subsection{Model Output}

The output of our code is a list of the galaxies that form in each
simulated halo at one or more redshifts. For each galaxy the output
lists: a flag which indicates whether the galaxy is the central galaxy
of the halo in which it is contained or a satellite galaxy; the mass
of cold gas in the disk; the mass of stars in the disk and bulge; the
luminosities in any chosen band of the stars in the disk and bulge;
selected emission line luminosities and equivalent widths; the
half-mass radius of the disk and bulge individually and combined; the
circular velocities at the half-mass radii of the disk and bulge; the
metallicity of the cold gas and also, if the galaxy is a central
galaxy, the metallicity of its hot gas halo; the
metallicities and age of the bulge and disk stars weighted by mass or
by luminosity in any selected band; the instantaneous star formation
rate in the disk; the mass and circular velocity at the virial radius
of both the halo in which the galaxy was last a central galaxy and the
halo in which it is contained at the chosen output redshift.  The
effect of dust within each galaxy on the luminosities and line
strengths is computed in an additional step by assuming an inclination
angle for each galaxy. Also, since we know the disk and
bulge sizes, we can compute surface brightness distributions and
isophotal magnitudes for each individual galaxy, assuming exponential
profiles for disks and $r^{1/4}$ profiles for spheroids. Since we also
know the number density of each of the halos we have simulated, it is
straightforward to estimate galaxy luminosity functions and galaxy
number counts (both using either total or isophotal magnitudes) or to
sample our output to build up either volume-limited or
magnitude-limited galaxy catalogues from which to draw galaxy samples
for comparison with observational datasets. In this work we
have used the halo abundance given by the Press-Schechter formalism,
but it is now possible to adopt improved analytic estimates 
(Sheth, Mo \& Tormen \shortcite{smt99}) which accurately match the 
abundance of halos found in large N-body simulations 
(Jenkins \etal \shortcite{jfwccey00}). We have checked that switching 
to these more accurate formulae changes the model results far less 
than varying some of the galaxy formation parameters.

Once we have calculated a model, we have the ability to select from
the output any particular galaxy and recompute its formation history,
this time choosing to output its properties more frequently and to
record its star formation and merger history.  In this way we can
generate the complete formation history of selected galaxies and, for
each, construct their own individual merger trees. Examples of galaxy
merger trees constructed in this way have been presented in Figure~9
of Baugh \etal (\shortcite{bcfl98}).

\subsection{Strategy}

Our adopted methodology is to select a cosmological model based on
constraints from large-scale structure and then vary the galaxy
formation parameters in order to match as best as possible a selection
of low-redshift observational data. Since galaxy formation is
undoubtedly a complex process, the simple model we have constructed
cannot aspire to be a complete and full description.  Thus, it is
inevitable that in some cases our models will only produce a moderate
level of agreement with certain observational data.

In the following section, we illustrate this process of defining the
parameters of the galaxy formation model for one particular cosmology. We
use this example to illustrate the way in which we apply the observational
constraints and to show how the model predictions depend on each of the
model parameters.

%% file: lcdm.tex
\section{Observational constraints on the model and effects of varying
the parameters}
\label{sec:cons}

In the following sub-sections, we compare models constructed within a
particular cosmology with a variety of statistics estimated from the
observed properties of the local galaxy population. For each
statistic, we illustrate how the predictions of the galaxy formation
model depend on the parameters, and present one model, our reference
model, which is the best compromise when measured against a full range
of observational data. The observational constraints used to fix each
of the main parameters of our reference model are as follows:
\begin{itemize}
\item $\alphahot$: faint end of luminosity function and Tully-Fisher
relation
\item $\Vhot$: faint end of luminosity function and sizes of low
luminosity spirals
\item $\epsilon_\star$: gas fraction for $L_\star$ spirals
\item $\alphastar$: variation of gas fraction with luminosity
\item $\fellip$: morphological mix for $L_\star$ galaxies
\item IMF: observations of solar neighbourhood
\item $\Upsilon$: $L_\star$ in luminosity function
\item $p$: metallicity of $L_\star$ ellipticals
\end{itemize}
The chosen parameters values of our reference model are listed in
Table~1.

We emphasize that not all of the observational data presented in this
section are used to fix model parameters - some of the data provide
tests of the model, and are shown in this section to illustrate the
effects of varying the parameters.

The cosmology that we have chosen in order to illustrate how
observational data may be used to constrain the galaxy formation
parameters is a flat, low-density cold dark matter model with a
cosmological constant. The parameters that we adopt for this
$\Lambda$CDM model are $\Omega_0=0.3$ and $\Lambda_0=0.7$.  Such a
model is currently favoured by quite a range of observational
evidence. For reasonable values of the Hubble constant ($h\sim 0.7$),
the shape of the mass power spectrum is in good agreement with
estimates from large-scale galaxy clustering (\eg \cite{mesl90}). The
value of $\Omega_0$ is consistent with the high baryon fraction in
clusters (\cite{WENF93,Wet96,me97}) and with the mild evolution in the
abundance of X-ray clusters (\cite{ecfh97}). The joint values of
$\Omega_0$ and $\Lambda_0$ are in accord with estimates from high
redshift SNe (\cite{perlmutter98,riess98,garn99}), while
$\Omega_0+\Lambda_0=1$ is in agreement with the detection of
the first CMB Doppler peak (\cite{Bernardis00,Lange00,Hannay00,Balbi00}).
%(\eg \cite{bartlett98,hancock98,lineweaver98}).
For this model, the normalization of the mass power spectrum,
$\sigma_8$, derived from the number density of X-ray emitting galaxy
clusters is consistent with that from the amplitude of CMB
fluctuations (\eg \cite{cwfr97}), and both are consistent with the
power spectrum of the mass at $z=2.5$, inferred by Croft \etal
(\shortcite{croft98}) and Weinberg \etal (\shortcite{weinberg98}) from
analysis of the Lyman-$\alpha$ forest.

\begin{table*}
\centering
\caption{The values of the model parameters for the reference model.}
\begin{center}
\begin{tabular}{lrrrrrrrrrrrrrrrrrr} %\hline
\multicolumn{1}{l} {$\Omega_0$} &
\multicolumn{1}{l} {$\Lambda_0$} &
\multicolumn{1}{l} {$\Omega_{\rm b}$} &
\multicolumn{1}{l} {$h$} &
\multicolumn{1}{l} {$\Gamma$} &
\multicolumn{1}{l} {$\sigma_8$} &
\multicolumn{1}{l} {$\epsilon_\star$} &
\multicolumn{1}{l} {$\alpha_\star$} &
\multicolumn{1}{l} {$V_{\rm hot}$} &
\multicolumn{1}{l} {$\alpha_{\rm hot}$} &
\multicolumn{1}{l} {$e$} &
\multicolumn{1}{l} {$f_{\rm ellip}$} &
\multicolumn{1}{l} {$f_{\rm df}$} &
\multicolumn{1}{l} {IMF} &
\multicolumn{1}{l} {$p$} &
\multicolumn{1}{l} {$R$} &
\multicolumn{1}{l} {$\Upsilon$} & \\
%\hline
\\ 
0.3 & 0.7 & 0.02 & 0.7 & 0.19 & 0.93 & 0.005 & -1.5 &
200.0 & 2.0 & 0.0 & 0.3 & 1.0 & Kennicutt & 0.02 & 0.31 & 1.38 \\
%\hline			        
\end{tabular} 			 
\end{center}
\end{table*}

The galaxy formation model as a whole is a complex system, with the
result that the dependence of a particular statistic on a given
parameter can be complicated. Thus, when one parameter is varied, the
behaviour of the model certainly depends on the other constraints that
have been applied. This fact, combined with the number of parameters,
makes it unfeasible to present the full range of possible model
behaviour. One should therefore be careful not to over-interpret the
trends discussed below: since the effects of varying one parameter can
depend on the values of the others, it is dangerous to assume that
these trends can be used to assess the result of varying more than one
parameter at a time.

As well as varying the parameters that regulate the physics of galaxy
formation, we allow some variation of the parameters, $\sigma_8$, $h$
and $\Omega_{\rm b}$, that define the cosmological model. However, these are
only allowed to vary within the ranges permitted for consistency with
estimates of the abundance of rich galaxy clusters, Hubble's constant, 
and primordial nucleosynthesis. Whenever we vary any of these
parameters, we consistently adjust the power spectrum shape parameter,
$\Gamma$, according to the fitting formula for CDM proposed by
Sugiyama (\shortcite{Sug95}):
\begin{equation}
\Gamma = \Omega_0 h \, 
\exp\left( - \frac{\Omegab}{\Omega_0} ( \sqrt{2h}+\Omega_0) 
\right).
\label{eqn:gamma}
\end{equation}
The estimated values and 1-$\sigma$ errors that we adopt for these
quantities are $\sigma_8 = 0.93 \pm 0.07$ (\cite{ecf96}), $h=0.7 \pm
0.1$ (\cite{freedman98,madore98}), and $\Omegab h^2 =0.0125 \pm 0.0025
$ (\cite{walker91}). We note that somewhat higher values of 
$\Omegab h^2$ are favoured by recent estimates of the D/H ratio 
from QSO absorption line systems (\cite{st98,bt98}),
and by models of the mean optical depth of the Lyman-$\alpha$ 
forest (\cite{rauch97,weinberg97}). We discuss the consequences of 
increasing our adopted value of $\Omega_{\rm b}$ in Section~\ref{sec:disc}.

\begin{figure*}
\centering
\centerline{\epsfbox{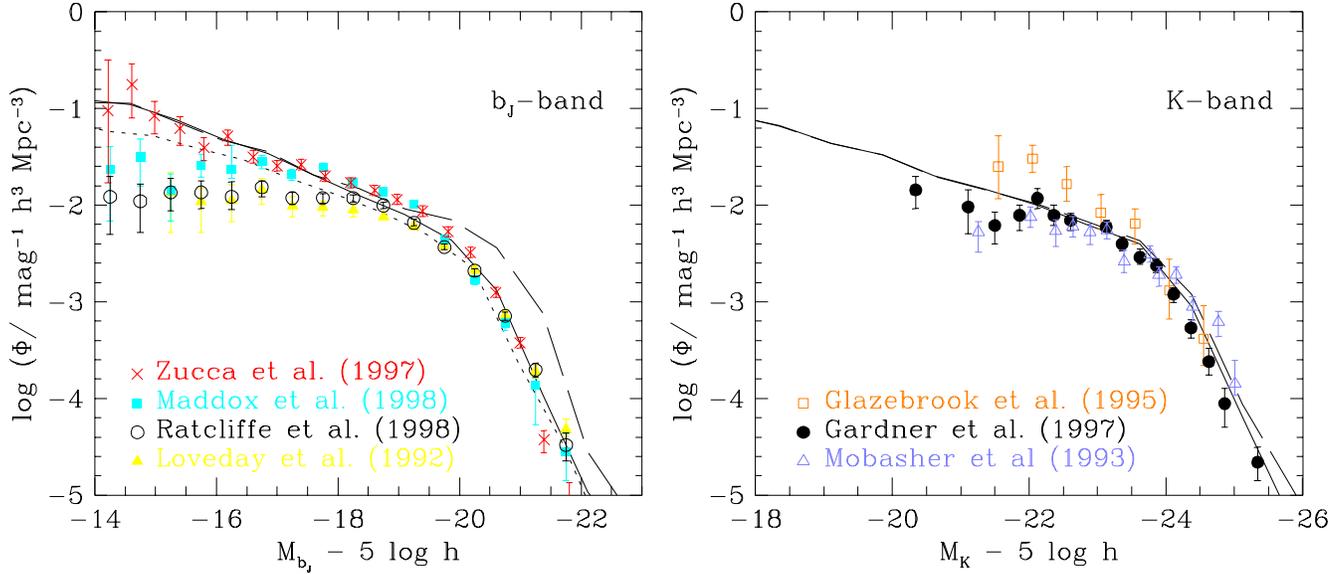}}
\caption{Comparison of the ${\rm b_J}$ and K-band galaxy luminosity functions
in the $\Lambda$CDM reference model with a compilation of observational
data. The solid line is the model including the effects of dust, with
$\Upsilon$ chosen so as to obtain agreement with the observed luminosity
functions at $M_{\rm b_J} - 5
\log h = -19.8 $. The dashed lines show the corresponding luminosity
functions before the effects of dust extinction are included.  The
dotted line on the left hand panel shows how the luminosity function
is modified if the isophotal magnitude within an isophote of 25 ${\rm
mag}/{\rm arcsec}^2$ is used instead of the true total magnitude. See
\S\protect\ref{sec:dsize} for details. 
}
\label{fig:lfone}
\end{figure*}

In a similar fashion, many of the parameters that describe the physics
of galaxy formation can only plausibly be varied by modest
amounts. Consider, for example, the threshold, $\fellip$, above which
galaxy mergers are termed violent and assumed to result in the
formation of an elliptical galaxy. By definition, $\fellip\leq 1$, and
based on simulations of galaxy mergers, a reasonable lower limit is
$\fellip \ga 0.3$ (Walker \etal \shortcite{walker96}; Barnes
\shortcite{barnes98}). Other parameters that are similarly constrained
to lie within relatively narrow ranges are $\fdf$ and, to a lesser
extent, $p$.  The merger timescale coefficient, $\fdf$, should be close
to unity if an infalling galaxy retains its dark matter halo
throughout most of the time when dynamical friction is removing angular
momentum and energy from its orbit. It could be larger than unity if
the dark matter halo is efficiently stripped off at early times, but
cannot plausibly be significantly smaller than unity. As described in
Section~\ref{sec:imf}, the choice of IMF quite accurately determines
the recycled fraction, $R$, and also sets some constraint on the
yield, $p$.

\begin{figure}
\centering
\centerline{\epsfxsize= 8.0 truecm \epsfbox[30 225 290 740]{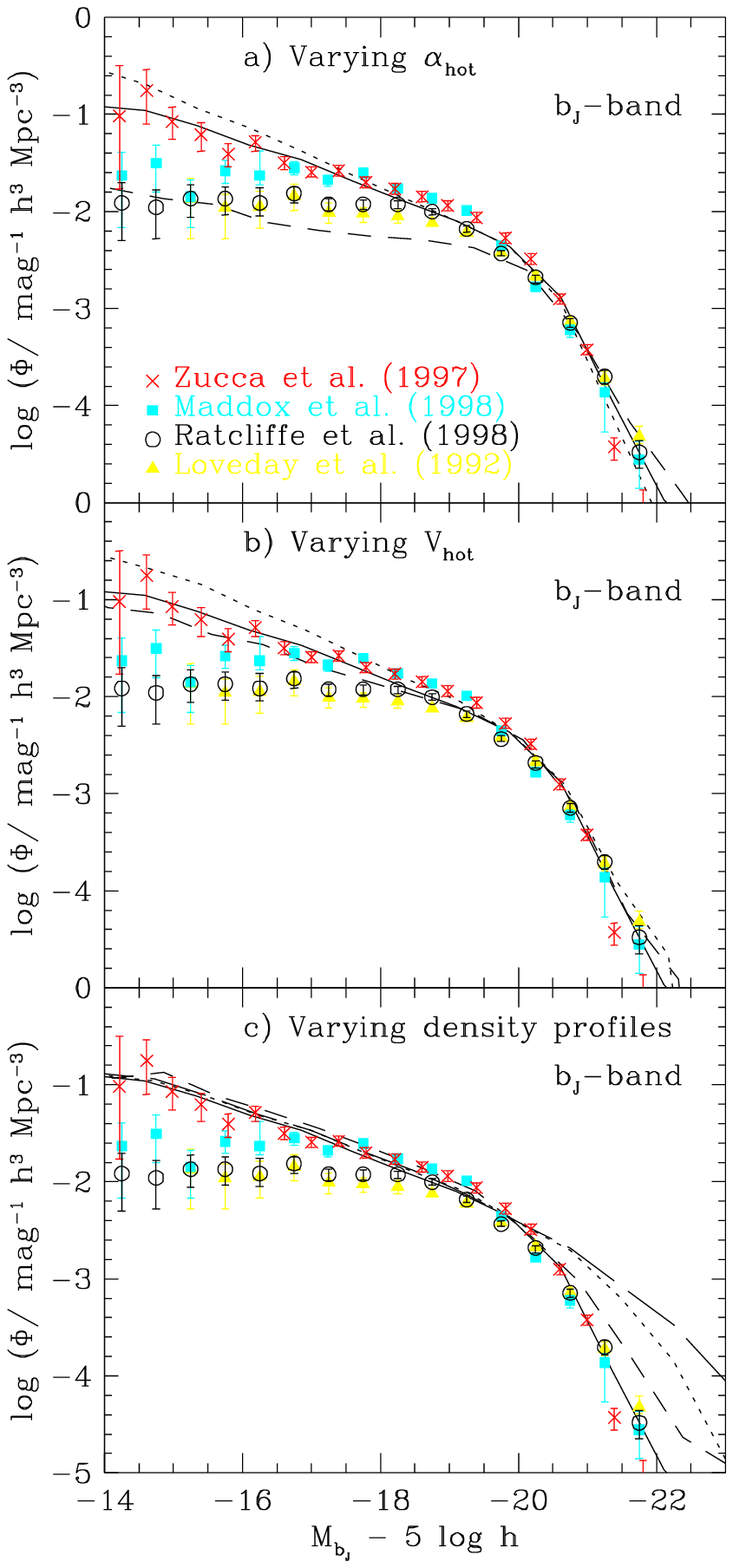}}
\caption{The effect on the 
${\rm b_J}$-band luminosity function of varying the star formation and
feedback parameters, $\alphahot$ and $\Vhot$, and the assumed halo DM
and hot gas density profiles. In all cases, the value of $\Upsilon$ has
been fixed by requiring the model luminosity functions to agree with
the observations at $M_{\rm b_J} - 5 \log h = -19.8 $.
a) Shows how increasing
$\alphahot$ suppresses the formation of low-luminosity
galaxies and so controls the faint-end slope of the luminosity
function.  The dotted curve is for $\alphahot= 1$, solid for
$\alphahot= 2$, and dashed for $\alphahot= 5.5$.  b)
Demonstrates how increasing $\Vhot$ lowers the faint end of the
luminosity function.  Results are shown for $\Vhot= 100\, \kms $ (dotted
curve), $\Vhot= 200\, \kms $ (solid curve),
and $\Vhot= 300\, \kms $ (dashed curve).
c) Shows how the bright end of luminosity function
depends on the model adopted for the density profile of the hot gas. 
The solid curve is our reference model which has an NFW profile for the DM and
the ``$\beta$-model'' for the gas, with a core radius that depends on the
fraction of gas that has previously cooled (see
Section~\ref{sec:hotgas}).  The dotted and long-dashed curves also
assume NFW profiles for the DM, but assume a fixed core radius
``$\beta$-model'' (dotted) or an NFW profile for the gas
(long-dashed).  The short-dashed line is for a singular isothermal
sphere ($\rho \propto r^{-2}$) model for both gas and DM.}
\label{fig:lftwo}
\end{figure}

\subsection{Galaxy Luminosity Functions} % B and K LF
\label{sec:lumfn}

The single most important constraint on our galaxy formation models is
the local galaxy luminosity function. This is one of the most fundamental
properties of the galaxy population and it is also one of the best
measured, at least over a restricted range of luminosities.  Matching the
galaxy luminosity function is a prerequisite for any realistic model of
galaxy formation, because the detectability of galaxies depends directly on
their luminosity. Thus, a model that fails to match the bright end of the
luminosity function can lead to misleading conclusions when tested, for
example, against samples selected by apparent magnitude.
 
Fig.~\ref{fig:lfone} shows estimates of the local galaxy luminosity
function in the blue optical ${\rm b_J}$-band and in the near infrared
K-band.  The ${\rm b_J}$-band data are taken from the APM-Stromlo
galaxy redshift survey (\cite{loveday}), the ESO slice project
(\cite{zucca97}), the DUKST survey (\cite{ratcliffe97}) and from preliminary
results from the 2dF galaxy redshift survey (\cite{maddox}).  The
K-band data are from Mobasher
\etal (\shortcite{mobasher93}), Glazebroook \etal (\shortcite{glazebrook})
and Gardner \etal (\shortcite{gardner}).  In both bands, there is
reasonably good agreement among the various estimates at the bright
end.  At the faint end, however, there is considerable dispersion
among the results from different surveys. In the \BJ-band these
differences are large compared to the statistical errors. We must,
therefore, conclude that either the galaxy luminosity function differs
in the different volumes surveyed, or that systematic differences in
the selection characteristics of the surveys give rise to these
differences.

The solid and dashed model
curves shown in Fig.~\ref{fig:lfone} correspond to the reference
model, both with and without the inclusion of dust. In the model with dust,
the vertical scaleheight of the dust distribution was taken to be the same
as that of the disk stars, but varying this assumption makes very little
difference to the luminosity functions.  The model without dust produces a
reasonable K-band luminosity function, but it gives rise to galaxies which
are systematically too bright in ${\rm b_J}$. By contrast, the model with
dust is a good fit to the bright end of both the ${\rm b_J}$ and K-band
luminosity functions.  We shall see below that the differential effect of
dust, which generates greater extinction at shorter wavelengths, is
important in providing a good match to the observed $B$--$K$ colour
distributions. It also results in a model that comes significantly closer
than our previous models to matching {\em simultaneously} the zero-point of the
observed I-band Tully-Fisher relation and the bright end of the ${\rm
b_J}$-band luminosity function, thus largely overcoming an important
shortcoming of our earlier models (\cite{cafnz}).  The importance 
of the role of dust in achieving this simultaneous match has also
been noted by Kauffmann \etal (\shortcite{kcdw98}) and 
Somerville \& Primack (\shortcite{sp98}).
The prescription for the
dust distribution assumed in Fig.~\ref{fig:lfone} will be retained in all
the following comparisons.

Many of the model parameters have a direct effect on the galaxy
luminosity function. Typically, varying any one of them on its own
causes only a minor change in the {\it shape} of the luminosity
function, but can cause a significant overall shift (either left or
right) in the luminosity scale.  Our normalization strategy separates
out these two effects by adjusting the parameter $\Upsilon$ so as to
keep the amplitude of all the luminosity functions fixed at $M_{\rm
b_J}= -19.8+5\logh$. Recall that $\Upsilon^{-1}$ is the initial
stellar mass fraction in luminous stars (equation~\ref{eqn:upsilon}),
and that the remaining fraction is assumed to be made up of
non-luminous brown dwarfs. Physically, one ought to vary the net
recycled fraction $R$ when adjusting $\Upsilon$, so as to keep the
value $R_1=\Upsilon R$ constant for the luminous stars alone 
(\S~\ref{sec:yield}). The reference model has $\Upsilon$ and $R$ chosen 
consistently to give $R_1=\Upsilon R = 0.42$ for the Kennicutt IMF, 
as found in Section~\ref{sec:yield}. To achieve this requires some 
iteration as the value of $\Upsilon$ is determined after the model has
been run by matching a point in the B-band luminosity function.
For this reason, when showing the effects of varying parameters
we choose to keep $R$ constant  rather than $R_1$.

The effect of parameter changes on the position of the luminosity
function is summarized in Table~2. Here, we give the values of
$\Upsilon$ required to shift each luminosity function into coincidence
with the Zucca \etal (\shortcite{zucca97}) luminosity function
at $M_{\rm b_J}= -19.8+5\logh$. We will see in Section~\ref{sec:m_l}
that in models which have realistic stellar mass-to-light ratios,
$\Upsilon \sim 1.3-2$ for the Kennicutt IMF.  
In a few cases, however, this normalization
procedure requires $\Upsilon<1$, which is unphysical. (It
corresponds approximately to removing from the IMF all of the brown
dwarfs and some of the low mass visible stars, by increasing the lower
mass limit above $0.1 \Msol$.) Nevertheless, we present these models
because they help to clarify some of the trends that occur when a
single parameter is varied. For the derived normalization of
$\Upsilon$, the table gives other properties of the galaxy
population. Several of the entries in this table are discussed further
in the relevant sections below.

\begin{table*}
\centering
\caption{The variation of the average colour, the mass-to-light ratio 
of the stellar populations and the zero-point of the Tully-Fisher
relation with model parameters.  In all cases, $\Upsilon$ is adjusted
so as to match the observed $\rm b_J$-band luminosity function at
$M_{\rm b_J}= -19.8+5\logh$.  The first column lists the parameter
that has been varied relative to the reference $\Lambda$CDM model. The
second column gives the required value of $\Upsilon$. The third lists
the median $B$--$K$ colour for galaxies in the range $-24.5 < M_{\rm
K}-5\logh< -23.5$.  The following four columns list the median B- and
I-band stellar mass-to-light ratios, in units of $h{\rm
M_\odot/L_\odot}$, for disk and elliptical galaxies with $-20<M_{\rm
B}-5\logh< -19.0$. These are compared with observed values 
in Section~\ref{sec:m_l}.
The last two columns show the offset in magnitudes
from the observed I-band Tully-Fisher relation at $V_{\rm disk}=160\,\kms$, 
and the median ratio of the circular velocity of the disk to that at
the virial radius of the halo in which it formed for galaxies with
$-20<M_{\rm I}-5\logh<-18$.}
\begin{center}
\begin{tabular}{lrrrrrrrrrrrrrrrrrr} %\hline
\multicolumn{1}{l} {Modified Parameter} &
\multicolumn{1}{c} {$\Upsilon$} &
\multicolumn{1}{c} {$B$--$K$} &
\multicolumn{1}{c} {Disk: $M/L_{\rm B}$} &
\multicolumn{1}{c} {Disk: $M/L_{\rm I}$} &
\multicolumn{1}{c} {Elliptical: $M/L_{\rm B}$} &
\multicolumn{1}{c} {Elliptical: $M/L_{\rm I}$} &
\multicolumn{1}{c} {$\Delta M_I$} &
\multicolumn{1}{c} {$V_{\rm disk}/V_{\rm halo}$} \\
%\hline 
\\
Reference Model         & 1.38 & 3.86 & 2.0 &  1.8 & 5.4 & 2.9 &  0.98 & 1.35 \\ % 0
No Dust                 & 2.30 & 3.60 & 2.1 &  2.4 & 8.7 & 4.9 &  1.48 & 1.41 \\ % 0 no dust
$ \alphahot=1 $         & 1.39 & 3.81 & 2.1 &  1.8 & 4.8 & 2.8 &  1.13 & 1.50 \\ % 1
$ \alphahot=5.5 $       & 1.1  & 4.01 & 1.4 &  1.2 & 4.3 & 2.3 &  1.02 & 1.02 \\ % 2
$ \vhot=100\, \kms$     & 1.32 & 4.11 & 2.5 &  2.0 & 6.3 & 3.1 &  1.73 & 1.71 \\ % 3
$ \vhot=300\, \kms$     & 1.15 & 3.56 & 1.2 &  1.2 & 2.5 & 1.8 &  0.75 & 1.20 \\ % 4
$ \alphastar=0.0  $     & 1.28 & 3.96 & 2.0 &  1.7 & 5.4 & 2.8 &  0.83 & 1.27 \\ % 5
$ \alphastar=-2.5 $     & 1.31 & 3.81 & 1.7 &  1.5 & 5.0 & 2.8 &  1.09 & 1.41 \\ % 6
$\Omegab=0.01$          & 0.45 & 3.62 & 0.5 &  0.5 & 1.0 & 0.7 &  0.23 & 1.13 \\ % 7
$\Omegab=0.04$          & 3.07 & 4.11 & 7.3 &  4.7 & 13.9& 6.9 &  2.12 & 1.96 \\ % 8
$ h=0.6 $               & 0.95 & 3.87 & 1.7 &  1.5 & 4.8 & 2.6 &  0.92 & 1.34 \\ % 9
$ h=0.8 $               & 1.77 & 3.86 & 2.1 &  1.9 & 5.4 & 3.0 &  1.19 & 1.38 \\ % 10
$ \epsilon_\star=0.01$  & 1.70 & 3.85 & 2.1 &  1.9 & 7.4 & 3.9 &  1.12 & 1.30 \\ % 11
$ \epsilon_\star=0.0033$& 1.13 & 3.90 & 1.8 &  1.5 & 4.5 & 2.5 &  1.03 & 1.40 \\ % 12
$ p=0.0075 $            & 1.66 & 3.28 & 1.6 &  1.7 & 3.9 & 2.7 &  1.01 & 1.36 \\ % 13
$ p=0.03 $              & 1.17 & 4.16 & 2.1 &  1.7 & 5.7 & 2.8 &  0.92 & 1.34 \\ % 14
$ R=0.19 $              & 1.31 & 3.80 & 2.1 &  1.9 & 5.9 & 3.2 &  1.07 & 1.38 \\ % 15
$ R=0.49 $              & 1.41 & 3.97 & 1.7 &  1.4 & 3.7 & 2.0 &  0.85 & 1.31 \\ % 16
$ \sigma_8=0.86 $       & 1.43 & 3.82 & 2.0 &  1.7 & 4.2 & 2.5 &  1.03 & 1.34 \\ % 17
$ \sigma_8=1.0 $        & 1.48 & 3.90 & 2.2 &  1.9 & 6.1 & 3.3 &  1.20 & 1.38 \\ % 18
IMF: Salpeter, $R=0.28$ & 0.79 & 3.85 & 1.9 &  1.7 & 5.5 & 3.0 &  1.00 & 1.35 \\ % 19
$f_{\rm form}=1.5$      & 1.27 & 3.82 & 1.9 &  1.6 & 5.1 & 2.8 &  0.93 & 1.42 \\ % 29
$f_{\rm df}=0.5$        & 1.34 & 3.89 & 2.0 &  1.7 & 4.4 & 2.5 &  1.03 & 1.36 \\ % 20
$f_{\rm df}=2.0$        & 1.16 & 3.80 & 1.5 &  1.4 & 4.8 & 2.6 &  0.84 & 1.34 \\ % 21
$f_{\rm ellip}=0.5$     & 1.38 & 3.90 & 2.1 &  1.8 & 5.8 & 3.1 &  0.99 & 1.35 \\ % 22
$r_{\rm core}=r_{\rm NFW}/6$   & 1.40 & 3.85 & 2.0 &  1.8 & 5.4 & 3.0 &  1.05 & 1.35 \\ % 25
Fixed gas core radius   & 1.23 & 3.97 & 2.3 &  1.8 & 2.9 & 1.9 &  0.87 & 1.34 \\ % 26
NFW gas traces DM       & 1.23 & 4.04 & 2.5 &  1.8 & 3.1 & 2.0 &  0.92 & 1.34 \\ % 27
SIS gas traces DM       & 1.08 & 4.24 & 2.4 &  1.7 & 3.4 & 2.0 &  0.95 & 1.41 \\ % 28
Unstable disks          & 1.50 & 3.91 & 2.1 &  1.9 & 4.7 & 2.8 &  1.11 & 1.28 \\ % 24
Accretion by disk       & 1.40 & 3.89 & 2.3 &  1.9 & 5.5 & 3.0 &  1.12 & 1.38 \\ % 23
%Salpeter BC1996        & 0.80 & 3.84 & 1.9 &  1.8 & 5.8 & 3.2 &  1.05 & 1.35 \\ % 31
%sigma_lam=0.35         & 1.43 & 3.85 & 2.2 &  1.8 & 5.6 & 3.1 &  0.99 & 1.34 \\ % 40
%Jenkins 2000 Ups fixed & 1.2 & 3.86 & 1.9 &  1.7 & 5.7 & 3.1 &  1.04 & 1.35 \\ % 30 
%Reference Model         & 1.38 & 3.86 & 2.0 &  1.8 & 5.4 & 2.9 &  0.99 & 1.35 \\ % 0

%\hline			        
\end{tabular} 			 
\end{center}
\end{table*}

Two parameters that strongly affect the shape of the
galaxy luminosity function are $V_{\rm hot}$ and $\alphahot$, the
quantities that define our model of stellar feedback (equation
\ref{eqn:feedback}). As can be seen in Fig.~\ref{fig:lftwo}a, the faint-end
slope of the luminosity function is very sensitive to $\alphahot$,
with large values producing a shallower slope.
Similarly, as Fig.~\ref{fig:lftwo}b shows, increasing $V_{\rm hot}$
also reduces the number of faint galaxies.  
%It does so by changing the
%amplitude rather than the slope at very low luminosities, but over the
%luminosity range plotted in the figures, this also appears as a change
%in the faint-end slope. 
Both these dependencies are easily understood:
stronger stellar feedback makes it increasingly more difficult for
luminous stars to form in low-mass halos.  To match the  very
shallow faint-end slope seen in the data of Loveday \etal
(\shortcite{loveday}) or Ratcliffe \etal (\shortcite{ratcliffe97}),
requires a high value of $\alphahot$, such as that adopted by Cole
\etal (\shortcite{cafnz}), who compared their models against the first
of these surveys.  Such a value, however, would lead to a disagreement
with the data of Zucca \etal (\shortcite{zucca97}).
The differing observational estimates of the
luminosity function indicate that the faint-end slope is not as
robustly determined as one might wish, perhaps because it depends on
the details of the survey selection criteria. We therefore do not use it
as a model constraint. Instead, we will see in
Section~\ref{sec:TF} that extreme values of $\alphahot$ are disfavoured
on other grounds and this leads us to favour models whose luminosity 
functions have quite steep faint end slopes.

The bright end of the luminosity function is sensitive to the density
profile assumed for the halo gas, because this controls how much of
the gas can cool. This effect is not important in low-mass halos, in
which the gas temperature $\Tvir$ is low enough that most of the gas
can cool anyway, but it becomes important in large groups and
clusters, in which only the dense central regions have time to cool.
Fig.~\ref{fig:lftwo}c compares the effects of using different gas
profiles. Our reference model (shown by the solid line) assumes an NFW
dark halo and a $\beta$-model for the gas (equation~\ref{eqn:rhogas}),
with a core radius that starts at $r_{\rm core}= \rnfw/3$ and grows
depending on how much gas has already cooled in progenitor halos. 
The model luminosity function and other properties are
not sensitive to the precise value of this initial core radius.
For example, if instead, we set $r_{\rm core}= \rnfw/6$ as the initial value, 
then the change in the luminosity
function is almost too small to be visible and  the other properties
listed in Table~2 also only vary slightly.
In principle, constraints can be placed on the initial gas
density profile from the observed X-ray emission profiles of groups and
clusters, but in practice this requires complex modelling to take
account of the emission associated with the central cooling flow.

Our model fits the observed bright end of the luminosity function well. It
is compared in the figure to a model in which the gas core radius 
is kept  fixed at $r_{\rm core}= \rnfw/3$ (dotted curve), 
another in which both gas and dark matter have the same NFW profile
(long-dashed curve), and finally to a model in which
both gas and dark matter have singular isothermal sphere profiles
(short-dashed curve), as
has been assumed in most previous work. The latter three models produce 
many more high luminosity $L\gsim L_\star$ galaxies than is
observed. This difference in the assumed halo gas profiles explains
most of the differences in the shape of the bright end of the
luminosity function between our reference model and the models of
Kauffmann \etal (\shortcite{kwg93},\shortcite{kcdw98}) 
and Somerville (\shortcite{s97}),
although the procedure in these papers of using the
Tully-Fisher relation rather than the luminosity function as the
primary observational constraint also has an effect. These authors
all invoke an artificial cut-off on the circular velocity of
halos in which gas is allowed to cool to form visible stars, in order
to ameliorate their problems in fitting the luminosity function. 
We believe that our standard model for cooling is the most physically
reasonable in this regard, for the reasons given in
Section~\ref{sec:hotgas}.
More recently, Somerville \& Primack (\shortcite{sp98}) have presented
models with no artificial cooling cut-off which nevertheless produce
a good match to the bright-end of the B-band luminosity function.
In this case, this improvement is achieved partly as a result of the empirical 
dust model they have adopted, which has an extinction that increases 
with increasing
galaxy luminosity. However, as dust has less effect in the K-band, 
they find that for some of their models, the shape of the bright-end of
the K-band luminosity function remains a poor match to observations.

As was shown in Cole \etal (\shortcite{cafnz}), the shape of the luminosity
function is also influenced by the efficiency of galaxy mergers, which is
controlled by $\fdf$. However, we now impose the constraint $\fdf
\gsim 1$, a limit suggested by numerical simulations (\cite{nfw95}), which 
also leads to an acceptable morphological mix in the model.  With this
bound, the residual variation in the shape of the luminosity function
with $\fdf$ is small compared to its dependence on the feedback
parameters $V_{\rm hot}$ and $\alphahot$ and on the halo gas
profile. Our treatment of feedback is, in fact, the main factor
responsible for the overall shape of the model luminosity function at
$L\lsim L_\star$, and our assumptions about cooling are the main
determinant of the shape at $L\gsim L_\star$.

\subsection{The Tully-Fisher Relation} % I-band TF
\label{sec:TF}

\begin{figure} \centering
\centerline{\epsfxsize= 8.0 truecm \epsfbox[30 225 290 740]{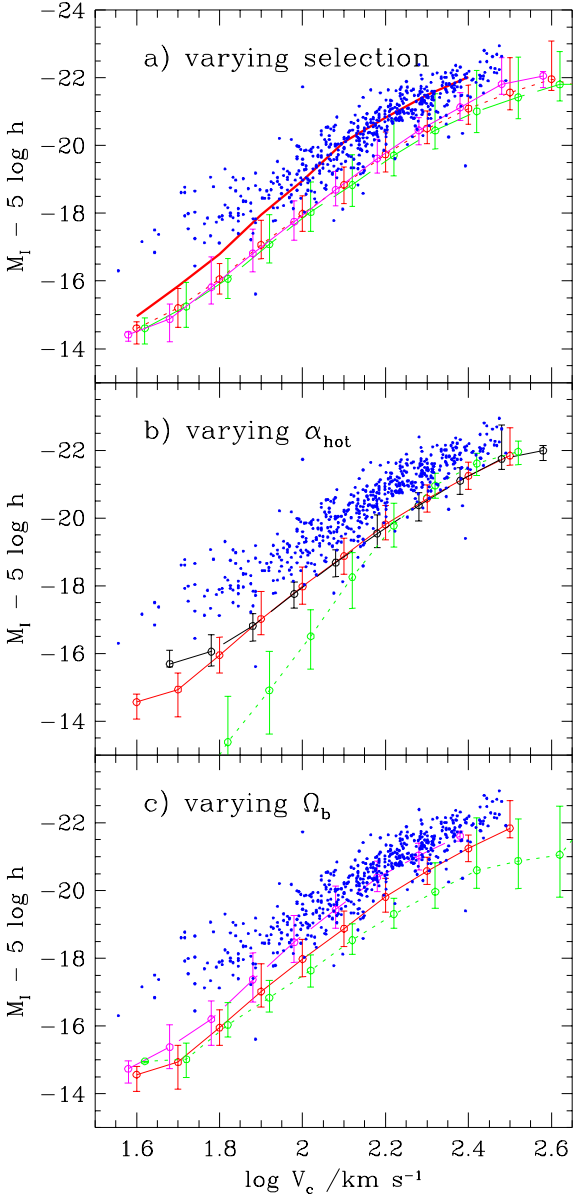}}
\caption{The dependence of the model I-band Tully-Fisher relation on
a) the sample selection criteria, b) the feedback parameter,
$\alphahot$, and c) the baryon density, $\Omegab$.
In each case, the model curves trace the median magnitude as
a function of circular velocity, and the errorbars show the 10 and 90
percentiles of the distribution. The magnitudes are face-on values,
including the effects of dust. The points show the observed
distribution for a subsample of Sb-Sd galaxies selected by de Jong \& Lacey
(\protect\shortcite{djl99}) from the Mathewson, Ford \&
Buchhorn(\protect\shortcite{mathewson}) catalogue and, again, all magnitudes
have been corrected to face-on values.  All the curves in the top
panel are for the reference model, but for different galaxy selection
criteria. The dotted line is for all spiral galaxies which are the
central galaxies in their halos, the dashed line for all spiral
galaxies (both central and satellite), and the solid line for all
star-forming spiral galaxies with gas fractions of 10\% or greater
(this final selection is retained in~b \&~c).
The thick solid line shows, for central galaxies, the result of using
the circular velocity at the virial radius of the halo in which the
galaxy formed, rather than the disk circular velocity.  In b),
 the solid line refers to the reference model (which has
$\alphahot=2$), while the dashed line is for $\alphahot=1$ and the
dotted line for $\alphahot=5.5$.  In c), the solid line
refers, again, to the reference model (which has $\Omegab=0.02$),
while the dashed line is for $\Omegab=0.01$ and the dotted line for
$\Omega_{\rm b}=0.04$.}
\label{fig:tf1}
\end{figure}

In Fig.~\ref{fig:tf1}, we compare our predicted I-band Tully-Fisher
(TF) relation with the observed relation defined by the complete
diameter-limited subset of spiral galaxies selected by de Jong \&
Lacey (\shortcite{djl99}) from the catalogue of Mathewson \etal
(\shortcite{mathewson}). The observed circular velocity plotted here
is the maximum, $V_{\rm max}$, of the measured rotation curve. The
observed I-band magnitudes have been corrected to face-on values.  For
the model, we use the dust-extincted I-band magnitude for galaxies
seen face-on and the circular velocity, $\Vdisk$, evaluated at the
half-mass radius of the disk. $\Vdisk$ includes the self-gravity of
the disk, and is evaluated in the disk mid-plane, as discussed in
Appendix~\ref{app:contract}. The peak of the rotation curve may occur
at a radius other than the half-mass radius, in which case the
quantity, $\Vdisk$, that we plot may be systematically low compared to
the measured $V_{\rm max}$.  However, we expect the difference to be
small since our model galaxies typically have reasonably flat rotation
curves, as indicated by the values of $\Vdisk/\Vhalo$ listed in
Table~2, which are close to unity.

The upper-panel in Fig.~\ref{fig:tf1}a shows how the model TF relation
depends on the sample selection criteria.  In all cases, we have selected
disk-dominated galaxies with dust-extincted I-band bulge-to-total
light ratios in the range $0.02$ to $0.24$, to match approximately the
range of galaxy types, Sb-Sd, selected in the Mathewson \etal
(\shortcite{mathewson}) catalogue.  This makes use of the approximate
conversion between Hubble T-type, (${\rm T}=3$--$7$ for Sb-Sd) and
bulge-to-disk ratio, described in Baugh \etal (\shortcite{bcf96b}) and
based on the data of Simien \& de Vaucoleurs (\shortcite{sv86}).  The
model TF relation for this complete galaxy sample is shown by the
dashed line in Fig.~\ref{fig:tf1}a. The slope of the
predicted TF relation is close to that of the data, and this remains
true for the sub-samples discussed below.  It has an offset of $1.2$
magnitudes relative to the zero-point of the observed relation and a
spread between the 10 and 90~percentiles of the distribution at $V_c=
160\,\kms$ of $1.7$ magnitudes.  This spread is significantly larger
than that for the observational sample, which is $1.1$ magnitudes.

In Cole \etal (\shortcite{cafnz}) the TF relation was plotted for
galaxies at the centres of halos only. Here the result of selecting
only central galaxies is shown by the dotted line. The exclusion of
satellite galaxies removes some galaxies which have exhausted their
reservoirs of cold gas and so have faded as their stellar populations
have aged. This has the effect of producing a
somewhat tighter TF correlation, with a spread of only $1.2$ magnitude,
and of reducing the offset in the zero point to $1.0$ magnitudes 
(at $V_{\rm c}= 160\, \kms$). A more realistic selection is 
to consider only disk galaxies with a
significant cold gas fraction, which we take to be $M_{\rm
gas}/(M_{\rm gas}+M_{\rm stars})> 10$\%.  This is reasonable, as
without ongoing star formation, disk galaxies will not have prominent,
recognizable spiral arm features. In addition, interstellar gas is
required for the measurement of the rotation velocity in TF datasets,
either to produce the emission lines from which optical rotation
curves are measured, or to produce the HI emission used in HI rotation
measurements. The TF relation for this subsample of star-forming
spiral galaxies is shown by the solid curve in Fig.~\ref{fig:tf1}a,
and is repeated as the solid curve in the lower two panels.  It has a
spread of $0.98$ magnitudes, which is slightly smaller than the
observed spread of $1.1$ magnitudes.  The offset in the zero point of
the relation is $0.98$ magnitudes,  
which is equivalent to a factor of approximately $1.3$ in circular velocity.
Since the effective 
mass-to-light ratio in our models is normalized (through $\Upsilon$)
by reference to the bright end of the \BJ-band luminosity function, we
find that both the zero-point and the scatter in the model
Tully-Fisher relation are insensitive to most changes in the galaxy
formation model parameters.

The parameters that do have an effect are $\alphahot$ and $\Omegab$.
This is illustrated by the curves in Fig.~\ref{fig:tf1}b and~7c
respectively.  Increasing the feedback parameter, $\alphahot$, makes
it increasingly difficult to form stars in low-circular velocity
galaxies. Consequently, the luminosity of low-$\Vdisk$ galaxies is
reduced, and the model Tully-Fisher relation bends away from the
observed correlation at faint magnitudes. A value of $\alphahot
\approx 2$ is required to produce a correlation that runs parallel to
the observed relation over the full range of magnitudes probed by the
data.  The effect of increasing $\Omegab$ (Fig.~\ref{fig:tf1}c) is to
cause the model Tully-Fisher relation to bend away from the
observations at bright magnitudes. The reason for this is that, in
order to maintain a match to the bright end of the \BJ-band luminosity
function, larger $\Omegab$ requires a larger $\Upsilon$ and thus
larger mass-to-light ratios for all the galaxies. The self-gravity of
bright spiral disks then plays a larger role in determining the
galaxy's rotation curve. This is quantified by the ratio,
$\Vdisk/\Vhalo$, listed in Table~2, which increases substantially as
$\Omegab$ is increased. It is this effect that leads us to favour a
relatively low value of $\Omegab$ as compared to the currently most
favoured numbers derived from primordial nucleosynthesis
considerations.  Note that a very low value, $\Omegab=0.01$, results
in a good match to the zero-point of the Tully-Fisher
relation. However, this apparent success is at the cost of an
unphysical value, $\Upsilon=0.49$, required to make galaxies bright
enough to match the \BJ-band luminosity function.

The failure of our model to produce a Tully-Fisher relation with a
zero-point that matches the observations well is reminiscent of a
similar shortcoming in the earlier $\Omega=1$ CDM model of Cole \etal
(\shortcite{cafnz}) and also, but to a lesser extent, the
low-$\Omega_0$ CDM models of Heyl \etal (\shortcite{hcfn}). However, this
discrepancy hides a significant improvement in our new models. In our
previous work we did not attempt to model the internal mass
distribution within a galaxy, and simply took the circular velocity of
the galaxy to be that at the virial radius in the halo in which it formed.
If we followed
this same procedure now, and plotted $\Vhalo$ rather than $\Vdisk$ as
a function of I-band magnitude, we would find a near-perfect match
to the observed Tully-Fisher relation, as indicated by the heavy solid
line in Fig.~\ref{fig:tf1}a.  The reason for this difference in the
$\Vhalo$--$M_I$ relation between our old and new models is largely the
inclusion of dust in the new models, which helps in two different ways
to simultaneously match the \BJ-band luminosity function and the I-band 
Tully-Fisher relation.  Firstly, dust makes galaxies dimmer in \BJ,
allowing a better match to the observed luminosity function with a
smaller value of $\Upsilon$, but it also makes them redder in \BJ--$I$. 
The net effect is that the I-band luminosities used in the
TF relation are increased. Secondly, dust affects the calculation of
the luminosity function and the Tully-Fisher relation in different
ways, because observational estimates of the luminosity function use
magnitudes uncorrected for dust, whereas observational estimates of
the Tully-Fisher relation partially correct for the effects of dust
through the correction to face-on magnitudes.  Some of these effects are
also discussed by Somerville \& Primack (\shortcite{sp98}).
Increasing the amount
of dust beyond that present in our reference model by, for instance,
increasing the assumed yield, can further improve the Tully-Fisher
zero-point. However, this is achieved at the expense of making the
galaxy population too red.

\subsection{Disk Sizes} % Size distribution at two magnitude cuts
\label{sec:dsize}

\begin{figure*}
\centering
\centerline{\epsfbox{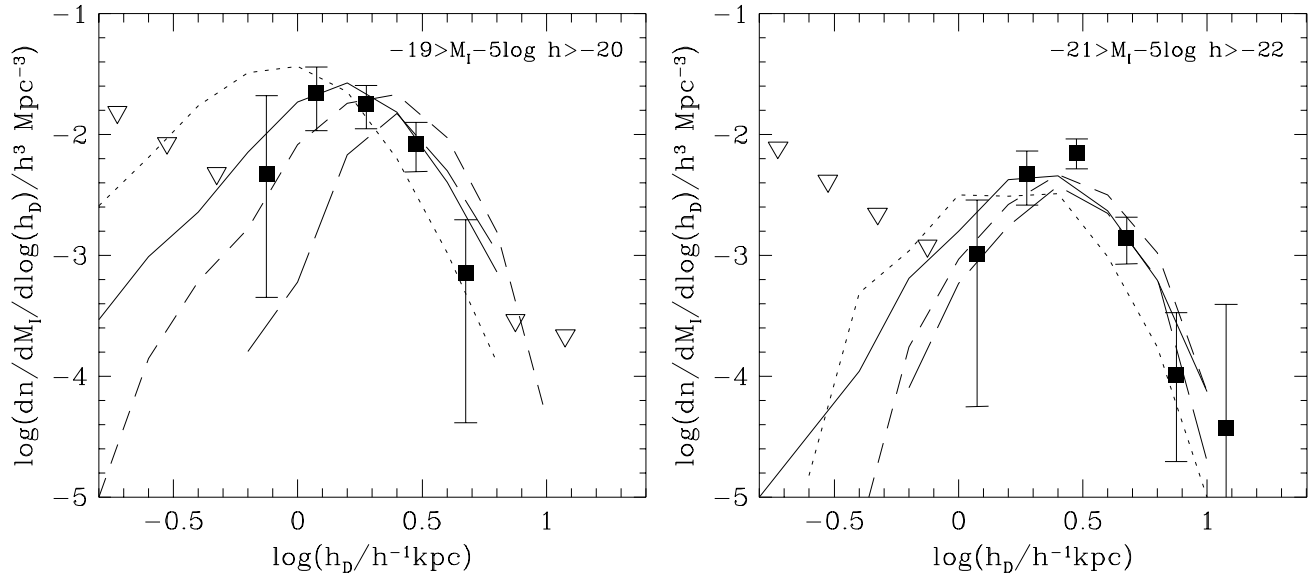}}
\caption{A comparison of predicted spiral galaxy disk sizes with
observations. The two panels show the distribution of disk exponential
scalelengths $\hD$ (number density as a function of scalelength and
absolute magnitude) in luminosity bins on either side of $L_\star$.
The points with error bars and the triangles are, respectively,
observational data and 95\% confidence upper limits for Sb--Sd
galaxies from the work of de~Jong \& Lacey (\shortcite{djl99}),
allowing for the dependence of the observational selection function on
galaxy size and luminosity. The lines are the model results for
varying $V_{\rm hot}$, for galaxies with $(B/T)_I<0.24$.  The
dotted, solid and short-dashed curves are for $V_{\rm
hot}=100$, $200$ and $300~\kms$ respectively.  The long-dashed
curve is for a variant of the reference model in which 
unstable disks have been converted to bulges, as described in the text.  }
\label{fig:sizes1}
\end{figure*}

In our model, the sizes of galaxy disks are fundamentally determined
by the angular momentum gained by the proto-disk material through the
action of tidal torques, which are most effective when halos are
turning around and collapsing, prior to becoming virialized. The
distribution of halo spin parameters is well understood, and is
reasonably accurately modelled by the distribution (equation
\ref{eqn:spin_prob}) that we have adopted. The main sources of uncertainty
are the distribution of angular momentum within a halo, which
determines the angular momentum of that fraction of the gas that cools
to form a disk, and whether the gas conserves its angular momentum
during the collapse. The assumptions that we 
have discussed in
Sections~\ref{sec:halo_rot} \&~\ref{sec:ang_mom} are reasonable, 
but are not directly supported by simulations (see the discussion in
Section~\ref{sec:ang_mom}) and warrant further investigation. 
Apart from this, the largest remaining influence on the
distribution of galaxy disk sizes is the strength of stellar
feedback. If feedback is weak, stars form efficiently in small, dense
halos at high redshift, while if feedback is strong, star formation
is suppressed until larger halos form at lower redshift. Thus,
increasing the value of $V_{\rm hot}$ results in galaxies having
larger disk scalelengths at a given luminosity.  This dependence is
shown explicitly in Fig.~\ref{fig:sizes1}, and is weak for $L\gsim
L_\star$, but is significant at lower luminosities.  A value of
$V_{\rm hot}= 200\, \kms$ produces a model for which the position of
the peak in the disk scalelength distribution of spiral galaxies at
different luminosities is close to what is found observationally by
de~Jong \& Lacey (\shortcite{djl99}). Moreover, the predicted width of
the distribution is quite similar to that observed, although somewhat
broader. Our model does not predict a large population of bright
galaxies with either extremely large or small scalelengths.

The long-dashed line in Fig.~\ref{fig:sizes1} shows
how the size distribution of disks in the reference model 
could be modified by the effects of disk instability. In this
variant, we have checked the disk stability criterion, 
equation~(\ref{eq:diskinstab}),
at each timestep and have taken the material from unstable disks with 
$\epsilon_{\rm m}<1$ and added it to the spheroid or bulge component. 
Like Mo, Mao \& White (\shortcite{mmw98}), we find
that this depletes the small disk scalelength side 
of the distribution and produces a slightly better match to the
observed distribution for $L \sim L_\star$ disks.

It is interesting to examine the effects of surface brightness
selection on estimates of the galaxy luminosity function. 
For galaxies in the reference model, we computed the
difference between the total magnitude and the magnitude
within an isophote of 25 ${\rm mag}/{\rm arcsec}^2$ in
\BJ, for a galaxy seen face-on, assuming that the dust attenuation
factors are constant with radius for the bulge and disk components.
We then assumed that this aperture
correction is independent of the actual inclination angle of the
galaxy, and estimated the resulting galaxy luminosity function.  Because of
these approximations, and because in real galaxy surveys some attempt
is made to extrapolate to total magnitudes, the resulting model
luminosity function cannot be quantitatively compared to
observations. Nevertheless, the difference between this estimate,
shown by the dotted curve in Fig.~\ref{fig:lfone}, and that based on
the true total magnitudes gives an indication of the sort of
systematic errors that could be present in real surveys. The main
effect of using isophotal magnitudes is to produce a small faintward
shift at the bright end of the luminosity function, and a larger
change in the faint-end slope. The dependence on the surface
brightness limit suggests that the faint-end slope derived from
surveys selected from photographic plates could be artificially
shallow (see also McGaugh \shortcite{mcgaugh96}).

\subsection{Morphology } % 
\label{sec:morph_mix}

In our model, bright elliptical galaxies form predominantly through
galaxy mergers. When two galaxies of comparable mass coalesce
($M_2> \fellip M_1$, where $M_1 > M_2$), a violent merger is assumed to
occur, leaving an elliptical galaxy as the remnant.  Spheroids can
also be built-up by the repeated accretion of smaller, gas-poor
galaxies, because accreted stars are assumed to add to the bulge
component of the accreting galaxy.  Thus, the parameters $\fdf$
and $\fellip$, which determine, respectively, the frequency of
galaxy-galaxy mergers and the threshold above which a merger is deemed
to be violent, are the primary parameters that influence the
production of elliptical galaxies. The morphological mix depends also
on the strength of stellar feedback. If feedback is weak, massive
disks form at high redshift and have a long interval of time during
which they can merge to form ellipticals.  Conversely, if feedback is
strong, the formation of massive stellar disks is delayed, they experience
fewer mergers and fewer ellipticals are produced.  We can therefore
constrain these parameters by comparing the relative abundances of
galaxies of different morphological types in the model with observations.

\begin{table}
\centering
\caption{The morphological mix of galaxies  brighter than
$M_{\rm B}-5\logh< -19.5$, for various values of the merger
parameters $\fdf$ and $\fellip$ and the feedback parameter $V_{\rm hot}$.
Also listed are two variants, which are described in the text.
In the first,$^\dag$ unstable disks are transformed to spheroids and in the
second,$^\ddag$ accreted stars are added to the disk rather than the bulge.}
\begin{center}
\begin{tabular}{lrrrrrr} %\hline
\multicolumn{1}{l} {$\fellip$} &
\multicolumn{1}{c} {$\fdf$} &
\multicolumn{1}{c} {$V_{\rm hot}/ \kms$} &
\multicolumn{1}{c} {S : S0 : E} \\
%\hline 
\\
0.3 & 1   & 200 & 61:08:31 \\ % 
0.5 & 1   & 200 & 70:07:23 \\ % 
0.3 & 0.5 & 200 & 53:09:38 \\ % 
0.3 & 2.0 & 200 & 79:06:15 \\ % 
0.3 & 1   & 100 & 38:05:57 \\ % 
$^\dag$0.3 & 1   & 200 & 46:09:45 \\ % unstable disks
$^\ddag$0.3 & 1   & 200 & 66:08:26 \\ % stars accreted by disk 
%\hline
\end{tabular} 
\end{center}
\end{table}

Our models do not strictly predict galaxy morphology, but rather the
relative masses and luminosities of the bulge and disk components. The
bulge-to-disk luminosity ratio is known to correlate with morphology,
albeit with quite a large scatter, and so we simply take cuts in this
ratio in order to define morphological classes. Ellipticals are
defined as galaxies for which the bulge contributes more than 60\% of
the B-band light, spirals as those whose bulge contributes less than
40\% of the B-band light, and S0s as galaxies in the intermediate
range.  The B-band magnitudes used here include dust extinction for
galaxies with random inclination angles.

Table~3 gives the predicted morphological mix of galaxies at the
present day that results from these definitions, for various values of
the parameters $\fdf$, $\fellip$ and~$V_{\rm hot}$. These ratios apply
to a volume-limited sample of galaxies with absolute magnitude
brighter than $M_{\rm B}=-19.5 + 5 \log h$, but the mix is very
similar if one instead constructs an apparent magnitude-limited
catalogue. For comparison, the morphological mix in the APM Bright
Galaxy Catalogue (which is apparent magnitude limited) is ${\rm
S+Irr:S0:E} =$ $67:20:13$ (\cite{loveday96}, table~10), when one
groups together spirals and irregulars, and assumes that the 90\% of
the galaxies in this survey that were classified are
representative. This agrees well with the estimate ${\rm S+Irr:S0+E}
=$ $66:34$ for galaxies brighter than $M_{\rm B}=-19.0 + 5
\logh$ in the SSRS2 redshift survey (Table~2 of \cite{marzke}). Comparing the
model results with the observational values indicates that the
morphological mix depends somewhat on the values of $V_{\rm hot}$,
$\fellip$ and $\fdf$. Reasonable agreement with the observed mix can
be obtained for a range of values of these parameters, indicated by
the examples given in the first three rows of Table~3.  For the
reference model (cf. Table~1), the values of $\fellip$ and $\fdf$ are
approximately the lowest that would seem reasonable, given the
physical processes that these parameters are trying to describe.
Considering the crude manner in which we have defined morphologies
the agreement between model and data is satisfactory, and it does not 
seem warranted to fine-tune these parameter values any further. 

The morphological mix may also be influenced by the disk instability
discussed in Section~\ref{sec:stability}. The penultimate row in 
Table~3 gives the mix found for the model with disk instability 
whose disk scalelength distribution was discussed in Section~\ref{sec:dsize}.
Here, it has been assumed that gas and stars in unstable disks are 
transferred to the bulge component, with the gas being consumed in a burst. 
We see that this significantly reduces the spiral fraction and boosts
the elliptical fraction, though the majority of ellipticals are still
formed by mergers. In fact, the effect on the morphological mix is quite
a strong function of luminosity. Brighter than 
$M_{\rm B}=-20.5 + 5 \log h$ disk instability has very little effect, 
but it becomes increasingly important in low luminosity systems.
We plan to discuss the effects of disk instability in detail 
in a subsequent paper.

One further assumption of our reference model is that stars that are
accreted in minor mergers are added to the bulge of the resulting
galaxies (see Secion~\ref{sec:galmergers}).  The last row in Table~3
shows how the morphological mix is modified if instead these accreted
stars are added to the galaxy disks. Such a change only
causes a modest increase in the spiral fraction. Also, we can see in Table~2
that this model (labelled ``Accretion by disk'') differs little from
the reference model.

\subsection{Cold gas in Spiral Galaxies}
\label{sec:lcdm_gas} 

\begin{figure}
\centering
\centerline{\epsfxsize=8.0 truecm \epsfbox[22 52 281 595]{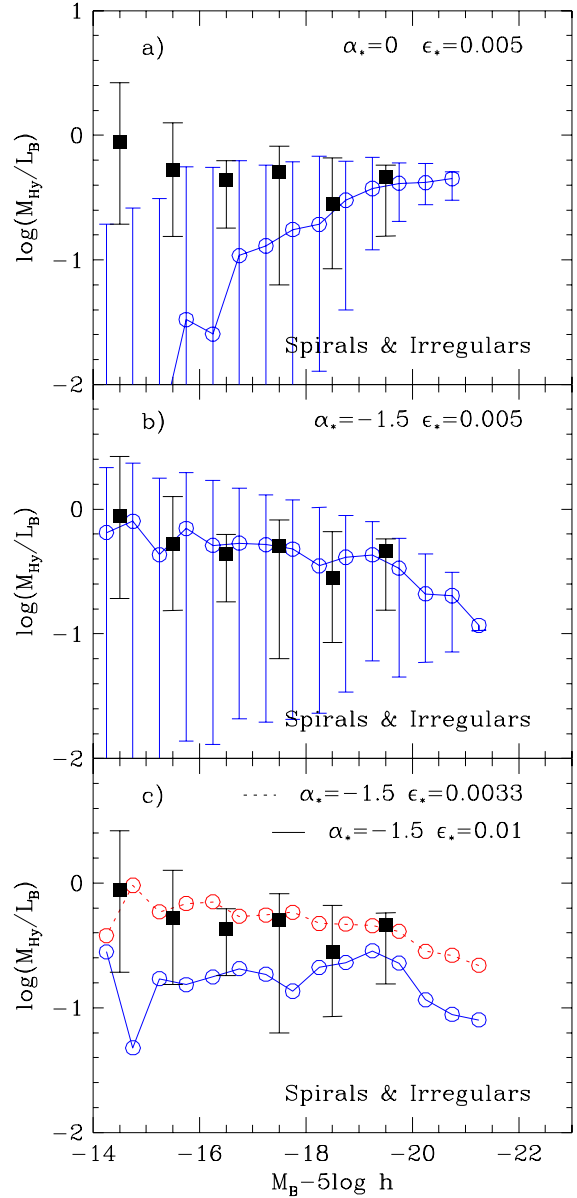} }
\caption{The cold gas content of spiral and irregular galaxies
as a function of luminosity. In each panel, the filled squares and
associated errorbars show observational estimates of the median, 10
and 90 percentile points respectively of the distribution of $M_{\rm
Hy}/L_{\rm B}$ for Sa to Im galaxies. Here, $M_{\rm Hy}=M_{\rm
HI}+M_{\rm H_2}$ is the mass of cold hydrogen which includes gas both
in atomic and molecular forms. We have made these estimates using a
combination of data from Huchtmeier \& Richter
(\protect\shortcite{hr88}) and Sage (\protect\shortcite{sage}), as
described in the text.  The model results are shown by the open
circles and their errorbars. For these, we select galaxies of
comparable morphological type by requiring the B-band bulge-to-total
luminosity ratios to be less than 0.4 . We express the model cold gas
mass, $M_{\rm cold}$, in the observational units, $h^{-2} \Msol$, and
set $M_{\rm Hy}=0.7 M_{\rm cold}$ to take account of the mass fraction
of He. The upper panel shows the model with $\alphastar=0$. The middle
panel shows the reference model, which has $\alphastar=-1.5$. 
The lower panel shows two models (the errorbars have been removed for 
clarity), each  with $\alphastar=-1.5$, but with the parameter
$\epsilon_\star$, which controls the star formation timescale, varied
up and down from the value in the reference model. }
\label{fig:gasfrac}
\end{figure}

In our model, there are two parameters, $\epsilon_\star$ and $\alphastar$, 
which determine the star formation timescale in galaxy disks (see equation 
\ref{eqn:sflaw}). The first, $\epsilon_\star$, determines the star 
formation timescale for spiral galaxies with circular velocities
comparable to those of $L_\star$ galaxies, while the second,
$\alphastar$, determines how this timescale varies with circular
velocity. The luminosity function (after rescaling by $\Upsilon$) is
fairly insensitive to $\epsilon_\star$ and $\alphastar$, but they do
strongly affect the cold gas content of galaxies.

Fig.~\ref{fig:gasfrac} shows how the amount of cold gas present in
spiral and irregular galaxies depends on galaxy luminosity.  The
observational data are taken from Sage (\shortcite{sage}) and
Huchtmeier \& Richter (\shortcite{hr88}).  The Sage (\shortcite{sage})
data come from a complete sample of Sa-Sd galaxies with measurements
of atomic HI and molecular $\rm H_2$, whereas the Huchtmeier \& Richter
(\shortcite{hr88}) data come from a complete sample of Sa-Im galaxies,
but with only HI measurements.  Brighter than $M_{\rm B}-5{\rm log}h<-16$,
Sa-Sd galaxies dominate over Sdm-Im, and so we simply plot the Sage
(\shortcite{sage}) data.  Fainter than $M_{\rm B}-5{\rm log}h>-16$, the mass
fraction of molecular hydrogen appears to be small ($M_{\rm
H_2}/M_{\rm HI} \lsim 0.2$), and so here we neglect the molecular $\rm
H_2$ contribution and simply plot the data of Huchtmeier \& Richter
(\shortcite{hr88}). In each case, the luminosity is corrected to
face-on.

The model plotted in Fig.~\ref{fig:gasfrac}a has $\alphastar=0$,
corresponding to the standard Kennicutt law in which the star
formation timescale is proportional to the disk dynamical time.  In
this case, the faint galaxies in the model typically contain less cold
gas than is observed. In fact, the trend of $M_{\rm gas}/L_{\rm B}$
with $L_{\rm B}$ is in the opposite sense to that observed.  The reference
model plotted in Fig.~\ref{fig:gasfrac}b, has $\alphastar=-1.5$, 
implying longer star
formation timescales for low circular velocity galaxies compared to
$\alphastar=0$. This has a beneficial effect, and produces a
correlation in $M_{\rm gas}/L_{\rm B}$ vs. $L_{\rm B}$ which is much
closer to the data. Fig.~\ref{fig:gasfrac}c shows the effect of
varying $\epsilon_\star$, and demonstrates how the observed gas
fraction in bright spirals constrains this model parameter.

\subsection{Galaxy Metallicities}
\label{sec:metals}

\begin{figure}
\centering
\centerline{\epsfxsize= 8.0 truecm \epsfbox[ 0 285 285 750]{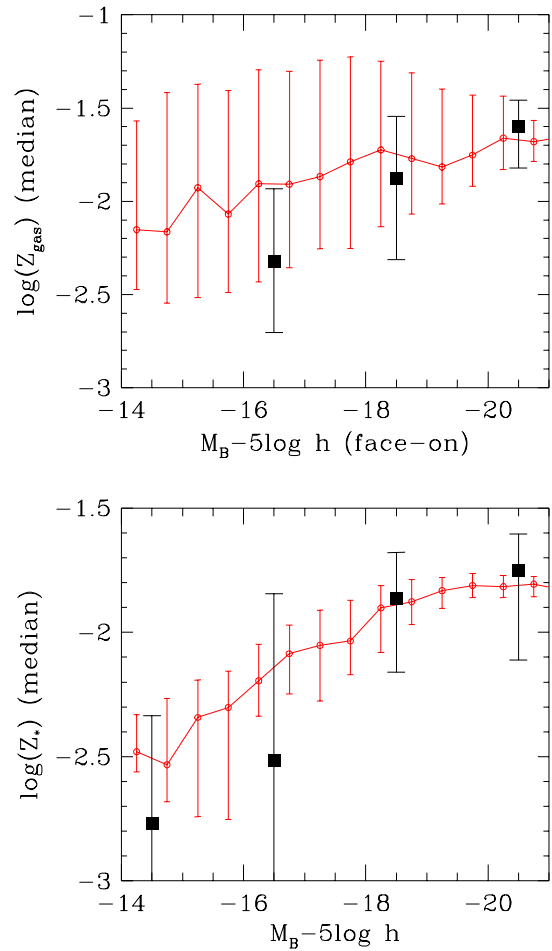}}
\caption{ The dependence of metallicity on luminosity in
our reference model, compared with observational data. In each panel, the
lines show the median metallicity in the model and the error bars indicate
the 10 and 90 percentiles of the distribution. The observational data,
taken from the compilation by Zaritsky \etal (\shortcite{zaritsky94}),
are indicated by filled squares, where again the error bars show the
10 and 90 percentiles. 
The upper panel compares
the metallicity of the cold star-forming gas in disk-dominated galaxies
which in our model are galaxies selected to have B-band bulge-to-total 
ratios of less than 0.4,
with observational data for spiral and irregular galaxies.  The lower
panel compares the stellar metallicity for bulge-dominated galaxies
(B-band bulge-to-total ratios greater than 0.6) with
observations for elliptical galaxies.  
}
\label{fig:zmet}
\end{figure}

Our model predictions for galaxy metallicities all scale linearly with
the yield $p$, aside from the effects of metallicity on the cooling of
halo gas. We fix the value of $p$ in our reference model by requiring
a good match to the mean stellar metallicity in $L_\star$ ellipticals.

In Fig.~\ref{fig:zmet}, we show what the reference model predicts for
the metallicity-luminosity relation for spiral and elliptical
galaxies, compared to observational data. Fig.~\ref{fig:zmet}a shows
the gas metallicity vs. luminosity for spirals. The model points are
derived from the metallicity of the cold, star-forming gas. The
observational data in this case, from Zaritsky \etal
(\shortcite{zaritsky94}), are based on HII region gas metallicities
measured at $r=0.4 R_{25}$ in each galaxy (where $R_{25}$ is the
isophotal radius at a B-band surface brightness of 25 ${\rm mag}/{\rm
arcsec}^2$), rather than being averages over the whole
galaxy. Fig.~\ref{fig:zmet}b shows the stellar metallicity
vs. luminosity for ellipticals. The model points are obtained from
mass-weighted mean stellar metallicities, but using V-band
luminosity-weighted stellar metallicities would give nearly identical
results. The observational data for the ellipticals are based on a
compilation of estimates from line-strengths, which Zaritsky \etal
have tried to put on a common metallicity scale. For both the spirals
and ellipticals, we have converted the Zaritsky \etal observational data
from relative to absolute metallicities assuming a solar metallicity
$Z_\odot=0.02$.

For both the spirals and ellipticals, our models predict a
metallicity-luminosity relation which is in the same sense as the
observed one, but is not as steep.  The origin of this correlation lies
in our model of stellar feedback.  Our treatment of chemical evolution
differs from the traditional ``closed-box'' model because gas reheated
by SNe is allowed to escape from the galaxy while cooling gas can flow
into it. Consequently, the appropriate expression for the effective
yield becomes $p_{\rm eff} = (1-e) p/(1-R+\beta)$
(equation~\ref{eqn:peff}), which depends on the strength of feedback
via the quantity $\beta$ given in equation~(\ref{eqn:feedback}).  Since
$\beta$ is smaller for larger, more massive galaxies with deeper
potential wells, their effective yield is large and this naturally
results in more metal rich stellar populations in more luminous
galaxies. We could obtain a steeper metallicity-luminosity relation,
in better agreement with the observed one, by assuming stronger
feedback, but this would have deleterious effects on our fits to other
properties. This issue is discussed further in 
Section~\ref{sec:col-mag} in connection with the colour-magnitude relation.

\subsection{Mass-to-Light Ratios} 
\label{sec:m_l}

While the luminosities of galaxies are easily measured, the masses of
the stellar populations they contain are less accurately determined,
mainly because of the difficulties in separating the contributions to
the mass from stars and dark matter in dynamical measurements. The
mass-to-light ratios, $M/L$, for stellar populations in galaxies are
correspondingly somewhat uncertain. For this reason, we do not use
them as primary constraints in determining the model
parameters. Nonetheless, they provide useful a consistency check, 
and can be used to exclude, for example, models which have
very large brown dwarf fractions (i.e. large $\Upsilon$), which might
otherwise be viable.

Table~2 lists the mass-to-light ratios of the stellar populations of
both spiral and elliptical galaxies for each of our models.  These
mass-to-light ratios, which include the contribution of brown dwarfs,
depend on the age and metallicity of the stellar populations, and also
on the value of the parameter $\Upsilon$ which has been fixed by
reference to the \BJ-band luminosity function, as described in
Section~\ref{sec:lumfn}.  Mass-to-light ratios are observed to depend
on galaxy luminosity, so to make a fair comparison, we compare
$M/L$ values for observed and model galaxies at the same
luminosity.  The model $M/L$ values given in the table are median values 
for $-20<M_{\rm B}-5 \log h <19$. 
For each of the observational estimates described
below, we have estimated the trend of $M/L$ with luminosity from the
values given for the individual galaxies in the paper concerned, and
used this to estimate the average $M/L$ for the galaxies at $M_{\rm B}
-5 \log h\approx -19.5$. For spiral galaxies, the
observed values are dust-corrected to face-on values, and so the model
$M/L$ values are also calculated from face-on luminosities including
the effects of dust.

For spiral galaxies, most estimates of $M/L$ are based on fitting
models to measured rotation curves. Buchhorn (\shortcite{buchhorn92})
finds $M/L = 3.4 h M_\odot/L_\odot $ in the I-band, and Broeils
(\shortcite{broeils92}) $4.9 h M_\odot/L_\odot $ in the B-band.  These
values are consistent with the mean colour, $B$--$I\approx 1.8$, found
for spirals by de Jong (\shortcite{dejong}).  Note, however, that
these numbers are based on maximum disk fits to the observed rotation
curves, and since a fraction of the rotation velocity may be produced
by dark matter, they should be viewed as upper limits to the stellar
mass-to-light ratios.  (For $L_\star$ spiral galaxies 
in our reference  model the mean contribution to the mass within 
the disk half-mass radius by non-baryonic dark matter is 62\%.)
Comparing with the model values, we see that
only the model with the high value of the baryon density ($\Omega_{\rm
b} =0.04$) comes close to violating these constraints. An alternative
method for measuring the $M/L$ of disks, which avoids contamination by
dark matter, is to combine measurements of the vertical scaleheights
and velocity dispersions of galaxy disks. Using this method, Bottema
(\shortcite{bottema97}) finds an average $M/L = 2.4 h M_\odot/L_\odot$
in the B-band, which is about a factor of two lower than the maximum
disk value.  The median $M/L$ of our reference model and most of the
variants are only slightly below this estimate, and so are quite compatible
with observations.

The comparison for elliptical galaxies is similar.  In the B-band,
Mobasher \etal (\shortcite{mobasher99}) and van der Marel
(\shortcite{vdmarel91}) find $M/L = 9.6 h M_\odot/L_\odot$ and~$8 h
M_\odot/L_\odot$ respectively, using stellar velocity dispersion
measurements.  Again, these values should be viewed as upper limits on
the stellar mass-to-light ratios, as they include the effect of any
dark matter within the effective radii of the galaxies.  With the
exception of the high baryon density model, all the models listed in
Table~2 are consistent with these data.

\subsection{Average Colours} 
\label{sec:av_colours}

Table~2 also lists the median $B$--$K$ colours of galaxies with $-24.5
<M_{\rm K}-5 \log h <-23.5$ for various parameter values. The observed
median colour for this luminosity range, calculated from the data of
Gardner \etal (\shortcite{gardner96},\shortcite{gardner}) 
that are presented in
Section~\ref{sec:colours}, is $B$--$K=3.8$. The largest effects on the
model colours result from varying the yield $p$, $\Omegab$,
$\epsilon_\star$ and $\vhot$.  A higher yield leads both to
intrinsically redder stellar populations and to greater amounts of
dust, which redden the observed galaxy colours still
further. Increasing $\Omegab$ increases the gas density in halos, and
thus shortens the cooling time. This then results in a greater
fraction of the gas cooling and forming stars at high redshift,
producing an older, redder stellar population.  Decreasing $\vhot$
reduces the strength of feedback, allowing more stars to form in
low-mass halos at high redshift, leading, again, to older, redder
stellar populations by the present. Somewhat surprisingly, increasing
the star formation efficiency, $\epsilon_\star$, (cf. equation~\ref{eqn:sflaw})
has little effect on the present star formation rate and galaxy colours. 
This happens because the
shorter star formation timescale allows more star formation to occur
at high redshift, and this leads to a reduction in the amount of cold
star-forming gas at low redshift which compensates for the shortened
star formation timescale.

The colours also depend on the choice of IMF.  However, changing from the
Kennicutt to the Salpeter form makes very little difference to the
$B$--$K$ colours, which become bluer by just $0.015$ magnitudes. This
is to be expected, as Fig.~\ref{fig:imf} demonstrates that the
differences in the spectral properties of the stellar populations for
these two IMFs are small. The metallicities of the stellar populations
and the dust content of the galaxies are very similar in the two
cases, as the adopted values of the yield are identical and of the
recycled fraction almost identical.

\subsection{Summary of Parameter Constraints} 

\label{sec:sumcons}

In this section we have demonstrated how the predictions of local
galaxy properties depend on each of the model parameters. We now summarise
the reasons for adopting the specific parameter values that define our
standard or reference $\Lambda$CDM model. 

The cosmological parameters, $\Omega_0=0.3$ and $\Lambda_0=0.7$, were
chosen without reference to galaxy properties, and simply define the
background cosmology in which we sought a viable model of galaxy
formation. The Hubble parameter, $h=0.7$, was chosen to be in
reasonable accord with recent estimates.  The baryon density,
$\Omegab$, was chosen as a compromise between constraints from
primordial nucleosynthesis and the need to prevent the disks of bright
galaxies becoming strongly self-gravitating and so distorting the
bright end of the Tully-Fisher relation. This choice also prevents
galaxies becoming too luminous, or equivalently, 
their $M/L$ ratios becoming too
large once $\Upsilon$ has been adjusted.  The shape, $\Gamma$, of the
mass fluctuation power spectrum was chosen to be consistent with the
above choices of $\Omega_0$, $\Omegab$ and $h$
(cf. eqn~\ref{eqn:gamma}). The resulting value, $\Gamma= 0.19$, is in
accord with the shape of the power spectrum inferred from studies of
large-scale galaxy clustering. The amplitude, $\sigma_8$, was chosen
for consistency with the observed abundance of galaxy clusters
(\cite{ecf96}).  For our chosen cosmology, this is consistent with the
value of $\sigma_8$ preferred by the COBE measurements of microwave
background anisotropies.

The stellar population parameters, we have essentially chosen to be
consistent with models constructed to account for solar neighbourhood
data. In particular, we have adopted the Kennicutt IMF. The fraction
of gas recycled via stellar winds and SNe, we have taken to be $R=
0.42/\Upsilon=0.31$, consistent with what is expected for this
IMF. (Recall that $\Upsilon$ is defined as the total mass in stars,
including brown dwarfs, divided by the mass in luminous stars.  The
value $\Upsilon=1.38$ was chosen by fitting the position of the bright
end of the \BJ-band luminosity function.)  We have adopted a yield,
$p=0.02 $, which results in roughly the observed metallicity for
galaxies similar to the Milky Way and for $L_\star$ ellipticals. This
may be seen in Fig.~\ref{fig:zmet}, where we also examine the model
prediction for the dependence of metallicity on galaxy luminosity. Our
adopted yield implies $p_1\equiv p \Upsilon= 0.028$, which is also
roughly consistent with theoretical estimates for our assumed IMF. The
yield also determines the metallicity and dust content of the cold gas
disk. Our adopted value gives rise to an amount of reddening which is
approximately that required to bring the model into good agreement
with the observed galaxy $B$--$K$ colour distribution. Varying the IMF
between the Kennicutt, Salpeter or Miller-Scalo forms 
has only a small effect on the galaxy colours
and mass-to-light ratios at $z=0$, and so the IMF is therefore not
well constrained by the data we have examined so far. However, the
small differences in these IMFs can have a significant effect on model
predictions at high redshift.

The dynamical friction parameter, $\fdf$, we simply set to its natural
value, $\fdf=1$.

The parameter $\fellip$, which sets the threshold for violent galaxy
mergers which produce ellipticals and bulges
(cf. Section~\ref{sec:galmergers}), was chosen so as to reproduce an
acceptable mix of morphological types.

Finally, the model requires parameters for the star formation and
feedback laws (cf \S~\ref{sec:sf_disk}). The feedback parameters,
$\Vhot$ and~$\alphahot$, we have constrained by the shape of the
\BJ-band luminosity function, the slope of the Tully-Fisher relation,
and the distribution of spiral disk scalelengths.  A low value of
$\alphahot$ is required to avoid curvature in the Tully-Fisher
relation, while a large value helps to reduce the faint end slope of
the galaxy luminosity function. Our adopted compromise, $\alphahot=2$,
results in a straight Tully-Fisher relation and a luminosity function
with a faint end slope in good agreement with the ESP survey. This is
steeper than  in several other surveys, including those by Loveday
\etal (\shortcite{loveday}) and Ratcliffe \etal (\shortcite{ratcliffe97}),
but we have demonstrated that the slope is sensitive to surface
brightness selection limits.  The value of $\Vhot$ influences both the
faint end of the luminosity function and the sizes of galaxies. Our
adopted value of $\Vhot = 200\, \kms$ appears to be a good compromise.
The parameter, $e$, which allows metals produced in SNe to escape
directly to the surrounding hot halo gas rather than first being mixed
with the cold gas in the galaxy disk, we have simply set to
zero. Regarding the star formation law (equations~\ref{eq:sfr}
and~\ref{eqn:sflaw}), the overall scaling of the timescale, set by
the efficiency $\epsilon_\star$, is constrained primarily by the cold
gas masses in $L_\star$ spirals.  The dependence of this timescale on
galaxy circular velocity is determined by $\alphastar$. We adopted the
value $\alphastar=-1.5$ in order to improve the correspondence between
model and observed cold gas masses in low-luminosity disk galaxies.

%% file: results.tex
\section{Further properties of the reference model}
\label{sec:results}

In this section we compare further properties of our reference model
with observational data that have not been used to set the model
parameters. We defer comparisons with observations at high and moderate
redshift to future papers.

\begin{figure*}
\centering
\centerline{\epsfbox{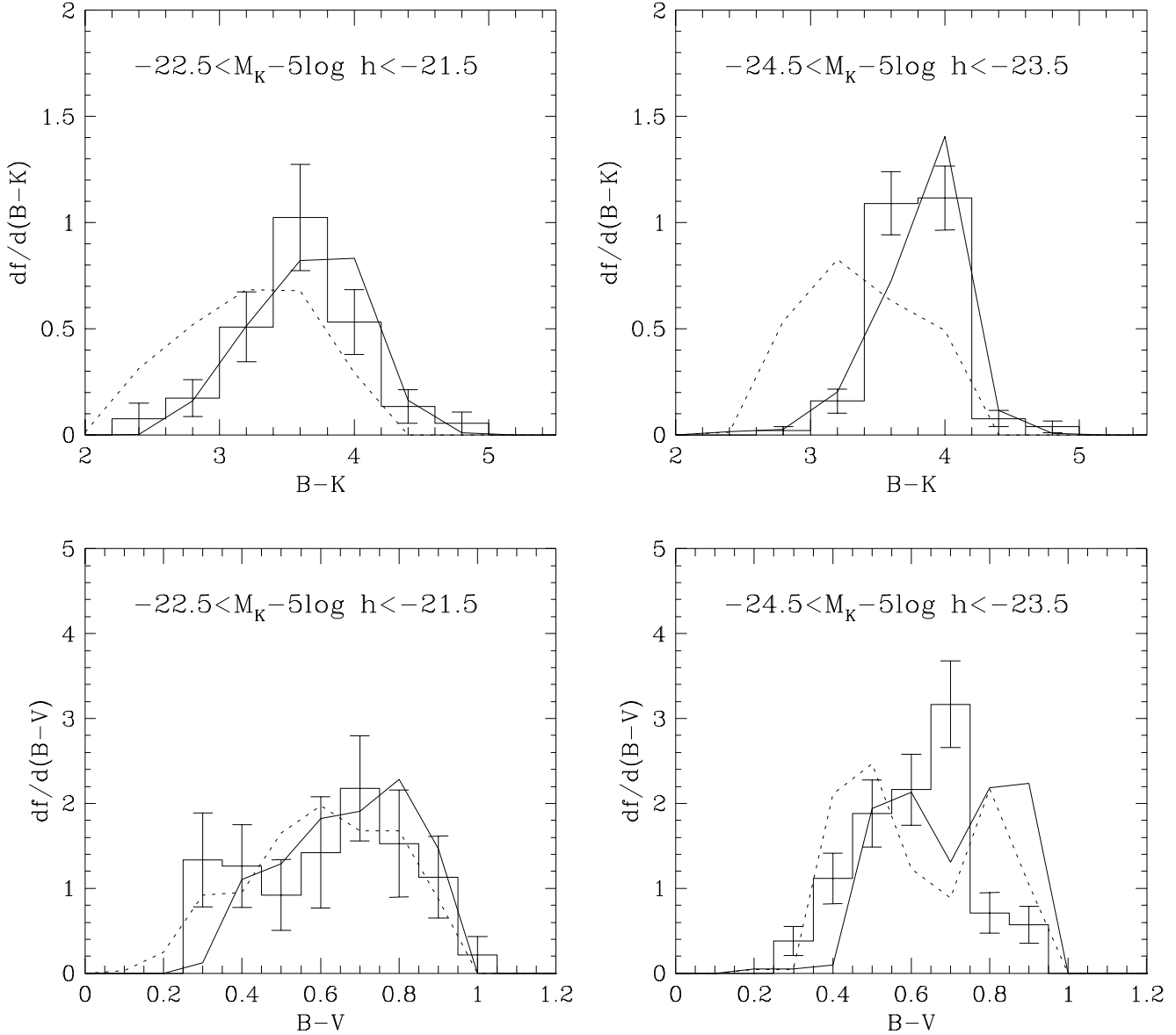}}
\caption{A comparison of galaxy colours in the reference model with
observations. The upper two panels compare rest-frame $B$--$K$ colour
distributions in volume-limited samples for two different ranges of
absolute $K$-band magnitude. The lower two panels compare the
rest-frame $B$--$V$ colour distributions for the same two ranges of
absolute $K$-band magnitude. In each case, the histograms with
errorbars show the observational distributions, derived from the
$K$-band redshift survey
of Gardner \etal
(\protect\shortcite{gardner96},\protect\shortcite{gardner}) using the
$1/V_{\rm max}$ method.  These
data have been k-corrected by matching each galaxy's observed colours
with one of a set of template spectra.  If the more uncertain
correction for evolution is also applied, then the observed
distributions typically shift redwards by $0.15$ magnitudes.  The
dotted lines show the colour distribution of the reference model
without including the effects of dust and the solid lines show the
distributions including the effects of dust.  }
\label{fig:col1}
\end{figure*}

\subsection{Colour Distributions} % B and K LF
\label{sec:colours}

We have already considered the average $B$--$K$ colours of $L \sim
L_\star$ galaxies in Section~\ref{sec:av_colours}.  In our reference
model, the mean galaxy $B$--$K$ colour is quite strongly constrained
by virtue of the requirement that the model give a reasonable fit to
the bright end of both the \BJ\ and K-band luminosity
functions. Nevertheless, it is still interesting to look at the full
distribution of predicted colours, and to compare them to other
observational data. In Fig.~\ref{fig:col1}, the reference model is
compared to the distributions of $B$--$K$ and $B$--$V$ colours in the
K-selected redshift survey of Gardner \etal
(\shortcite{gardner96},\shortcite{gardner}). This contains more than
500 galaxies, covers 10 square degrees and was imaged in the B, V, I
and~K bands. The redshift and colour information allow accurate
k-corrections to be derived by matching each galaxy's observed colours
with one of a set of template spectra. Thus, the histograms in
Fig.~\ref{fig:col1} show rest-frame colours.  If the more uncertain
correction for evolution is also applied, then the observed
distributions typically shift redwards by $0.15$ magnitudes. The model
with dust matches both the median colours and the widths of the colour
distributions quite well. If the effects of dust are not taken into
account, then the resulting model median colours are too blue by
approximately 0.3~magnitudes in $B$--$K$ and approximately
0.1~magnitudes in $B$--$V$.  Thus, we see again how modelling the
effects of dust is an important factor in producing acceptable galaxy
colours.

\subsection{The Colour-Morphology Relation}
\label{sec:col-morph}

\begin{figure}
\centering
\centerline{\epsfxsize= 9 truecm \epsfbox[42 175 525 620]{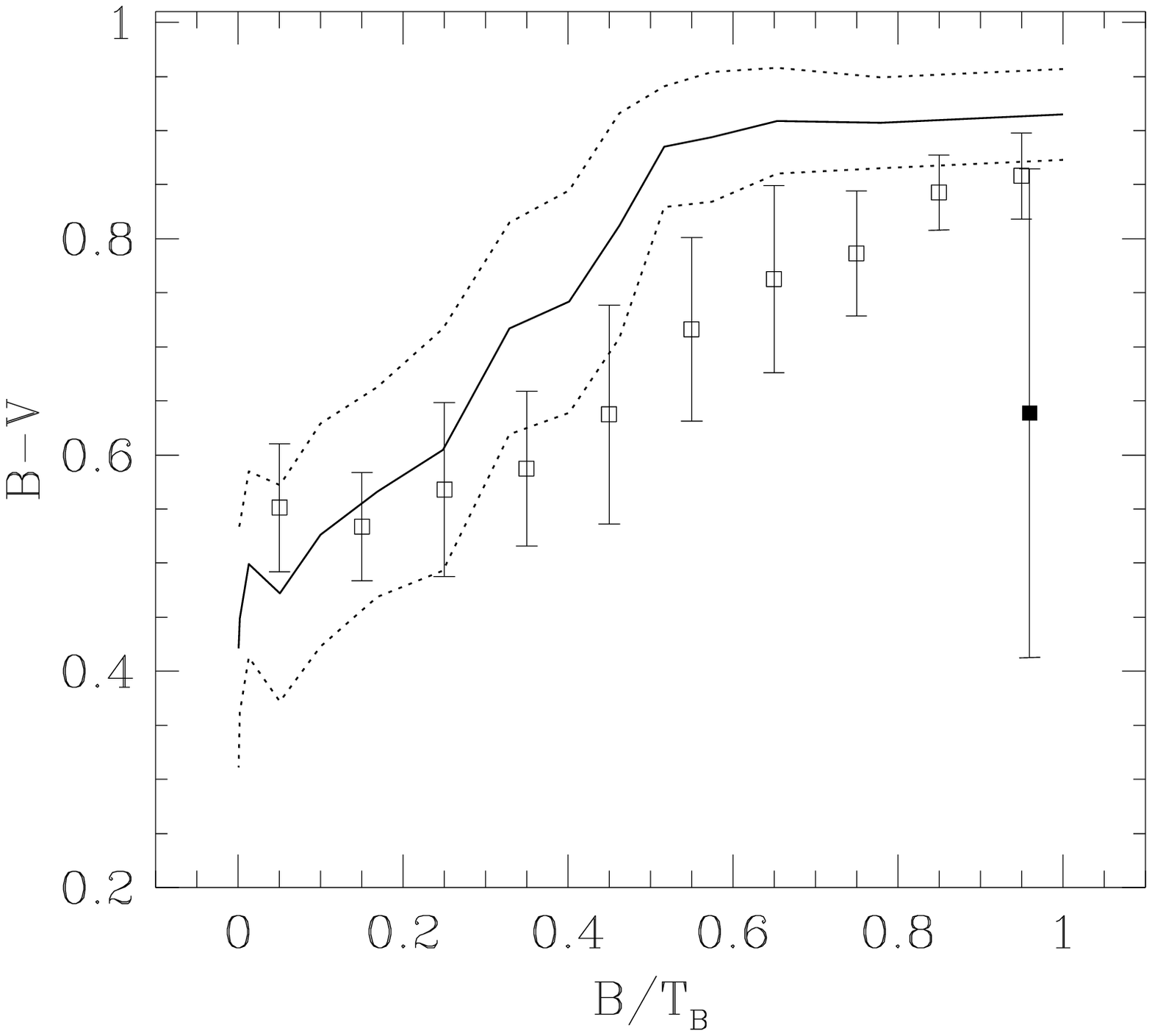}}
\caption{ 
Average galaxy $B$--$V$ colour as a function of bulge-to-disk ratio,
compared to observations. The open squares and error bars show the
model mean $B$--$V$ colour and rms dispersion as a function of B-band
bulge-to-total ratio for galaxies brighter than $M_{\rm V}-5\logh =
-20.5$.  Face-on colours are plotted, including the effects of
extinction.  The solid square shows the mean colour of model galaxies
which have experienced a merger and the associated burst of star
formation in the last 1~Gyr. The solid and dotted lines show the
observed mean and rms dispersion of the colour from the data of Buta
\etal (1994).   Hubble T-type has been converted to bulge-to-total ratio
using the data of Simien \& de Vaucouleurs (1986), as described in
Baugh \etal (1996b).  }
\label{fig:btcol}
\end{figure}

Fig.~\ref{fig:btcol} shows the predicted correlation of galaxy
$B$--$V$ colour, for face-on inclination, with bulge-to-disk ratio.  A
strong correlation is predicted, with bulge-dominated galaxies
typically being much redder than disk-dominated galaxies.  However,
galaxies which have experienced a major merger and the associated
burst of star formation within the last 1.0~Gyr have large
bulge-to-disk ratios, but are much bluer than bulge-dominated galaxies
that have not had a recent major merger.  Observationally, ``normal''
galaxies show a similar trend in colour to the model galaxies which
have not had recent bursts, as is illustrated by the observational
data of Buta
\etal (\shortcite{buta94}) which are also plotted, with the solid line
showing the mean galaxy colour and the dotted lines showing the
dispersion. Note that Buta \etal have removed from their sample some
galaxies viewed as outliers or having strong emission lines.  To plot
these data on Fig.~\ref{fig:btcol}, we have converted the Hubble
T-type given in Table~6 of Buta \etal to bulge-to-total ratio, using
the fit given in equation~(5) of Simien \& de Vaucouleurs
(\shortcite{sv86}).  This fit represents an extrapolation from
$T=-3$ (which corresponds to $B/T \approx 0.6$) to $T=-5$ (which
corresponds to $B/T \approx 1.0$).  Given the considerable scatter
that exists between bulge-to-total ratio and T-type (e.g. see
Figure~1. of Baugh \etal \shortcite{bcf96b}), the level of agreement
between the predictions of the model and the data is encouraging.  It
will be interesting to perform more quantitative comparisons of this
prediction with observations when a suitable data set exists in which
bulge-to-disk decompositions have been done for a complete survey.

\subsection{The Elliptical Colour-Magnitude Relation}
\label{sec:col-mag}

\begin{figure}
\centering
\centerline{\epsfxsize= 9 truecm \epsfbox[0 40 550 400]{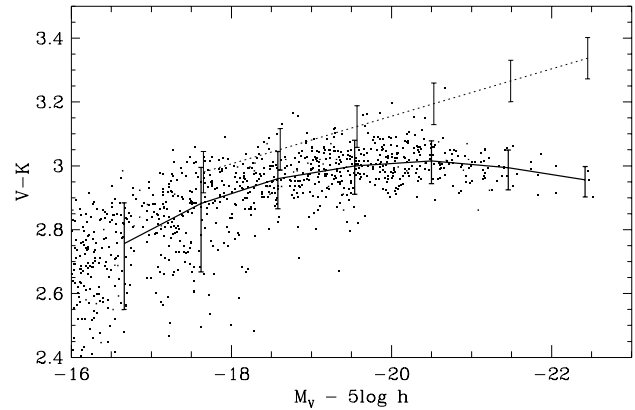}}
\caption{ The colour-magnitude relation for cluster elliptical
galaxies in the reference model, compared to observations.  The points
give the predicted distribution of V-K colour versus V-band magnitude
for elliptical galaxies in clusters with circular velocity greater
than $1000 \kms$.  The heavy line and error bars indicate the median
and the 20 and 80 percentiles of this distribution. The observed
correlation and scatter, from Bower, Lucey \& Ellis
(\shortcite{ble92}), are indicated by the dotted line and associated
error bars.  }
\label{fig:col_mag}
\end{figure}

A second correlation that we have examined is the colour-magnitude
relation of elliptical galaxies. This is closely related to the
metallicity-luminosity relation for elliptical galaxies, already
discussed in Section~\ref{sec:metals}. Fig.~\ref{fig:col_mag} compares
the distribution that we predict for cluster ellipticals with that
determined observationally for the Coma cluster (Bower, Lucey \& Ellis
\shortcite{ble92}). The model predicts quite a small spread in the
colours of bright ellipticals, consistent with the observed scatter,
but it does not reproduce the strength of the observed correlation of colour
with luminosity. This is the same problem that we found when we made
this comparison in Baugh \etal (\shortcite{bcf96b}), even though our
new model includes chemical enrichment, which the earlier model did
not.

In Kauffmann \etal (\shortcite{kwg93}),
Baugh \etal (\shortcite{bcf96b}) and Kauffmann (\shortcite{k96}),
it was argued that the inclusion of chemical enrichment (which was
neglected in the early models) might give rise to the desired
correlation. This was explicitly demonstrated for certain models by
Kauffmann \& Charlot (\shortcite{kc98}). In our present models,
chemical enrichment does appear to have the desired effect at low
luminosities, where the models produce a similar gradient in the
colour-magnitude relation to the one observed.  However, the
correlation between metallicity and luminosity flattens at the
brightest magnitudes (see Figure~\ref{fig:zmet}b), and this gives rise
to a flattening in the predicted colour-magnitude relation for bright
ellipticals. Our models are capable of producing a significant
gradient in this diagram if we assume very strong feedback and a large
yield $p$, just as Kauffmann \& Charlot (\shortcite{kc98}) did, but
this is then at the expense of other successes of the model. In
particular, strong feedback gives rise to excessively large disk
scalelengths. This is clearly an issue that deserves further
investigation.

\subsection{The cosmic star formation history}

\begin{figure}
\centering
\centerline{\epsfxsize=8.0 truecm \epsfbox[0 0 550 750]{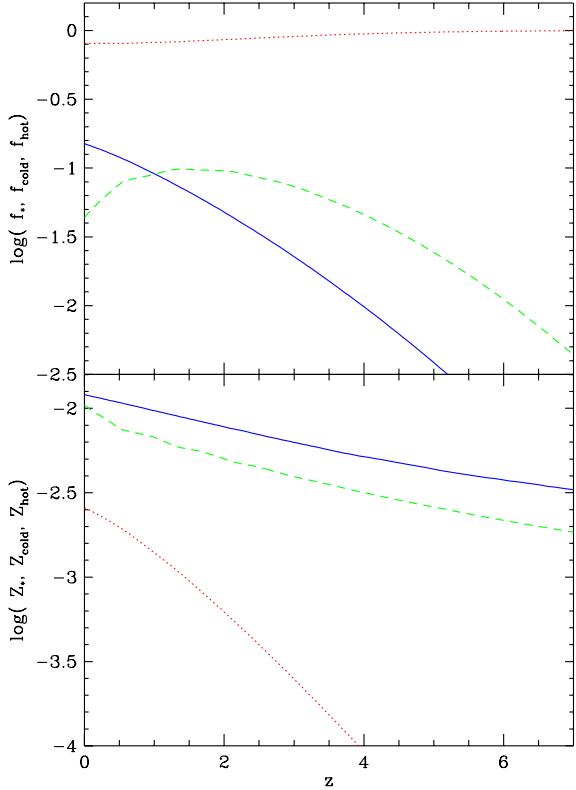}}
\caption{Evolution of baryonic mass fractions and average metallicities. The
upper panel shows the fraction of baryons in stars, cold gas and hot
gas as a function of redshift, and the lower panel shows their
corresponding mean metallicities.  The solid lines are for stars, the
dashed lines for cold gas and the dotted lines for hot gas.}
\label{fig:frac}
\end{figure}

\begin{figure}
\centering
\centerline{\epsfxsize=9.0 truecm \epsfbox[0 385 600 775]{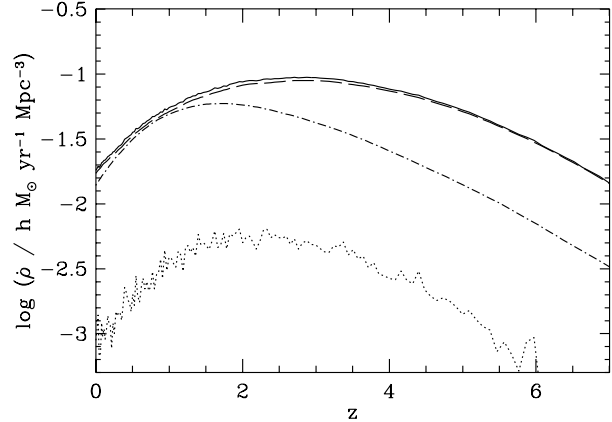}}
\caption{ 
The luminous star formation rate (i.e. excluding brown dwarfs) 
per unit comoving volume as a function of redshift. The solid
line shows the total star formation rate per unit volume in the
reference model. The dashed line shows the contribution from quiescent
star formation in galactic disks. The dotted line shows the
contribution from bursts of star formation that occur during major
mergers.  The dot-dashed line is the star formation history in model G
of Baugh \etal (1998) (for the same cosmological parameters as in our
reference model). In the current reference model, the star formation
rate per unit volume is higher at high redshift than in the old model,
because of the weaker feedback in low mass halos and the shorter star
formation timescale at high redshift that are assumed in the new
model.}
\label{fig:sfh}
\end{figure}

It is interesting to examine how the improvements in our modelling
techniques affect our predictions for the global history of star
formation, first presented in Cole \etal (\shortcite{cafnz}).
Fig.~\ref{fig:frac} shows the redshift evolution of the mass fractions
of baryons in the forms of hot gas, cold gas and stars, and the mean
metallicities of each of these components.  Consistent with our
feedback model, we have assumed that baryons contained in halos with
mass below our resolution limit are in the hot, diffuse phase. We
remind the reader that by ``hot'' gas we simply mean diffuse gas in
halos, whatever its physical temperature, and by ``cold'' gas we mean
all the gas that has cooled and collapsed into galaxies. The mass of
stars increases steadily towards the present, with just over half of the
present stellar mass having been formed since redshift $z<1.5$. This
is similar to our earlier results in Cole \etal (\shortcite{cafnz})
and Baugh \etal (\shortcite{bcfl98}), despite the various changes
in model parameters that we discuss below.
The evolution of the cold gas
mass shows a broad peak between redshifts $1<z<2$.  The fall-off at
high redshift reflects the efficiency of stellar feedback, which keeps
gas in low mass halos in the hot phase. At low redshift, the cold gas
fraction begins to decline as cooling becomes increasingly less
efficient in the large mass, high virial temperature groups and
clusters that form at late times, while, at the same time, the
reservoirs of cold gas are depleted by ongoing star formation. The
mean metallicity of the hot gas grows at a rate which mirrors the
growth in total stellar mass. In contrast, the mean metallicity of the
stars and cold gas reaches $1/3$ of its present value even at very high
redshift, and then increases only gradually between redshift $z=6$ and
$z=0$.

Fig.~\ref{fig:sfh} shows the star formation history in differential
form.  The solid line is the average formation rate of luminous stars
per unit comoving volume (\ie the total star formation rate divided by
$\Upsilon$). The dashed and dotted lines show separately the
contributions to the total rate from quiescent star formation in disks
and from bursts of star formation induced by galaxy
mergers. Approximately 10\% of the stars formed at any redshift are
formed in bursts.

The dot-dashed line in Fig.~\ref{fig:sfh} shows our earlier result,
presented as model~G of Baugh \etal (\shortcite{bcfl98}). Model~G had
very similar cosmological parameters to the present reference model,
but was calculated with a slightly earlier version of our code. The
differences between the two results are easy to understand and provide
a nice illustration of how various aspects of our galaxy formation
modelling are closely interconnected. The main difference between the
two model predictions is the lower star formation rate at high
redshift obtained in model~G compared to the present model. This
difference mainly reflects the different values of the feedback
parameters that we adopted in the two models. In our older model 
we set these parameters by the requirement that the faint-end of the
present-day galaxy luminosity function reproduce as closely as possible
that measured by Loveday \etal(\shortcite{loveday}). This led us
in Baugh \etal (and in Cole \etal \shortcite{cafnz}) to assume very 
strong feedback. Here, we have argued that the uncertainty
in the faint end of the luminosity function as indicated by the wide
range  of observational  esimates, including the very steep slope found by
Zucca \etal (\shortcite{zucca97}), suggests that this is not a
robust constraint. We have instead adopted a weaker feedback that
avoids introducing curvature in the faint end of the Tully-Fisher relation.
Thus, in our models, the behaviour of the star formation rate
at high redshift is intimately tied in with our assumptions about the
strength of stellar feedback and the 
faint-end of the present-day galaxy luminosity function. A second
difference between the present model and that of Baugh \etal
(\shortcite{bcfl98}) is the assumed dependence of the star
formation timescale on galaxy properties. We now incorporate a scaling
with the galaxy dynamical time, which results in all star formation
timescales at high redshift being smaller than in the older
model. We must emphasize, however, that in spite of the uncertainties
in the predicted star formation rate at high redshift, our prediction
of a late epoch for the majority of star formation is robust
(cf. Figure~21 of Cole \etal
\shortcite{cafnz} and Figure~14 of Baugh \etal
\shortcite{bcfl98}). This is because in all the versions of our model,
only a small fraction of stars form at $z>3$, and it remains the case
that we expect half the stars in the universe to have formed at
redshift $z\lsim 1.5$.

%% file: disc.tex
\section{Discussion}
\label{sec:disc}

In this paper, we have presented a new semi-analytic model of galaxy
formation, based upon the one developed by Cole \etal (\shortcite{cafnz}).
Our new model contains a number of additions and improvements. For example,
we have designed and implemented a new algorithm to generate halo merger
trees with arbitrary mass resolution, and we have extended our modelling to
include realistic descriptions of the density profiles of dark matter halos
and their gas content, as well as calculations of chemical evolution, dust
extinction, and galaxy sizes. We applied this model to a specific
cosmology, the $\Lambda$CDM model, which has $\Omega_0=0.3$, $\Lambda_0=0.7$
and a primordial power spectrum whose amplitude is consistent with both the
local abundance of galaxy clusters and with the COBE anisotropies in the
microwave background radiation. We then compared the results with a range
of observational data for the local galaxy population.

In principle, a model like ours can predict virtually any simple
property of the galaxy population over a large range of
redshift. However, not all predictions are equally reliable. For a
given cosmological model (e.g.  $\Lambda$CDM), the evolution of the
population of dark matter halos is known with high accuracy and can be
calculated without any free parameters. The initial internal structure
of these halos is also completely specified if one adopts the results
of recent high-resolution N-body simulations (\cite{nfw97}).  The
process of gas cooling is more uncertain but can be calculated also
without free parameters (other than the value of the gas metallicity),
using a model based on simple assumptions about the geometry and
initial configuration of the gas. We assume an initial spherically
symmetric distribution of gas at the virial temperature of the halo in
which it is contained, and a gas density profile given by the
``$\beta$-model''. In practice, the dynamics of the cooling gas are
likely to be subtantially more complex than this simple model
implies. Nevertheless, as Benson \etal (\shortcite{benson00}) 
have shown, the simple
model does predict global fractions of hot and cold gas, and their
distribution in halos of different mass, that are broadly in agreement
with the results of gasdynamics simulations.

The formation of stars from the gas that has cooled and the associated
feedback processes are the most uncertain components of the model.  As
in Cole \etal (\shortcite{cafnz}), we have represented them using
simple scaling laws, but we have adopted a more flexible treatment
than we had done previously. In addition to specifying the IMF and the
associated fraction of brown dwarfs, our model of star formation and
feedback requires four free parameters. Our treatment of feedback is
simplified and neglects potentially important sources of energy such
as active galactic nuclei and quasars. It also neglects the dynamical
response of the hot halo gas to the heating generated by the injection
of feedback energy. In principle, this energy can modify the structure
of the gaseous halo and hence its cooling rate. Thus, the detailed
treatment of feedback impacts on all aspects of the galaxy formation
model. For example, as we discussed in Section~7, our model only works
well if we assume a value of the mean cosmic baryon density,
$\Omega_b$, which is lower than recent determinations based on the
deuterium abundance at high redshift by Schramm \& Turner (\shortcite{st98}) 
and Burles \& Tytler (\shortcite{bt98}), although it is consistent with 
older determinations by Walker \etal (\shortcite{walker91}) and Copi \etal 
(\shortcite{copi95}). A larger value of
$\Omegab$ causes too much gas to cool, leading to a poor match to the
Tully-Fisher relation and to unacceptably large mass-to-light
ratios. A successful model with larger values of $\Omegab$, however,
is likely to be possible if feedback raises the entropy of the hot
halo gas, thereby strongly suppressing cooling, as argued recently by
Bower \etal (\shortcite{Bower00}) in the context of X-ray clusters. 

Intimately linked to the processes of star formation and feedback is the
chemical evolution of the gas. Although the basic principles of chemical
enrichment are well-understood, important aspects, such as the mixing of
metals in the interstellar medium, remain uncertain. Our model of chemical
evolution requires specifying 3 parameters, two of which, however, (the
yield and the fraction of stellar mass ejected by stellar winds and SNe)
are determined by the choice of IMF. The remaining parameter is related to
the mixing of metals. Once the metallicity of the gas is obtained, the
spectrophotometric properties of the stars are calculated using a
population synthesis model which has no free parameters.

These five ingredients: gravitational evolution of halos, gas cooling, star
formation and feedback, chemical evolution, and stellar population
synthesis make up the core of our model of galaxy formation. To predict
observable quantities, however, still requires a model for dust
extinction. We assume that the mass of dust that forms is proportional to
the mass in metals, and derive the extinction in any passband using a
simple model for the distribution of dust in the disk. Our dust model has
one free parameter (the ratio of the scaleheights of dust and stars), but
our results are insensitive to its value. The core galaxy formation model
can now be used to predict a wide range of visible galaxy properties
including basic quantities such as the luminosity function in different
passbands or the distribution of colours, as well as their evolution with
redshift.

To go beyond estimates of luminosity requires the addition of more
physical ingredients into the model. As the model becomes increasingly
complex, it encompasses a fuller range of galaxy properties and this,
inevitably, requires a growing number of physical assumptions and
additional parameters. For example, it is possible to distinguish
between disk and spheroidal stellar configurations by introducing
simple but plausible assumptions. Here we have assumed that cooling
gas settles into a centrifugally supported disk, that the distribution
of halo and gas angular momentum has a particular form (consistent
with results of N-body simulations), and that major mergers or disk
instabilities produce spheroidal stellar systems. Our model requires
one further free parameter to define what a major merger is. With
these assumptions, the scalelengths of disks can be computed without
further free parameters, as can the sizes of spheroids, assuming that
energy is conserved in mergers.
 
In summary, a model of galaxy formation consists of a mixture of
assumptions about the physical processes at work, together with
adjustable parameters that reflect our lack of knowledge about certain
complex astrophysical processes such as star formation.  It is
important to recognize that these parameters are not statistical
variables describing a particular dataset (\eg the faint-end slope of
the luminosity function), but genuine physical quantities that
describe a model for a specific physical process (\eg the conversion
of cold gas into stars). In many instances, the parameters can vary
only over a relatively narrow range of physically sensible values. In
our model, just as in models by other groups, we strive to make the
simplest possible assumptions at every stage and to introduce the
minimum number of adjustable parameters, the majority of which, in
fact, are required to describe the poorly understood processes of star
formation, feedback and the mixing of metals. Given the current
theoretical and observational understanding of these processes, it is
not possible to build a realistic model of galaxy formation which has
substantially fewer parameters than ours.

Although the number of parameters is small, in any case, compared to the
vast array of properties that the model can predict, it is important to
adopt a well-defined, {\it a priori} methodology for fixing their values
and for testing the validity of physical assumptions. In our case, this
strategy is straightforward: we fix the values of all the parameters by
attempting to match a small subset of the local galaxy data which we regard
as the most fundamental. In order of the weight we give them, these are:
the luminosity functions in the B- and K-bands; the relative fractions of
ellipticals, S0s, and spirals; the slope of the Tully-Fisher relation at
faint magnitudes; gas fractions in disks as a function of B-band luminosity; 
the distribution of disk sizes; and the
metallicity of $L_\star$ ellipticals. Once the model parameters have been
set by these comparisons, we test our model predictions against a wide
range of other data, without any further adjustments.

The galaxy formation model presented here differs in several
significant respects from that of \CAFNZ (\shortcite{cafnz}).  That
model was based on a cruder method for generating halo merger trees,
but that alone makes little difference to the resulting galaxy
properties. Similarly, the extensions we have included which allow us
to predict more galaxy properties, such as sizes and mean
metallicities, make little difference to the properties that we
were able previously to predict.  There are three differences which do
lead to significant changes in our results. Firstly, the adoption of a
cosmological model with a low value of $\Omega_0$ reduces the number
of galaxy-sized halos, and this helps to reduce the offset in the
Tully-Fisher relation for models normalized to the B-band luminosity
function (see \cite{hcfn}).  Secondly, the inclusion of dust in the
calculation of luminosities and colours makes a typical galaxy colour
slightly redder and this helps to match the B and K-band luminosity
functions simultaneously. Furthermore, since the luminosities that
enter into the I-band Tully-Fisher relation are partially corrected
for the effects of extinction, the inclusion of dust also helps reduce
the offset between model and data seen in \CAFNZ
(\shortcite{cafnz}). Thirdly, we have adopted a weaker feedback law
for low circular velocity galaxies, since some recent determinations of
the galaxy luminosity function indicate that the very flat faint-end
which we strived hard to match in \CAFNZ (\shortcite{cafnz}) is not a
robust observational constraint and may be affected by survey
selection criteria. This, in turn, implies a higher star formation
rate at redshifts $z\gsim 3$.

Although in this paper we have focussed on the $\Lambda$CDM model, we have
also investigated models with other values of the cosmological
parameters. For reasons of space, we do not present our results in any
detail here, but confine ourselves instead to a few general remarks,
applicable to cluster-normalized models. A standard CDM model (SCDM,
$\Omega_0=1$) still fails to match the Tully-Fisher relation (even using
halo circular velocities) and the luminosity function simultaneously. An
$\Omega_0=1$ model with the same power spectrum shape as $\Lambda$CDM (the
$\tau$CDM model of Jenkins \etal \shortcite{jenkins98}) shares this problem
and, in addition, the smaller amount of small-scale power compared to SCDM
results in a later epoch for the onset of galaxy formation and, thus, in
both a lower star formation rate density and a lower abundance of
Lyman-break galaxies at high redshift. (This problem does not
afflict the $\Lambda$CDM model because the effect due to the shape of the
power spectrum is compensated for by the difference in the linear growth
rate and the higher initial amplitude required by the cluster
normalization.)

There are now in the literature a number of semi-analytic galaxy formation
models of varying sophistication, based on similar principles to ours
(e.g. Avila-Rees \& Firmani \shortcite{a-rf99}, Guiderdoni \etal 
\shortcite{guiderdoni99}, Roukema \shortcite{roukema98}, 
Wu \etal \shortcite{wfn98}, Kauffmann \etal \shortcite{kcdw98},
Somerville \& Primack \shortcite{sp98}).  It is beyond the scope of
this paper to carry out a detailed comparison of all these models, but
we refer the reader to the recent paper by Somerville \& Primack
(\shortcite{sp98}) which has compared some of the different
approaches, including those of the Durham, Munich, and Santa-Cruz
groups. For the most part, results from different models tend to agree
well when similar assumptions are made. In practice, however, it is
not uncommon for different groups to make somewhat different
assumptions and, most importantly, to include different physical
effects in their models. This naturally leads to different results. An
example of the former are the different strategies for constraining
the star formation and feedback laws adopted by Kauffmann
\etal (\shortcite{kcdw98}) and ourselves. We give most weight to the
local B-band luminosity function and do not make any further
adjustments when calculating the zero-point of Tully-Fisher
relation. By contrast, Kauffmann \etal (\shortcite{kcdw98}) give most
weight to the zero-point of the Tully-Fisher relation. Since the
models do not match both these observables perfectly, there is a
difference in the luminosity normalization of the two models that
propagates to other observables. An example of different physical
processes is the treatment of the structure and angular momentum
transport of gas cooling onto galaxies adopted by Wu
\etal (\shortcite{wfn98}). Their model leads to a completely different
mechanism for making ellipticals and spirals to the one operating in
our own model or in that of Kauffmann \etal (\shortcite{kcdw98}).

\section{Conclusions}
\label{sec:conc}

We have presented a new semi-analytic model of galaxy formation which
contains several novel features. It employs a state-of-the art
Monte-Carlo algorithm for calculating the merging evolution of dark
matter halos and it incorporates, for the first time, detailed
prescriptions for calculating the sizes of disks and spheroids. We
used this model to calculate observable properties of galaxies in the
$\Lambda$CDM cosmology ($\Omega_0=0.3$, $\Lambda_0=0.7$, $h=0.7$, and
$\sigma_8=0.93$) and focussed primarily on galaxy properties at the
current epoch, with the following main conclusions:

\begin{itemize}

\item[1.] A pleasing agreement can now be obtained between the model and
observed galaxy luminosity functions in the B-band and the K-band,
over at least 8 magnitudes. This is a non-trivial success.  In the
B-band, the model was tuned to fit the ESP luminosity function of
Zucca \etal (\shortcite{zucca97}), which has a steep faint-end.
Unfortunately, there is still a large uncertainty in the observational
estimate of the number of galaxies fainter than $L_\star$. This is
disappointing because the faint-end slope of the luminosity function
is extremely sensitive to feedback processes which are therefore only
crudely constrained. A flatter faint-end, like that measured, for
example, by Loveday \etal (\shortcite{loveday}) in the Stromlo-APM
survey, could be obtained by increasing the strength of feedback. Our
inability to constrain the feedback model better has a knock-on effect
on our ability to predict the cosmic star formation rate at high
redshift since this too is strongly influenced by feedback.  Dust
extinction has a relatively modest effect, dimming the bright end of
the B-band luminosity function by about 0.5~mag. We showed that
surface brightness effects can be important for faint galaxies, and
this could help explain some of the discordant estimates of the faint
end of the luminosity function.

\item[2.] Our model reproduces both the observed mean galaxy
colours and the spread in colour over a large range of galaxy
luminosity. The stellar mass-to-light ratios of both stellar disks and
ellipticals match the observational values well.  Inclusion of
reddening is important for this comparison, which does not involve
adjusting any model parameters. The mean and scatter in the colours of
galaxies of different morphological types, as measured by blue
bulge-to-total light ratio, are also reproduced well. 

\item[3.] The current cold gas content of galaxies of different luminosity
is related to the efficiency of past and current star formation. Our
adopted star formation model (which is consistent with the
observations analysed by Kennicutt \shortcite{kennicutt98}) leads to
excellent agreement with the observed ratio of cold gas mass to blue
luminosity over 7 magnitudes.

\item[4.] The predicted distribution of disk sizes is sensitive to the
strength of feedback. Our model agrees well with the data,
particularly for bright galaxies. 
%At fainter magnitudes, it gives the
%correct median size, but it predicts a slightly broader distribution
%than is observed.

\item[5.] The more realistic treatment of various properties and processes
(e.g. dark halo and gas density profiles, dust, etc) leads to a better
match to the I-band Tully-Fisher relation than was possible with our
earlier, simpler model.  If, as in most previous work of this kind,
the circular velocity of a galaxy is identified with the circular
velocity at the virial radius of the halo in which it formed, then our
model gives an excellent fit to the zero-point, slope and scatter of
the Tully-Fisher relation. However, in the model, the rotation
velocities of galaxy disks at their half-mass radii are typically 30\%
higher than the circular velocities of the halos at the virial
radius. This results in an offset of $+30$\% in the velocity
zero-point of the Tully-Fisher relation when the calculated disk
velocities are used. It remains unclear whether this disagreement
reflects a fundamental shortcoming of the cold dark matter theory or
whether it is simply a reflection of various physical uncertainties in
the calculation. For example, the derived disk rotation velocity
depends on the assumptions of angular momentum conservation and
adiabatic invariance during the collapse and formation of a galactic
disk. If the collapse were, in fact, clumpy, then angular momentum
would be transferred from the disk to the halo (Frenk \etal
\shortcite{frenk85}, Navarro \& White \shortcite{nw94}, 
Navarro \& Steinmetz \shortcite{ns97}). 
In this case our calculation may have overestimated the amount by
which the inner part of the halo contracts.
This would help reduce the Tully-Fisher offset, but at the same
time the loss of angular momentum from the disk would make the
disks physically smaller and could act in the opposite direction,
compressing the halo more strongly.
These issues are worthy of further investigation, but 
they are best addressed using numerical simulations (see e.g. \cite{ns99}).

\item[6.] Our model calculates chemical evolution, taking into account the
effects of gas loss due to winds and gas accretion due to cooling in a
self-consistent way. The model predicts a trend of increasing
metallicity with luminosity, similar to that observed, for
star-forming gas in disk-dominated galaxies and for stars in
bulge-dominated galaxies. However, the colour-magnitude relation for
ellipticals in clusters is significantly flatter than observed at
bright magnitudes, although the scatter is about right.  Kauffmann \&
Charlot (\shortcite{kc98}) have shown that 
a steeper slope for the colour-magnitude relation can be
obtained by simultaneously increasing the strength of the feedback and
the value of the yield. However, in our $\Lambda$CDM model, a much
stronger feedback than we have assumed would result in disk sizes that
are much too large (because the accretion of gas onto galaxies is
delayed), and would degrade the fit to the Tully-Fisher relation. We 
intend to carry out a more thorough investigation of this conflict in
a later  paper.

\item[7.] Our more sophisticated modelling techniques do not change our
earlier conclusion (\CAFNZ \shortcite{cafnz}, Baugh \etal
\shortcite{bcfl98}) that half of the stars in the universe formed since
$z\lsim 1.5$. However, the relaxation of the requirement for strong
feedback (arising from the fact that we now fit the steep faint-end
slope of the ESP luminosity function rather than the flat slope of the
Stromlo-APM survey) allows a somewhat higher star formation rate at
$z\gsim 3$ than we had predicted previously.  The fraction of baryons
in cold gas has a broad peak at $1<z<2$. The evolution of the mean
metallicity of the hot gas mirrors the growth of stellar mass but, as
noted also by Kauffmann (\shortcite{k96}), the mean metallicity of the
stars and cold gas builds up very rapidly: it is already about one
third of the present value at $z=5$.

\end{itemize}

In summary, the model we have presented is broadly successful in
matching a large range of galaxy properties. There remain, however,
some interesting discrepancies, for example, the Tully-Fisher relation
and the colour-magnitude relation for cluster ellipticals. Although
the discrepancies are relatively small, further work is required to
assess whether they point to incorrect assumptions or to the neglect
of important physical processes in our modelling procedure.

%% file: app_halo.tex
\section[]{Halo Rotation Velocity}
\label{app:halo}

In this appendix we relate the halo rotation velocity, $\vrot$, (assumed
constant) to its spin parameter $\lambdah$.

The total angular momentum of the halo is given by
\begin{equation}
\Jh(\rvir) = \int_0^\rvir \frac{\pi}{4}\vrot \, r^{\prime} \,  \rho(r^\prime)
\, 4\pi r^{\prime 2} dr^\prime. 
\label{eqn:J}
\end{equation}
The total energy of the halo within the virial radius 
is the sum, $ \Eh = \Wh + \Th$, of the potential and kinetic
energies.
The self-binding energy of the material within the virial radius, $\rvir$, is
\begin{eqnarray}
\Wh(\rvir) &=& \frac{1}{2} \int_0^\rvir 
     \phi(r^\prime) \, \rho(r^\prime) \, 4\pi r^{\prime 2} \, dr^\prime \nonumber \\
     &=& - \frac{1} {2 \G}  \int_0^\infty \vert\grad\phi(r^\prime)\vert^2  
     r^{\prime 2} \, dr^\prime \nonumber \\
    &=& - \frac{\G}{2} \left[ \int_0^\rvir \frac{M(r^\prime)^2}{r^{\prime2}} 
\, dr^\prime + \frac{M^2(\rvir)}{\rvir} \right] ,
\label{eqn:W}
\end{eqnarray}
where $\phi(r)$ is the gravitational potential.  Assuming hydrostatic
equilibrium with an isotropic velocity dispersion $\sigma(r)$, the
corresponding kinetic energy of material inside the virial radius can
be expressed as
\begin{equation}
\Th(\rvir) = \int_0^\rvir \frac{3}{2} \sigma^2(r^\prime) \, \rho(r^\prime) \, 
4\pi r^{\prime 2} \, dr^\prime .
\end{equation}
With the same assumptions, the velocity dispersion obeys the Jeans
equation $d(\rho \sigma^2)/dr = -\rho\, G M(r)/r^2$. Provided that
$r^3\rho(r)\sigma^2(r)$ vanishes as $r \to 0$ we obtain,
\begin{eqnarray}
\Th(\rvir) 
    &=&2 \pi \Big[ r_{\rm vir}^3 \, \rho(\rvir) \, \sigma^2({\rvir}) \nonumber \\ 
     &+& \int_0^\rvir \G M(r^\prime) \, \rho(r^\prime) \, r^\prime \, dr^\prime
\Big] .
\label{eqn:T}
\end{eqnarray}

For our standard case of halos with the NFW density profile,
equation~(\ref{eqn:nfw}), we simply integrate the Jeans equation out to
$r = \infty$ to derive $\sigma(r)$, assuming that the NFW profile
and hydrostatic equilibrium apply at all radii (Cole \& Lacey
(\shortcite{cl96}), equation~(2.14)). This is an
approximation, since in principle we should not expect the NFW halo
model to be valid beyond the virial radius, where material is still
infalling. However, the velocity dispersion within the halo derived
in this way using the NFW model has been found to be in good agreement with
numerical simulations (\eg figures 4, 5 and~6 of Cole \& Lacey
\shortcite{cl96}). Note also that truncating the NFW profile at the
virial radius implies that $2\Th(\rvir)+\Wh(\rvir) \ne 0$. If the
integrals in equations (\ref{eqn:W}) and~(\ref{eqn:T}) were extended
to $r=\infty$ then the NFW halo model would exactly satisfy the virial
theorem, but for the truncated model $2T(\rvir)/|W(\rvir)|$ is
slightly greater than unity and varies slowly with the NFW
scalelength, $\anfw$. This behaviour was also found for the \nbody
halos in Cole \& Lacey (\shortcite{cl96}) and our definitions are
fully consistent with the way in which they defined the spin parameter
$\lambdah$.

Inserting the above definitions of 
$\Jh(\rvir)$ and $\Eh(\rvir)$ into
equation~(\ref{eqn:lambda}) for $\lambdah$  defines the
coefficient $A(\anfw)$ in the relation
\begin{equation}
	\vrot =   A(\anfw) \lambdah \Vh ,
\label{eqn:vrot_app}
\end{equation} where $\Vh \equiv (G M /\rvir)^{1/2}$ is the
circular velocity of the halo at the virial radius. For the
limited range $0.03<\anfw<0.4$ the result is well fit by
$A(\anfw) \approx 4.1+1.8\, \anfw^{5/4}$.

For the non-standard case of an isothermal density profile for the
halo (with or without a core radius), we follow a slightly different
approach. If, as above, the kinetic energy, $\Th(\rvir)$, is calculated
by integrating the Jeans equation to derive $\sigma(r)$, then
$2\Th(\rvir)/|\Wh(\rvir)|$ is found to be considerably greater than
unity.  The discrepancy is large because we have extrapolated the halo
density profile beyond the virial radius with a model whose mass does
not rapidly converge.  Thus, for these profiles, we prefer to define
the coefficient, $A$, in equation~(\ref{eqn:vrot_app}) by evaluating the
binding energy, expression~(\ref{eqn:W}), with the appropriate density
profile but then assuming the virial theorem,
$2\Th(\rvir)/|\Wh(\rvir)|=1$, to estimate the kinetic energy and hence
the total energy $\Eh(\rvir)$. This is then identical to the assumption
made in Mo \etal (\shortcite{mmw98}) to define the energy and spin 
parameter. In the range $0.01<a<0.4$ the resulting dependence
is well fit by $A \approx 3.66-0.83\, a$.

%% file: app_sf.tex
\section[]{Star Formation}
\label{app:sf}

The set of coupled differential equations,(\ref{eqn:sff}-\ref{eqn:sfl}),
describing an episode of star formation has the following analytic
solutions. The mass of gas that has cooled and been accreted
in a time $t$ since the start of the timestep is
\begin{equation}
\Delta \macc = \crate \ t .
\end{equation}
The increase in the mass of long lived stars
\begin{eqnarray}
&&\Delta M_{\star} = M^0_{\rm cold} \frac{1-R}{1-R+\beta} \ 
\left[1 -\exp(-t/\tauf)\right]
\nonumber \\
&& - \crate \tauf \frac{1-R}{1-R+\beta} \ 
\left[1 -t/\tauf -\exp(-t/\tauf)\right] 
\end{eqnarray}
where $\tauf=\tau_\star/(1-R+\beta)$ .
In terms of these quantities the changes in the masses of cold and
hot gas are 
\begin{equation}
\Delta M_{\rm cold} = \Delta \macc - \frac{1-R+\beta}{1-R} \Delta M_{\star} 
\end{equation}
and
\begin{equation}
\Delta M_{\rm hot} = -\Delta \macc + \frac{\beta}{1-R} \Delta M_{\star} .
\end{equation}
The corresponding changes in the masses of metals are
\begin{equation}
\Delta M^Z_{\rm cold} = \Delta M^Z_{\rm acc} 
                      + \frac{(1-e)p}{1-R} \Delta M_{\star}
                      - \frac{1-R+\beta}{1-R} \Delta M^Z_{\star}
\end{equation}
\begin{equation}
\Delta M^Z_{\rm hot} =  - \Delta M^Z_{\rm acc} 
                      + \frac{ep}{1-R} \Delta M_{\star}
                      + \frac{\beta}{1-R} \Delta M^Z_{\star},
\end{equation}
where
\begin{equation}
\Delta M^Z_{\rm acc} =  \crate Z_{\rm hot} \ t
\end{equation}
and
\begin{eqnarray}
&&\Delta M^Z_{\star} =  \frac{1-R}{1-R+\beta} 
\Big( M^{Z0}_{\rm cold} \left[1 - \exp(-t/\tauf)\right] 
\nonumber \\
&&- \crate \tauf Z_{\rm hot} \left[1-t/\tauf - \exp(-t/\tauf) \right] 
\nonumber \\
&&+ \frac{(1-e)p}{1-R+\beta} \Big[ M^0_{\rm cold} \left[1-(1+t/\tauf)\exp(-t/\tauf)\right]
\nonumber \\
&& -\crate \tauf \left[2 \negthinspace
- \negthinspace
t/\tauf\negthinspace
 - \negthinspace
(2 \negthinspace
+\negthinspace
t/\tauf)\exp(-t/\tauf)\right]\Big]\Big)
\end{eqnarray}

For the case where there is no supply of cooling gas, $ \crate =0 $,
the above equations show that when $ t \gg \tauf$, 
the mean metallicity of the stars that have formed is
\begin{equation}
Z_\star = Z_{\rm cold}^0 + \frac{(1-e) p}{1-R+\beta} .
\label{eqn:peff}
\end{equation}
Thus our model of star formation and feedback produces an effective yield
$p_{\rm eff} = (1-e) p/(1-R+\beta) $ which, through $\beta$, 
and possibly also $e$, is a function
of the potential well depth of the galaxy disk or bulge in which
the star formation is occuring.

%% file: app_con.tex
\section[]{Adiabatic Contraction of Halo, Disk and Spheroid}
\label{app:contract}

This appendix describes how we use the adiabatic contraction model to
calculate the dynamical equilibrium of the disk, bulge and halo. The
outputs from the calculation are the disk and bulge radii and the halo
density profile after deformation by the gravity of the disk and
spheroid. 

\subsection{Adiabatic Contraction of Halo}
%\label{sec:contract}

To model the effect on the halo density profile of a galaxy condensing at
its centre, we start by assuming that baryons and dark matter have the same
initial density distribution, with total mass profile, $\MHz(r_0)$
(e.g. given by the NFW profile). A fraction $1-\fH$ of the total mass
condenses to form a galaxy at the centre of the halo, leaving a fraction
$\fH$ of the mass still in the halo component. This fraction includes any
baryons that have not yet cooled and also satellite galaxies. For
simplicity, any baryons left in the hot component are assumed to be
distributed like the dark matter.  We now assume that in response to the
gravity of the disk and the bulge, each shell of halo matter adjusts its
radius to conserve its pseudo specific angular momentum $r \VcH(r)$, \ie
that $r \VcH(r)$ is an adiabatic invariant. Thus, 
\begin{equation}
r_0 \VcHz(r_0) = r \VcH(r) \label{eq:halor}
\end{equation}
where $\VcH(r)$ is the total circular velocity at radius $r$,
$r_0$
is the radius of a shell before condensation of the galaxy,
and $r$ is the final radius of the same shell after condensation of
the galaxy.
The initial and final halo masses interior to the shell are related by
\begin{equation}
\MH(r) = f_H \MHz(r_0), \label{eq:halom}
\end{equation}
where $\MH(r)$ is the final mass halo profile. For the purpose of
computing the circular velocity of the halo (averaged over spherical
shells), we treat the mass distribution (including the disk) as
spherical. 
This should be a better approximation for estimating the
gravitational influence of the disk on the halo, than using the
circular velocity due to the disk in the disk plane, which is somewhat
larger. Thus, 
\begin{equation}
\VcH^2(r) = G \left[ \MH(r) + \MD(r) + \MB(r) \right]/r , \label{eq:halov}
\end{equation}
where $\MD(r)$ and $\MB(r)$ are the disk and bulge masses interior to
radius $r$. For consistency, the total masses should be related by
$\MHz = \fH M_{\rm H0} + \MD + \MB$, so that the outer radius of the halo
is unchanged. Combining (\ref{eq:halor}), (\ref{eq:halom})
and (\ref{eq:halov}) we have,
\begin{equation}
r_0 M_{H0}(r_0) = r \left[ f_H M_{\rm H0}(r_0) + M_D(r) + M_B(r)\right], 
\label{eq:halor2}
\end{equation}
which relates the final radius of any halo shell to its initial
radius, once the galaxy disk and bulge profiles, $\MD(r)$ and $\MB(r)$,
are known. The accuracy of this approach has recently been tested in
Navarro \& Steinmetz (\shortcite{ns99b}).

\subsection{Dynamical Equilibrium of Disk and Bulge}
%\label{sec:radii}

Application of the galaxy rules described in
Section~\ref{sec:diskform} give the total mass, $\MD$, and specific
angular momentum, $\jD$, of a galaxy disk, but, in order to use
(\ref{eq:halor2}), we require the complete mass profile, $\MD(r)$. To
obtain this we make the following simplifying assumptions.  First, we
assume that all disks are well-described by an exponential surface
density profile,
\begin{equation}
\Sigma_{\rm D}(r) = {\MD\over 2\pi \hD^2} \, \exp(-r/\hD),
\end{equation}
for which
\begin{equation}
\MD(r) = \MD \left[ 1- (1+r/\hD)\, \exp(-r/\hD) \right] .
\end{equation}
The half-mass radius, $\rD$, is related to the scalelength,
$\hD$, by $\rD = 1.68 \hD$.  
We also assume that the specific angular momentum of the disk is given by
\begin{equation}
\jD = \kD \rD \VcD (\rD) \label{eq:jd},
\end{equation}
where $\VcD(\rD)$ is the circular velocity in the disk plane at the
disk half-mass radius and $\kD$ is a constant. We adopt $\kD=1.19$ as
is appropriate for a flat rotation curve. The value of $\kD$ is only
weakly dependent on the assumed rotation curve. For example, if
$\VcD(r)$ is taken to be the circular velocity of a self-gravitating
exponential disk (Binney \& Tremaine \shortcite{BT87} eq. 2-169), then
$\kD=1.09$, while if $\VcD(r)\propto 1/r^{1/2}$, then $\kD=1.03$.

The radius of the disk is then related to its angular momentum by
\begin{eqnarray}
\jD^2 & = & \kD^2 \rD^2 \VcD^2(\rD) \nonumber \\
& = & \kD^2 G r_D \left[ \fH \MHz(\rDz) + \frac{1}{2}\kh \MD +
\MB(\rD)\right] ,
\label{eq:diskeq}
\end{eqnarray}
where $\rDz$ is the initial radius of the shell whose final radius is
$\rD$.  The factor $\kh$ arises from the disk geometry; if the disk
contribution to $\VcD(r)$ is computed in the spherical approximation,
$\kh=1$, but here instead we calculate the circular velocity in the
midplane of the exponential disk, using equation~(2-169) from Binney
\& Tremaine (\shortcite{BT87}), giving $\kh=1.25$.

The disk half-mass radius, $\rD$, must satisfy this equation and
also equation (\ref{eq:halor2}) evaluated at $\rD$, \ie
\begin{eqnarray}
\rDz \MHz(\rDz) &=& \nonumber \\ 
  &\rD& \negthinspace \negthinspace \negthinspace \negthinspace
\negthinspace \negthinspace
\Big[ \fH \MHz(\rDz)  + \frac{1}{2}\MD + \MB(\rD) \Big] . 
\end{eqnarray}
Using (\ref{eq:diskeq}), this can be written as
\begin{equation}
r_{D0} M_{H0}(r_{D0}) = \frac{j_D^2}{k_D^2 G} - \frac{1}{2}(k_h-1)r_D
M_D. 
\label{eq:rd0}
\end{equation}

To derive the size of the spheroidal component of a galaxy, we assume
that the projected density profile is well described by the de
Vaucouleurs $r^{1/4}$-law (\eg Binney \& Tremaine \shortcite{BT87},
(\S1-13)).  The effective radius, $r_{\rm e}$, of the $r^{1/4}$-law
(the radius that contains half the mass in projection) is related to
the half-mass radius, $\rB$, by $\rB=1.35 r_{\rm e}$.  We define a
pseudo-specific angular momentum for the spheroid:
\begin{equation}
\jB  =  \rB \VcB(\rB) \label{eq:jb},
\end{equation}
where $\VcB(r)$ is the circular velocity at $r$.
This pseudo-specific angular momentum, $\jB$, is assumed to be
conserved, except during galaxy mergers, when its value is determined
by the properties of the merger remnant (see Section~\ref{sec:merge}).

This model leads to the following two equations for the bulge radius,
in analogy to equations (\ref{eq:diskeq}) and (\ref{eq:rd0}) for the
disk radius,
\begin{eqnarray}
\jB^2 & = & \rB^2 \VcB^2(\rB) \nonumber \\
& = & G \rB \left[ \fH \MHz(\rBz) + \MD(\rB) +
\frac{1}{2}\MB\right]
\label{eq:bulgeeq}
\end{eqnarray}
and
\begin{equation}
\rBz \MHz(\rBz) = \frac{\jB^2}{G} .
\label{eq:rb0}
\end{equation}
Comparing this last equation to the analogous equation for the disk,
(\ref{eq:rd0}), we see that the 2nd term on the RHS has vanished. This
results from assuming that the mean effect of the disk on a spherical
shell of the spheroid can be estimated by spherically averaging the
disk. 

To compute the disk and bulge radii we must solve equations
(\ref{eq:diskeq}), (\ref{eq:rd0}), (\ref{eq:bulgeeq})
and~(\ref{eq:rb0}) given the disk and bulge masses, $\MD$ and $\MB$,
the initial halo profile, $\MHz(r_0)$, and specific angular momenta,
$\jD$ and $\jB$. The two pairs of equations are coupled, but with care
they can be solved with a simple iterative scheme.

\subsection{Adiabatic Adjustment following Mergers or Disk
Instabilities}

When a spheroid is formed by a galaxy merger, we first calculate the
radius of the new spheroid $r_{\rm new}$ from (\ref{eq:rm}). We then
compute $\jB = \rB \VcB(\rB)$, with $r_B=r_{\rm new}$
and $\VcB^2(\rB) = G (M_1+M_2)/2 r_{\rm new}$.  Then, using this value
of $\jB$, we calculate the value of $\rB$ given by the adiabatic
contraction model for the disk-bulge-halo equilibrium, and take this
be the true value of $\rB$. If a spheroid is formed via disk
instability, we follow the same procedure, starting from $r_{\rm new}$
given by equation~(\ref{eq:rinstab}), and assuming that the dark
matter mass involved in the initial calculation of $\VcB^2(\rB)$ is
twice that within the initial half-mass radius of the galaxy.